# Data Justice Stories:

# A Repository of Case Studies

Prepared by: The Alan Turing Institute

In collaboration with:

GPAI / THE GLOBAL PARTNERSHIP ON ARTIFICIAL INTELLIGENCE

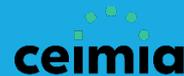 ceimia | International Centre of Expertise in Montreal on Artificial Intelligence

# Data Justice Stories: A Repository of Case Studies

By David Leslie, Morgan Briggs, Antonella Perini, Smera Jayadeva, Cami Rincón, Noopur Raval, Abeba Birhane, Rosamund Powell, Michael Katell, and Mhairi Aitken

*Acknowledgements*

*This research was supported, in part, by a grant from ESRC (ES/T007354/1), Towards Turing 2.0 under the EPSRC Grant EP/W037211/1, and from the public funds that make the Turing's Public Policy Programme possible.*

*The creation of this material would not have been possible without the support and efforts of various partners and collaborators. The authors would like to acknowledge our 12 Policy Pilot Partners—AfroLeadership, CIPESA, CIPIT, WOUGNET, GobLab UAI, ITS Rio, Internet Bolivia, Digital Empowerment Foundation, Digital Natives Academy, Digital Rights Foundation, Open Data China, and EngageMedia—for their extensive contributions and input. The research that each of these partners conducted has contributed so much to the advancement of data justice research and practice and to our understanding of this area. We would like to thank James Wright, Thompson Chengeta, Noopur Raval, and Alicia Boyd, and our Advisory Board members, Nii Narku Quaynor, Araba Sey, Judith Okonkwo, Annette Braunack-Mayer, Mohan Dutta, Maru Mora Villalpando, Salima Bah, Os Keyes, Verónica Achá Alvarez, Oluwatoyin Sanni, and Nushin Isabelle Yazdani whose expertise, wisdom, and lived experiences have provided us with a wide range of insights that proved invaluable throughout this research. We would also like to thank those individuals and communities who engaged with our participatory platform on decidim and whose thoughts and opinions on data justice greatly informed the framing of this project. All of these contributions have demonstrated the pressing need for a relocation of data justice and we hope to have emphasised this throughout our research outputs. Finally, we would like to acknowledge the tireless efforts of our colleagues at the International Centre of Expertise in Montréal and GPAI's Data Governance Working Group. We are grateful, in particular, for the unbending support of Ed Teather, Sophie Fallaha, Jacques Rajotte, and Noémie Gervais from CEIMIA, and for the indefatigable dedication of Alison Gillwald, Dewey Murdick, Jeni Tennison, Maja Bogataj Jančič, and all other members of the Data Governance Working Group.*

**Cite this work as:** Leslie, D., Briggs, M., Perini, A., Jayadeva, S., Rincón, C., Raval, N., Birhane, A., Powell, R., Katell, M.,and Aitken, M. (2022). Data justice stories: a repository of case studies. The Alan Turing Institute in collaboration with The Global Partnership on AI..

# Contents

# Introduction

The idea of "data justice" is of recent academic vintage. It has arisen over the past decade in Anglo-European research institutions as an attempt to bring together a critique of the power dynamics that underlie accelerating trends of datafication with a normative commitment to the principles of social justice—a commitment to the achievement of a society that is equitable, fair, and capable of confronting the root causes of injustice.

However, despite the seeming novelty of such a data justice pedigree, this joining up of the critique of the power imbalances that have shaped the digital and "big data" revolutions with a commitment to social equity and constructive societal transformation has a deeper historical, and more geographically diverse, provenance. As the stories of the data justice initiatives, activism, and advocacy contained in this volume well evidence, *practices of data justice across the globe* have, in fact, largely preceded the elaboration and crystallisation of the idea of data justice in contemporary academic discourse.

If anything, this latter labour of articulating the notion of data justice (from the pioneering work of early scholars like Johnson, Heeks, Dencik, Taylor, and others to our own "Advancing data justice research and practice" project) may be viewed both as an effort to play catch up and as an exercise of historical reflection and conceptual reconstruction: In elaborating the normative shapes and ethical priorities of data justice, academic researchers have, in an important sense, merely unearthed the emancipatory and moral logics that have already resided for decades, and intrinsically, in the actual work that has been done by social justice and digital rights advocates who have focused their transformative practices in the domains of data innovation and information and communication technologies (ICT). In other words, it may be argued that, rather than arising in conference papers and academic lectures, the true roots of the data justice perspective can be found in the real-world struggles that have long been undertaken around the world by digital and data rights activists, advocates, and affected communities to speak truth to power and to realise more equitable forms of access, recognition, representation, participation, and knowledge in data innovation ecosystems.

The data justice stories we present here are intended to introduce the reader to this longer journey of contestation, advocacy, and social transformation. It is a journey that encompasses a generation of activism and traverses the entire planet. It can be witnessed, for instance, in the work of the First Nations Technology Council (founded in 2002) to expand the equitable participation of Indigenous communities in the Canadian information economy and in the advocacy of the Progressive Technology Project (founded in 1998) and the May First Technology Movement (founded in 2005) to bolster the access of activists of colour in the US and Mexico to digital affordances that have been denied to them at the hands of race-based exclusion. It can be seen in the efforts of Derechos Digitales (founded in 2005) to develop, defend, and promote human rights in the Chilean digital environment and in the labours of the Digital Empowerment Foundation (founded in 2002) to actively engage with local communities in India to advance information empowerment and improve digital literacy. It can be witnessed in the work of EngageMedia (founded in 2005) to defend human rights and democracy in the online and offline environments of the Pacific by providing advocacy groups and grassroots organisations with accessible knowledge, documentaries, and resources for effective communication. It can be observed as well in the efforts of the Collaboration on International ICT Policy in East and Southern Africa (founded in 2004) to enhance the capacity of African stakeholders to participate in ICT policymaking processes and in the work of the Centre for Information Technology and Development (founded in 2000) to advance civic participation in data and digital innovation that supports sustainable development and good governance in Nigeria.

What is common to all these transformational stories of data justice advocacy is an active response to the set of challenges raised, amidst rapid digital transformation, to achieving a society that is fair, democratic, open, and able to address the root causes of inequity. And, as the convergence of the emancipatory concerns of these organisations demonstrate, this responsiveness has been oriented to the way that long-term sociohistorical conditions of inequality, discrimination, exploitation, and power asymmetry have been drawn into digitisation, datafication, and the proliferation of data-intensive information technologies.

Notwithstanding this unity of purpose, it is important to note that each of the organisations whose stories are told in this repository have emerged from unique sociocultural histories, ways of knowing, and lived experience. They have likewise had to respond to a range of data justice challenges that have emerged from their own distinctive social, political, and economic circumstances. For this reason, it is critical to acknowledge that the ethical beliefs, values, and practical knowledge upon which they have drawn—and the meanings they have given to words like "justice" and "equity"—are diverse and need to be understood with a recognition of ethical and cultural pluralism and of the importance of context. The stories presented here aim to demonstrate both the value of this plurality of understandings but also the importance of the unity of purpose that connects these differences. Throughout our work in this "Advancing data justice research and practice" project, and in the collaborations that we have been fortunate to have with our Policy Pilot Partners, we have found that the common pursuit of the liberating power of data justice is only strengthened by the many and varied ways that the cause of "data justice" has been taken up around the world.

## Organisation of the repository

We have organised the stories contained in this repository into two groups. The first group, 'Challenges to Data Justice: Stories of Data Discrimination and Inequity", poses the question: What are the sorts of problems and challenges to which data justice practitioners are responding? This section is intended to orient the reader to the range of empirical problems faced by data justice researchers and practitioners the world over. We have provided examples of data practices that have been criticised as posing risks of moral injury and that have been identified as leading to inequitable or discriminatory outcomes. Case studies include a national ID card that serves as a government payment system in Venezuela, a courier service/digital technology company in Colombia, and a digital registry of 'rights, tenancy, and crops' in India.

The second group, 'Transformational Stories of Data Justice: Initiatives, Activism, and Advocacy', poses the questions: What do responses to the range of challenges posed to data justice look like? What are the kinds of transformation that such responses are trying to bring about? The purpose of this section is to orient the reader to the 'moral grammar' intrinsic to boots-on-the-ground struggles for data justice.[1] To be sure, the initiatives and instances of activism and advocacy that are covered are intended to provide insight into the sources of normativity and liberation that inhere *pre-reflectively*[2] in the actual social and historical practices of resistance that organisations undertake. Case studies relating to these transformative data justice practices include a movement for Indigenous data sovereignty in Aotearoa, social mobilisation against violence done to trans people across Eastern Europe and Central Asia, and legal advocacy for public accountability in data use and algorithmic decision-making in the United Kingdom.

---

[1] Honneth, 1993, 1995, 2009
[2] When we refer to sources of normativity and liberation inhering 'pre-reflectively' in data justice practices, we are pointing to the ways that real-world activities can manifest ethical and emancipatory understandings without necessarily depending on (or drawing upon) theoretical frameworks or conceptual articulations. In other words, the ethics manifest in the practices of resistance themselves and indicate sources of normativity that operate within the social world of lived experience rather than being a product of reflection or of theoretical construction.

Ultimately, by bringing the first and second sets of data justice stories into high relief, we hope to provide the reader with two interdependent tools of data justice thinking: First, we aim to provide the reader with the critical leverage needed to discern those distortions and malformations of data justice that manifest in subtle and explicit forms of power, domination, and coercion. Second, we aim to provide the reader with access to the historically effective forms of normativity and ethical insight that have been marshalled by data justice activists and advocates as tools of societal transformation—so that these forms of normativity and insight can be drawn on, in turn, as constructive resources to spur future transformative data justice practices.

Below the title of each story, we have signalled the pillars of data justice research and practice they most directly relate to. These pillars are the guiding priorities of power, equity, access, identity, participation, and knowledge, and intend to orient critical reflection and prompt the generation of constructive insights to advance data justice globally. An extensive description of the pillars can be found [here at the GPAI website](#).

As a final methodological note, these cases have been identified through a variety of sources. We have used our own independent research, our participatory engagement platform, *decidim*, and Cardiff Data Justice Lab's very helpful *Data Harms Record*.[3] To signal where each of the cases have come from, we have created a legend which can be found below. Additionally, we would like to acknowledge the Association for Progressive Communications, APC, which has 62 organisational members across 74 countries. Many of these members carry out data justice and data justice adjacent work, and the membership list was integral to researching these organisations and writing transformational use cases about their work.[4] Lastly, we acknowledge that there are many other ways that data justice practices are being carried out in practice, and so the stories we offer here are meant to serve merely as illustrative, albeit important, examples.

---

[3] Redden et al., 2017
[4] https://www.apc.org/en/members

Data Justice Stories: A Repository of Case Studies    5

## Legend

Turing team

decidim*

Data Harms Record**

Policy Pilot Partner***

## Abbreviations

| | |
|---|---|
| **AGR** | Automated Gender Recognition |
| **APC** | Association for Progressive Communication |
| **CSO** | Civil Society Organisation |
| **DHS** | US Department of Homeland Security |
| **eHAC** | Electronic Health Alert Card |
| **EPA** | US Environmental Protection Agency |
| **FRT** | Facial Recognition Technology |
| **GDPR** | General Data Protection Regulation |
| **ICCPR** | International Covenant on Civil and Political Rights |
| **ICE** | US Immigration and Customs Enforcement |
| **ICT** | Information and Communication Technology |
| **LGBTQI+** | Lesbian, Gay, Bisexual, Transgender, Queer/Questioning, Intersex, and other identities |
| **NGO** | Non-Governmental Organisation |
| **NIIMS** | Kenya's National Integrated Identity Management System |
| **NPO** | Non-Profit Organisation |
| **OCI** | Online Compliance Intervention |
| **OPM** | US Office of Personnel Management |
| **OTT** | Over The Top |
| **TFA** | Technology-Facilitated Abuse |
| **TNC** | Tentative Non-Confirmation |
| **UDHR** | Universal Declaration of Human Rights |
| **UN** | United Nations |
| **UNCRC** | United Nations Convention on the Rights of the Child |
| **VPN** | Virtual Private Network |
| **WHO** | World Health Organisation |

# Challenges to Data Justice: Stories of Data Discrimination and Inequity

## Africa

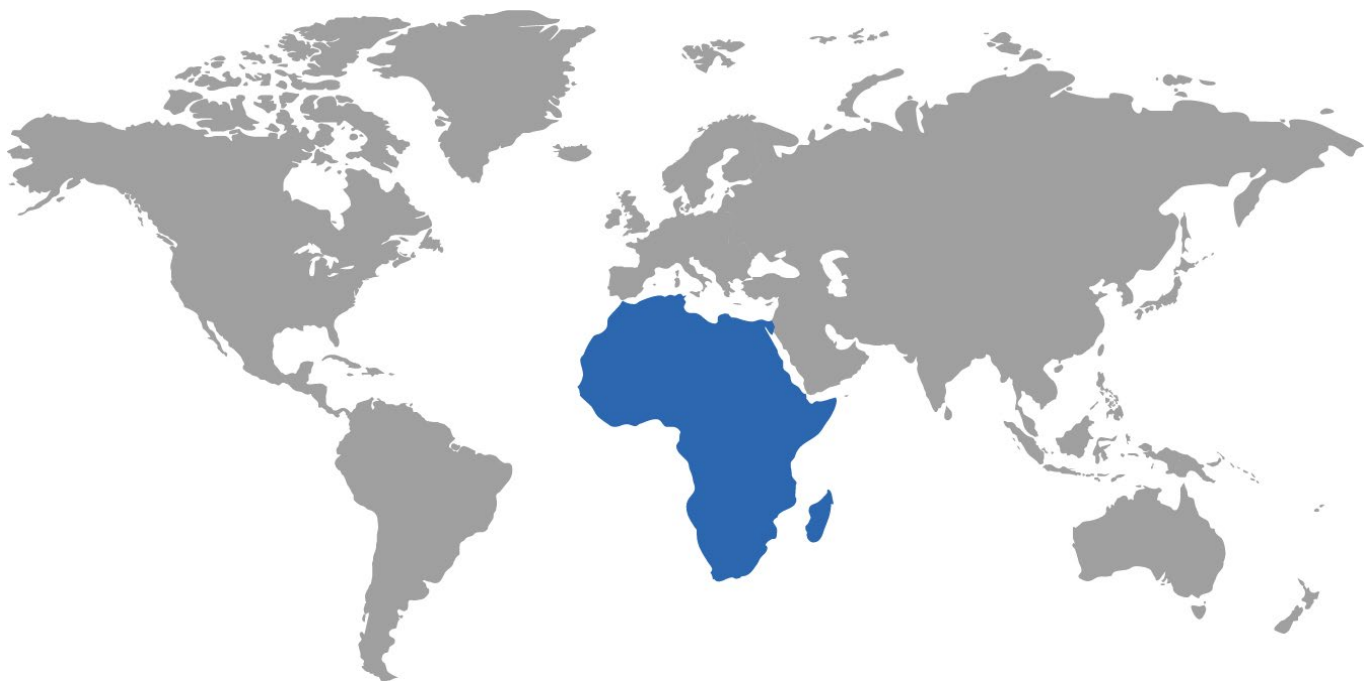

**Legislation Undermining Encryption, Africa***

*Pillars: Power and Participation*

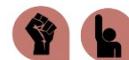

The Collaboration on International ICT Policy for East and Southern Africa (CIPESA) released a policy brief in 2021 which notes that encryption laws in certain countries of Africa raise concerns about privacy and freedom of expression. The use of effective encryption technology is critical to maintaining anonymity and ensuring digital communication is accessible only for the intended recipient. However, certain legislation allows for government authorities to decrypt and access communications to intercept communications related to crime or terrorism.

While the right to anonymity is emphasised by the African Commission on Human and Peoples' Rights, policies undermining encryption has been found to be in violation of the International Covenant on Civil and Political Rights (ICCPR). The brief highlights the concerns about privacy including the prohibition or limitation of encryption usage, compelled assistance by service providers, mandatory SIM card registration, and data localisation. Across the continent, legislation has mandated service providers to limit encryption keys, disclose cryptology means, and even ban Virtual Private Networks (VPNs).

Despite claims that such legislation is aimed at mitigating national security threats, countries including Ethiopia and Rwanda have actively accessed private communications of individuals purportedly connected to dissident movements and activists.

You can read more at:
https://cipesa.org/?wpfb_dl=477

**Challenges to data sharing and open data, Africa**

*Pillars: Identity, Knowledge, and Power*

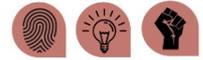

Within the Western scientific community, the calls for data sharing and open data have been gaining momentum. Data sharing is often regarded as a hallmark of transparency, scientific advancement, and generally a good practice that contributes to better science. These Western based initiatives for data sharing and open data also find their way into the African continent. While data sharing and open data indeed open opportunities for better science, a justice-oriented approach to data sharing/open data practices recognises that such practices are highly complex and contentious issues.[5] Without careful consideration and mitigation of concerns such as unjust historical pasts, structural challenges, colonial legacies, and uneven power structures, such initiatives risk exacerbating existing inequalities and benefiting the most powerful stakeholders.[6]

"Parachute research", the practice of Global North researchers absconding with data to their home countries, for example, is one of the concerns that arise with open data in the absence of safeguards that protect data workers (data collectors, data subjects, and other individuals and groups that deal with the task of data management and documentation).[7] Without safeguards in place, non-African researchers not only benefit from African generated data, but they are also afforded the opportunity to narrate African stories—at times contributing to deficit narratives. In a systematic review that examined African authorship proportions in the biomedical literature published between 1980 – 2016 where research was originally done in Africa, scholars found that African researchers are significantly under-represented in the global health community, even when the data originates from Africa.[8]

You can read more at:
https://dl.acm.org/doi/abs/10.1145/3442188.3445897

---

[5] Abebe et al., 2021
[6] Ibid.
[7] Ibid.
[8] Mbaye et al., 2019

# Asia

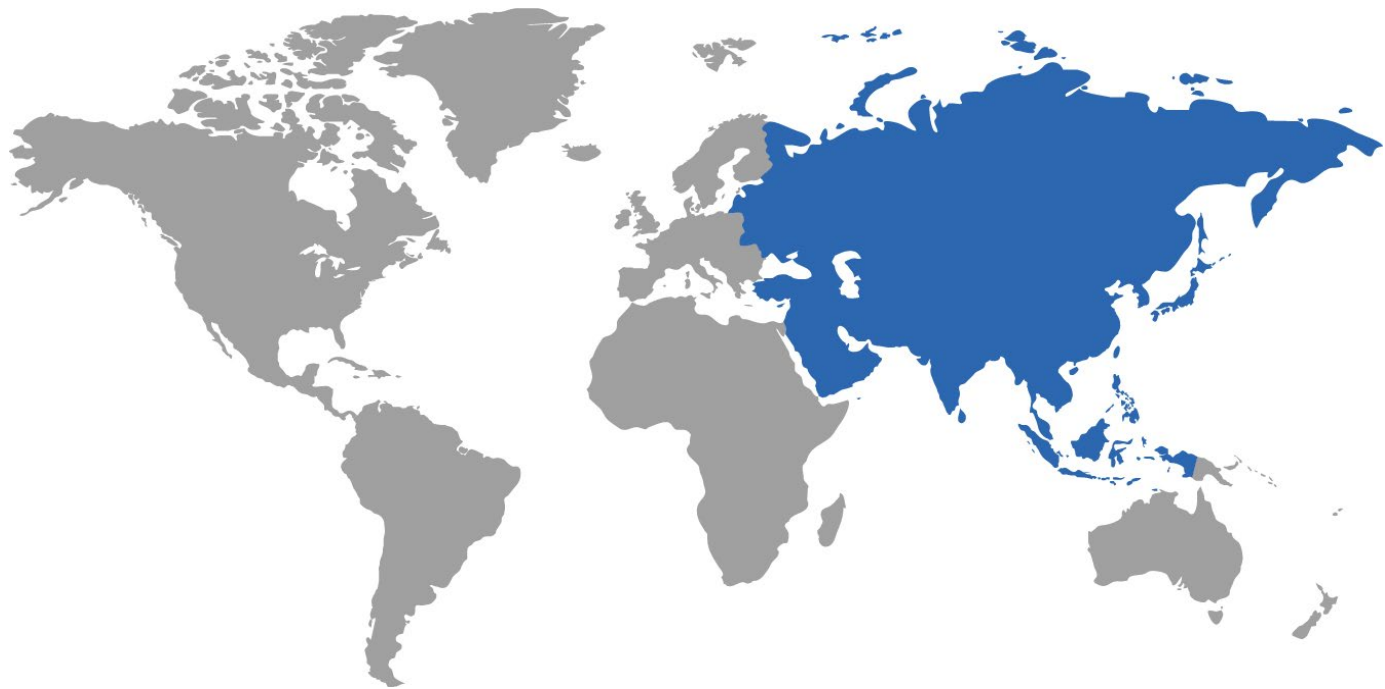

**Algorithms of Food Delivery Apps, China***

*Pillars: Equity, Participation, and Power*

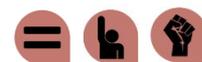

As the demand for food deliveries through apps and digital platforms has increased during the pandemic, there has also been a rise in concerns surrounding the unchecked connection between algorithms and driver safety in China. A report by People Magazine highlighted how food delivery platforms pose threats to driver safety and financial security, owing to limited or delayed responses to algorithmic harms in addition to insubstantial labour protections.[9]

The top food delivery platforms in China—Ele.me and Meituan—have utilised algorithms that prioritise rapid delivery while overlooking factors that influence delivery time such as weather conditions or traffic. By setting hard limits of around 30 minutes for drivers to complete a delivery—including the preparation time for restaurants—the app effectively penalises drivers for obstacles beyond their control. In some instances, even when a delivery is completed on time, a glitch can result in fines or penalties. Keeping in mind the dominance of gig-workers in the industry, the prevalence of algorithmic inequities, inefficiencies and faults further expose the limited safeguards that are present for drivers in a data-driven and digitally determined market. Moreover, platforms have been criticised for not protecting labour rights, not providing work and food safety, and not providing access to insurance. While Ele.me, owned by Alibaba, and Meituan have indicated a move to address these concerns, the current landscape is evidently subjecting gig workers to labour conditions that are shaped by technological parameters beyond their control.

You can read more at:
https://read01.com/AzAkPkm.html#.YfvPD_XP2w1
https://mp.weixin.qq.com/s/Mes1RqIOdp48CMw4pXTwXw
https://www.reuters.com/business/china-market-regulator-boosts-food-delivery-worker-protections-2021-07-26/

---

[9] Lai, 2020

**Bhoomi: Land Records Management System, India**
*Pillars: Access, Equity, Identity, Knowledge, and Power*

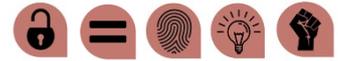

Bhoomi is a digital registry of 'rights, tenancy, and crops' produced by the state government of Karnataka, India. The registry is part of an open data effort to increase both the uniformity and availability of official records including land-ownership records. While initially developed by the department of revenue for taxation purposes, the information can be viewed publicly online and at internet kiosks. It is reportedly used extensively by real estate developers.

Critics have argued that the Bhoomi registry has disenfranchised members of the Dalit caste, considered to be the lowest class in the social hierarchy system of Hinduism,[10] whose claims are often not documented in official records but are well supported by other means. Research has found that Dalits face discrimination in Indian society and often live in poverty.[11] They nevertheless have longstanding claims to land. However, the informal and historical knowledge that supports these claims cannot be easily accommodated in the flattened landscape of a relational database, such as the Bhoomi registry, and so are more easily dismissed or overruled.

Furthermore, Bhoomi may be an example of 'open data under conditions of unequal capabilities'.[12] Like many digital resources, the Bhoomi registry is more likely to be accessible to people with computational and interpretive skills, who are also more likely to hold greater social and political power in society. In Karnataka, there have been mass evictions of Dalits and others in urban slums that were deemed desirable for redevelopment and in which the ability to present conflicting ownership claims based in local knowledge was diminished by the Bhoomi registry.

You can read more at:
https://doi.org/10.1007/s10676-014-9351-8 (pages 263-274)

---

[10] Sankaran et al., 2017
[11] Das & Mehta, 2012
[12] Johnson, 2018, p. 30.

**Social Exclusion and Cybersecurity Issues with Aadhaar, India****
*Pillars: Access, Equity, and Power*

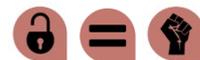

Regarded as one of the world's most ambitious yet controversial digital identity system, the Aadhaar database in India, covering around 89% of the population,[13] has disproportionately impacted India's poor. The system was established to detect and prevent welfare fraud, but for millions reliant on the public distribution system, access to the documentation and processes necessary to link government-issued ration cards to the Aadhaar system is arduous at best and discriminatory at worst. Many must travel long distances away from their residence in India's vast rural expanse, usually by foot or public transport, to places processing Aadhaar applications. Critics highlight that beyond the corruption-riddled procedure, oftentimes the failure of computer systems has required many individuals to undertake the long journey again. In an illustrative case, an individual was forced to reapply as the system did not accept a three-digit age.[14] Replacing forms of ID for food subsidies or financial transactions with the Aadhaar card has been found to culminate in numerous instances of death from starvation and poverty.[15]

The use of a digital identity system within an environment where internet connections are faulty, digital and linguistic literacy is low, and data protection legislation is weak has caused ostensibly insurmountable challenges for vulnerable groups. Sociohistorical marginalisation has entrenched certain classes and castes in poverty while financial, agrarian, and political crises further endanger communities on the brink of the poverty line. Informal systems of identification and limited government-issued documentation has highlighted the challenges of historically discriminated groups. Furthermore, it was reported that access to Aadhaar database and a software to print existing Aadhaar cards or generate fake ones were sold by an anonymous group on WhatsApp for less than £6. Before the vulnerability was fixed, more than 100,000 people had illegal access to the Aadhaar database. Combined with the nature of data and cross-platform use, the security of Aadhaar has been contested as data on millions were exposed.

You can read more at:
https://www.bbc.co.uk/news/world-asia-india-43426158
https://www.newyorker.com/news/dispatch/how-indias-welfare-revolution-is-starving-citizens
https://www.tribuneindia.com/news/nation/rs-500-10-minutes-and-you-have-access-to-billion-aadhaar-details/523361.html

---

[13] Tiwari, 2018
[14] Biswas, 2018
[15] Bhardwaj, 2018

### COVID-19 App Data Breach, Indonesia*

*Pillars: Identity and Power*

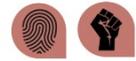

To track and prevent the spread of COVID-19, the Government of Indonesia used an electronic Health Alert Card (eHAC) track-and-trace app. In 2021, it was reported that the personal data of over a billion individuals collected and stored on the app was leaked online. Independent security researchers maintain that the breach was due to lack of privacy protocols.[16] The breach points to a larger threat of weak cybersecurity infrastructure and the potential for wrongdoers to exacerbate existing public distrust and catalyse misinformation. Moreover, similar vulnerabilities have been identified in tracing apps across the globe.

In the case of eHAC, sensitive personal data, from contact details and travel history to hospital identification (ID) numbers, were vulnerable to exploitation alongside data from 226 hospitals and clinics across the country. Unsecured and unencrypted platforms like eHAC pose a significant threat to individuals' rights to data protection and cybersecurity, particularly when the policies to implement tracking tools for medical data instituted mandatory use of such apps during the global pandemic. Reports also note that the replacement app, PeduliLindungi, was found not only to lack clarity both on the use of centralised servers and on the duration of data storage but also to fail to provide clear limitations on purpose and access.[17]

You can read more at:
https://www.reuters.com/technology/indonesia-probes-suspected-data-breach-covid-19-app-2021-08-31/
https://www.zdnet.com/article/passport-info-and-healthcare-data-leaked-from-indonesias-covid-19-test-and-trace-app-for-travellers/
https://www.nortonrosefulbright.com/-/media/files/nrf/nrfweb/contact-tracing/indonesia-contact-tracing.pdf?revision=1c30d2b8-e883-4878-beee-f6fc5a6eb7eb

---

[16] Rotem & Locar, 2021
[17] Norton Rose Fulbright, 2020

# Americas

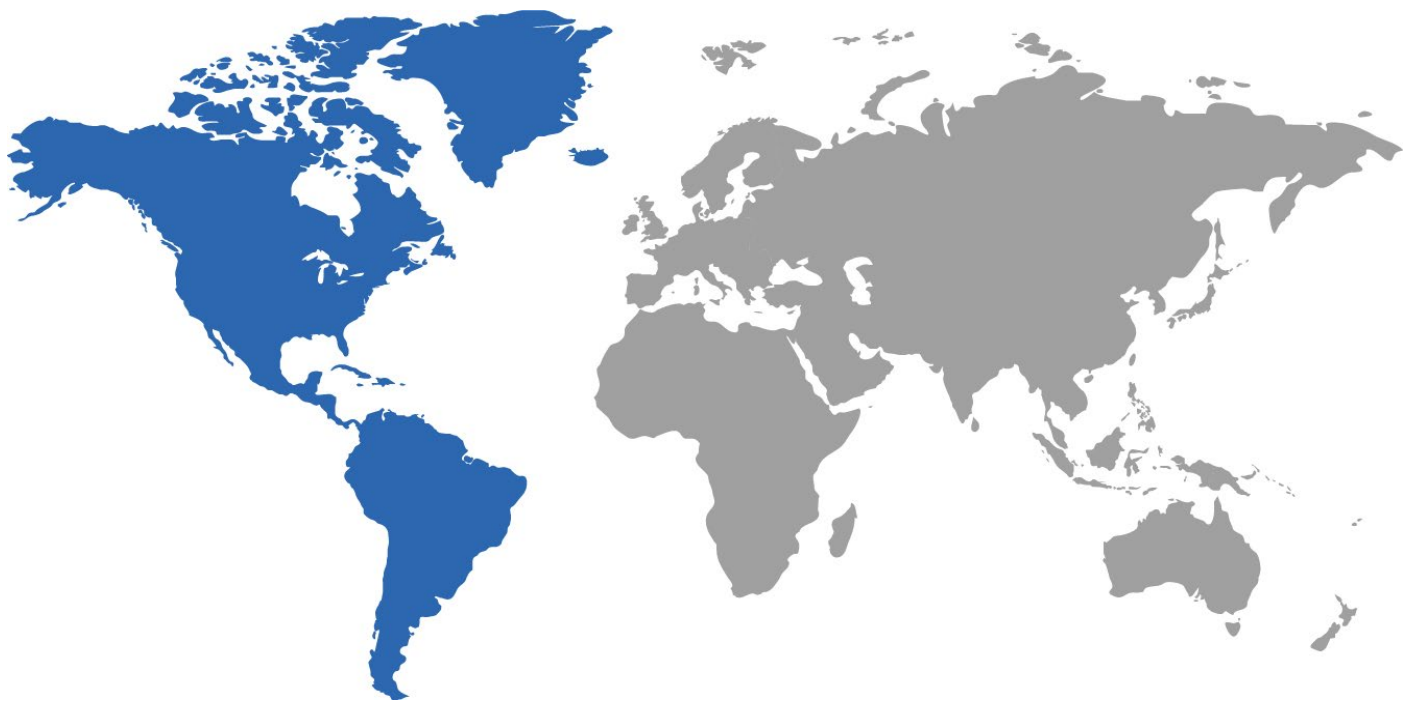

**Rappi: Gig work, Colombia**

*Pillars: Equity and Power*

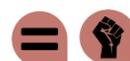

Rappi is a Colombian mobile application offering on-demand courier services including restaurant, grocery, and drugstore delivery. Over the past six years, Rappi has developed into one of the most valued digital technology companies in Latin America and is owned by international shareholders. Rappi couriers are considered to be independent contractors within the platform economy and as such are not protected by regulatory frameworks for employee relations, which govern labour and social protections.

Rappi couriers have protested their working conditions, stressing that they earn wages below the Colombian minimum wage, experience high rates of work-related accidents and health issues, and are subject to a strenuous points system that governs which drivers are entitled to work in high demand zones.[18] These complaints have been noted as examples of greater patterns of labour exploitation via digital platforms, especially given the fact that most couriers are migrants who are not entitled to employment-based social security in Colombia and rely on their courier work as a main source of income.[19] In this context, there exist power asymmetries between shareholders and workers and job creation is noted to be largely concentrated in low status and low standard digital service work.[20]

You can read more at:

https://www.wits.ac.za/media/wits-university/faculties-and-schools/commerce-law-and-management/research-entities/scis/documents/7%20Velez%20Not%20a%20fairy%20tale%20Colombia.pdf
https://www.reuters.com/article/us-rappi-colombia-idUSKCN25B0ZV

---

[18] Griffin, 2020
[19] Jaramillo Jassir, 2020
[20] Velez Osorio, 2020

Data Justice Stories: A Repository of Case Studies    13

**Algorithmic Profiling in Lending Practices, United States****

*Pillars: Equity and Power*

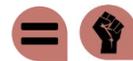

Corporate and financial surveillance has extracted and amassed large volumes of data that may be used to economically and racially profile individuals through discriminatory pricing practices. Not only have advertisers made use of behavioural profiling to provide differential pricing for goods, but research also suggests that vulnerable populations are subject to predatory lending practices.[21] However, variations in lending practices that lead to disparate treatment of social groups predate the use of data-driven technology. For instance, Black and Hispanic borrowers have, in the past, been offered discriminatory sub-prime loans, rates, or excessive fees compared to those offered to equally qualified white borrowers.

Regarded as 'reverse redlining', predatory subprime mortgage loans are targeted towards individuals identified as vulnerable, reversing the practice of redlining wherein goods are not made available to minority neighbourhoods.[22] Non-white, non-wealthy, and less-tech savvy individuals and communities are often disadvantaged through current credit-scoring systems. Subsequent late payments are accepted as objective indicators for credit scoring, despite originating from discriminatory practices. This then feeds forward into a loop that further impacts credit scores.

The equity of impact is dependent on the prevalence of bias in the data and software. While bias in lending is illegal, proxy datasets—available in abundance—and biased correlations can be used in discriminatory practices. Similarly, although exploitative loans are banned or restricted, some lenders continue to operate through online platforms which have actively solicited their advertising. Moreover, when machine learning models lack interpretability, the reasons behind lending practices cannot adequately be presented.

You can read more at:
https://www.ftc.gov/system/files/documents/public_comments/2014/08/00015-92370.pdf
https://www.brookings.edu/research/reducing-bias-in-ai-based-financial-services/
https://ssrn.com/abstract=2376209

---

[21] Bartlett et al., 2019
[22] Gilman, 2020

**Differential Pricing, United States****

*Pillars: Equity, Access, Identity, and Power*

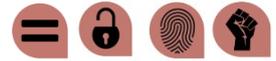

Large volumes of data collated through multiple sources of commercial surveillance have allowed corporate entities to model new forms of pricing that often vary across regions and communities. While forms of differential pricing have provided benefits for select consumers and subsequent profits for businesses, such models have been found to target low-income and protected classes within society. Importantly, consumers are often unaware of the use of data on online behaviour and transaction as well as price differentials.

In many cases of differential pricing, the variation is dependent on the ZIP codes of consumers. For SAT[23] materials and tutorials, the highest prices were observed in New York and California. While the Princeton Review maintained that the pricing is determined by 'competitive attributes', the model was found to disproportionately present prices to neighbourhoods with a greater density of Americans of Asian descent despite a much lower median income.[24] A journal examination found that Staples, the stationery retail chain, had used ZIP codes to establish higher prices for individuals in lower-income neighbourhoods.[25] In a similar vein, auto insurers have historically charged higher premiums for individuals residing in regions with a greater density of minority and protected classes even when risks are found to be the same for non-minority and white drivers.[26] Although laws have been instituted to prevent discrimination, the prevalence of differential pricing is essentially tantamount to redlining, or the practice of denying goods and services to minority neighbourhoods.

You can read more at:
https://www.propublica.org/article/minority-neighborhoods-higher-car-insurance-premiums-white-areas-same-risk
https://www.propublica.org/article/asians-nearly-twice-as-likely-to-get-higher-price-from-princeton-review

---

[23] The SAT is a standardised test commonly used for college admissions in the United States.
[24] Angwin et al., 2015
[25] Valentino-DeVries et al., 2012
[26] Angwin et al., 2017

**E-Verify, United States**\*\*

*Pillars: Equity, Identity, Access, and Power*

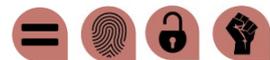

Employers in the US have turned to E-Verify, an online database run by the US Department of Homeland Security (DHS), to assist in evaluating whether an employee is eligible to work in the US. Under the veil of objective verification, the website has led to pre-employment discrimination or even the loss of jobs predominantly and disproportionately amongst minority groups, including 'lawful permanent residents and other authorized immigrants'.[27] It is argued that the root issue here is the continued use of an automated system noted for numerous technical and operational issues.[28]

The E-Verify website captures data from local, state, and federal agencies to cross-check information from an individual's I-9 form to establish whether they have the right to work. When the system identifies an inconsistency, a Tentative Non-Confirmation (TNC) is produced, and the relevant party may face an expensive and time-consuming appeals process to gain the necessary work authorisation. Some have observed that, as the automated system is trained on larges datasets, dominant modes of spelling and linguistic features are accepted as the norm while non-Americanised names are potentially flagged without adequate reason. The American Civil Liberties Union (ACLU) noted that non-US English names and spellings are 20 times more likely to be flagged.[29] Additionally, if a name change has not been updated on the source databases, the website will issue a TNC. For low-income and legal migrants who have limited to no knowledge of the bureaucratic procedures, correcting errors in their personal information is an arduous task. Additionally, when employers input erroneous information and symbols, such as in the case of a double space between names, a TNC or termination notice may be produced.

You can read more at:
http://www.datacivilrights.org/pubs/2014-1030/Employment.pdf
https://bigdata.fairness.io/wp-content/uploads/2015/04/2015-04-20-Civil-Rights-Big-Data-and-Our-Algorithmic-Future-v1.2.pdf

---

[27] Robinson et al., 2014, p.12
[28] Rosenblat et al., 2014
[29] ACLU, 2013

### Facebook's "Real Name" Policy, United States**
*Pillars: Access, Identity, and Power*

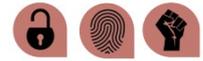

Facebook's former "Real Name" policy to prevent fake or anonymous accounts prevented numerous groups from creating profiles as their names did not fit in within the narrow standards on "real names". Native Americans, drag queens, Irish, and Tamil individuals were among those discriminated against in a policy that required government-issued IDs to continue using their accounts or create new ones.[30]

Following protests in 2014, Facebook implemented updates to their policy wherein individuals could provide other forms of identity proof that match the name by which they are generally identified in place of government ID. However, even with the new policy, many groups still face discrimination. For instance, those individuals who use an alias for personal reasons—including protection—as in the case of LGBTQ+ folk who choose not to reveal their identity on the social media platform are unable to successfully verify their identity. Currently, other accepted IDs include employment verification, diplomas, and student cards. Individuals can choose to explain their exceptional circumstances, but nevertheless, these mechanisms, while aiming to prevent abuse online, require the divulgence of additional personal information in the face of potential risks stemming from government surveillance or data sharing. Furthermore, the policy is enacted within a general environment of large volumes of reports, delayed responses, and a lack of actionable recourse and appeals processes.

You can read more at:
https://www.eff.org/deeplinks/2015/12/changes-facebooks-real-names-policy-still-dont-fix-problem
https://www.theguardian.com/technology/2015/feb/16/facebook-real-name-policy-suspends-native-americans?source=post_elevate_sequence_page

---

[30] Sampath, 2015; Holpuch, 2015

**Home Care Hours and Costs, United States**\*\*

*Pillars: Equity, Identity, and Access*

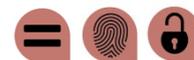

The introduction of algorithmic decision-making systems in healthcare has adversely affected those in need of support services in numerous states across the US, particularly low-income seniors and people with disabilities. Earlier reliance on periodic reviews and assessments of home care needs conducted by human assessors is now being replaced by computer programmes under the guise of "objectivity" and "efficiency". Yet critics highlight that, in multiple cases, the algorithms have severely restricted home care hours leading to a plethora of issues from body sores and skipping meals to other trade-offs related to curbs on personal care.[31]

These kinds of home care algorithms can categorise patients based on predicted degrees of need that determine corresponding hours of care. The points that comprise these scales of need can be calculated based on electronic medical records that may not account for ailments and challenges like diabetes, disabilities, or other health indicators associated with socioeconomic deprivation. Moreover, details on the functioning of these algorithm and associated outputs can be withheld from public scrutiny in certain cases. It should be noted that the developer of one such algorithm, Brant Fries of the University of Michigan, emphasises that the programme is meant for equitable resource allocation rather than mandating necessary hours of care.

It is argued that such developments fit within a larger trend observed in the US where the fraught outcomes of low healthcare budgets are exacerbated by the introduction of disruptive technologies, often without notice or explanation of the algorithm's decision-making process. Beyond this, benefit-allocation is also being automated. According to some critical scholars, this is presenting new issues as the data utilised in these cases is often riddled with racial and economic biases (owing to historical discrimination, restriction, and exclusion of certain communities from healthcare).[32] Notwithstanding judicial intervention in some cases,[33] the challenges of entirely replacing human assessors with algorithms in healthcare persist.

You can read more at:
https://www.theverge.com/2018/3/21/17144260/healthcare-medicaid-algorithm-arkansas-cerebral-palsy
https://themarkup.org/ask-the-markup/2020/03/03/healthcare-algorithms-robot-medicine
https://www.aclu.org/blog/privacy-technology/pitfalls-artificial-intelligence-decisionmaking-highlighted-idaho-aclu-case

---

[31] Lecher, 2018
[32] To learn more on the intersection of racial bias and healthcare algorithms, read Obermeyer et al., 2019
[33] Stanley, 2017

**Criminal sentencing algorithms and predictive policing, United States****

*Pillars: Equity, Identity, Participation, and Power*

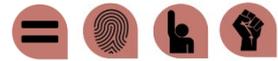

Predictive policing involves various applications that are utilised for the stated purposes of preventing future crime, uncovering past crimes, as well as informing police interventions. In their 2016 article *Machine Bias*, authors from ProPublica, an investigative journaling company, uncovered discriminatory patterns in software being used to predict defendants' likelihood of committing future crimes. ProPublica tested the algorithm being used in this prediction against 7,000 risk scores assigned to people arrested in Broward County, Florida. They discovered that 'only 20 percent of the people predicted to commit violent crimes actually went on to do so',[34] demonstrating the lack of reliability of these predicted risk scores. In addition to lack of reliability, the algorithm illustrated significant racial disparities, wrongly labelling Black defendants as future criminals at almost two times the rate of white defendants, along with simultaneously assigning lower risk scores for white defendants on average. Northpointe, the company that created the algorithm which generated the risk scores in Florida, denies ProPublica's methodology and the efficacy of their analysis.

Instances of predictive policing show evidence of a range of challenges to data justice, including violations of privacy, perpetrations of discriminatory behaviour, inequitable targeting of specific groups, and a lack of transparency on behalf of the organisations using these algorithms. Other examples of predictive policing occurring are outlined in Cathy O'Neil's 2016 book *Weapons of Math Destruction*[35] including Pennsylvania police's use of PredPol, Philadelphia police's use of Hunchlab (now ShotSpotter), and Compstat used by the New York Police Department (NYCPD).

You can read more at:
https://www.propublica.org/article/machine-bias-risk-assessments-in-criminal-sentencing
https://www.predpol.com
https://www.shotspotter.com/law-enforcement/patrol-management/

---

[34] Angwin et al., 2016
[35] O'Neil, 2016, p. 84-87

Data Justice Stories: A Repository of Case Studies    19

**Privacy and Poverty, United States**\*\*

*Pillars: Access, Equity, Identity, and Power*

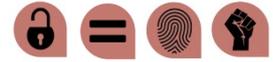

Research conducted by Madden et al. (2017) highlights how certain groups of low-income US adults are subject to varying degrees of surveillance and challenges to privacy. The results revealed the complex and multifaceted nature of surveillance mechanisms against the poor. Such mechanisms have either perpetuated patterns of historic marginalisation or served to further entrench disadvantaged groups.

Historically, the US has enforced various laws and policies to 'oversee' or monitor low-income communities. From Colonial America to the modern welfare state, surveillance tools have been used to keep watch on and influence the 'undeserving poor', or able-bodied individuals from socioeconomically deprived groups who possess the capacity to work. To access welfare and public assistance schemes, low-income individuals and households are subject to an array of intrusive verification policies including drug testing and invasive questioning about private relationships.[36] Low-wage employers utilise surveillance methods from closed-circuit television (CCTV) to monitoring breaks, calls, and emails.[37] Furthermore, there exists digital divides wherein the poor not only lack the capacity to access technology but are also less likely to be digitally literate or make use of more robust privacy protection mechanisms. Data gathered through such surveillance can then be shared across government and commercial entities which, in turn, potentially leads to other forms of discriminatory activity. It has been broadly argued that this wider class differential system of surveillance has caused psychological and physical damage to low-income individuals and households.[38]

You can read more at:
https://ssrn.com/abstract=2930247

---

[36] Gustafson, 2011
[37] Zickuhr, 2021
[38] Jacobson, 2009

**Social Sorting: Data Brokers and Credit Card Companies, United States****

*Pillars: Equity, Access, and Power*

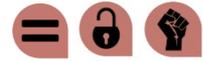

Many different kinds of datasets are accessed, combined, and sold by data brokers, or information brokers such as Acxiom, CoreLogic, and Epsilon, to third parties who use the information to 'socially sort' and target individuals with predatory practices.[39] Personal, sensitive, and biometric data are gathered from an array of sources including transaction history, public records, and bodily and geospatial movements recorded on smart phones. These are then categorised in ways that guide predatory practices that target certain groups of individuals. When vulnerable individuals and communities are subject to social sorting through algorithmic profiling, the outcomes have been found to raise concerns about a range of rights and freedoms—including privacy and data protection rights, rights to dignity and autonomy, and freedoms of expression and assembly—and the lack of legislation to prevent such potentially harmful practices.

Research by the World Privacy Forum alongside congressional testimony by Pam Dixon brought the scale of the data broker industry to US government's attention in 2013. Data brokers had been found to release lists of personal information grouped by characteristics. Data is essentially gathered from marketing sources and released lists have included the location of domestic abuse shelters, victims of rape, disease history, and so forth.

In a similar vein, credit card companies use personal information to target individuals through 'behavioural analysis' wherein credit scores and limits are determined by the linking of individual purchasing history to general purchase trends within recorded geospatial and temporal behavioural patterns, thereby leading to instances of 'creditworthiness by association'.[40]

There are limited legal frameworks that mandate data brokers to provide consumers a choice to opt-out. Not only are consumers generally unaware of the sale of their personal data, but most avenues to opt-out are nearly impossible to access. Currently, no federal legislation has yet to be introduced in the US to regulate data brokers. While the Federal Trade Commission (FTC) has charged data brokers for illegally selling financial information and payday loan applicants' personal data, it is widely maintained that the industry continues to function in a legal vacuum.

You can read more at:
https://www.worldprivacyforum.org/2013/12/testimony-what-information-do-data-brokers-have-on-consumers/
https://www.wired.com/story/opinion-data-brokers-are-a-threat-to-democracy/
https://openyls.law.yale.edu/bitstream/handle/20.500.13051/7808/Hurley_Mikella.pdf?sequence=2&isAllowed=y

---

[39] Sherman, 2021
[40] Hurley & Adebayo, 2016

### Welfare Automation, Indiana, United States**

*Pillars: Access, Equity, and Identity*

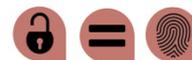

The 'modernisation' of welfare programme applications by moving them online in the US state of Indiana has been subject to critiques that they have been mired in faulty technology leading to many people losing necessary access to state benefits.

The issues are multipronged. First, where documentation is unreadable, improperly indexed, or even lost in application processing, this results in rejection. Second, as telephone interviews have replaced in-person visits to welfare offices, those who miss their interview call—often due to prohibitive costs of cellular plans, running out of call minutes, or the inability to answer without support—face the denial of their application. Third, the system does not account for individuals with sensory challenges as applications cannot be translated into Braille and calls must be answered regardless of conditions like deafness. While proponents claim that technology can serve to alleviate fraud or waste, the introduction of the system has caused a substantial increase in denials that disparately impact marginalised groups despite the increasing volume of applications.

You can read more at:
https://www.thenation.com/article/archive/want-cut-welfare-theres-app/

### Homeland Card: Identity Document, Venezuela

*Pillars: Access, Equity, Identity, and Power*

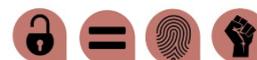

Venezuela's Homeland Card (Carnet de La Patria) is a national ID card that serves as a digital payment system and is used by the Venezuelan government to provide access to food, healthcare, pensions, and other social benefits to citizens. Citizens are incentivised to enrol in the card programme via rewards including bonuses, with over half of the population being enrolled in the card programme.

The Homeland Card has been critiqued by activists who stress that it is being used as a surveillance tool aided by digital telecommunication corporations.[41] It has been reported that the card is linked to databases storing cardholders' personal information including medical history, social media presence, residential addresses, and political party membership.[42] The use of the Homeland card in the context of a humanitarian emergency where most citizens depend on benefits for survival has raised concerns for opponents that citizens' information is being used to exclude individuals from accessing vital services based on their behaviours and political affiliations and that the ID system is serving as a method of control and electoral coercion.[43]

You can read more at:
https://www.reuters.com/investigates/special-report/venezuela-zte/

---

[41] Berwick, 2018
[42] Ibid.
[43] Ibid.

# Oceania

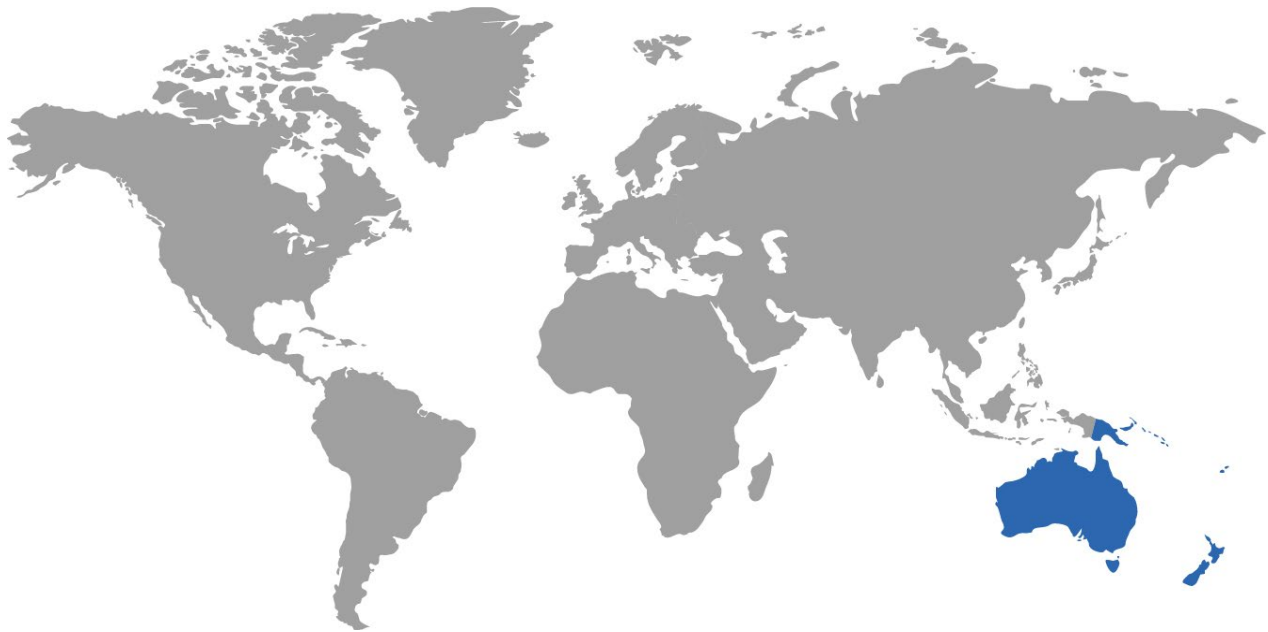

**Centrelink's Automated Debt Recovery System, Australia\*\***

*Pillars: Access and Equity*

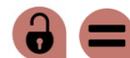

Since 2016, Services Australia has employed an automated debt assessment and recovery scheme called Online Compliance Intervention (OCI) through Centrelink compliance. The objective of the scheme is to automate the formerly manual system of calculating and issuing debt notices to welfare recipients through a process of matching Centrelink data and records with averaged data on income from other agencies in Australia. 470,000 debts, amounting to over A$1billion, were accrued through the automated process leading to the filing of a class-action lawsuit. Similarly, incidents of 'zombie debt', or notices for expired debts, have been identified in the United States through automated procedures.[44]

The way OCI removed in-person or face-to-face services led the system to disadvantage welfare recipients who did not have access to the internet. It also led to the elimination of decisions based upon discretionary and compassionate grounds.[45] In some cases, letters were not received which led to subsequent assumptions of debt even where letters had been posted to the wrong address or dated back many years.[46] Debts were not only found to be non-existent, but miscalculations meant that some people were charged more than what was owed. Other recipients continued to make payments despite contesting the charges.[47] Vulnerable groups, including those with histories of enduring abuse or mental illness, were subject to debt payments which, in some cases, led to deaths.[48] In 2021, many debts 'vanished' from Centrelink's database during ongoing government action to repay the debts.[49]

You can read more at:
https://www.sacoss.org.au/sites/default/files/SACOSS%20Fact%20Sheet%20-%20Centrelink%20Robo-Debt%20campaign%20and%20background%20information.pdfhttps://undocs.org/A/74/493

---

[44] Eubanks, 2019
[45] Henrique-Gomes, 2019
[46] South Australian Council of Social Service, 2017
[47] Belot, 2017
[48] Medhora, 2019
[49] Henrique-Gomez, 2021

# Europe

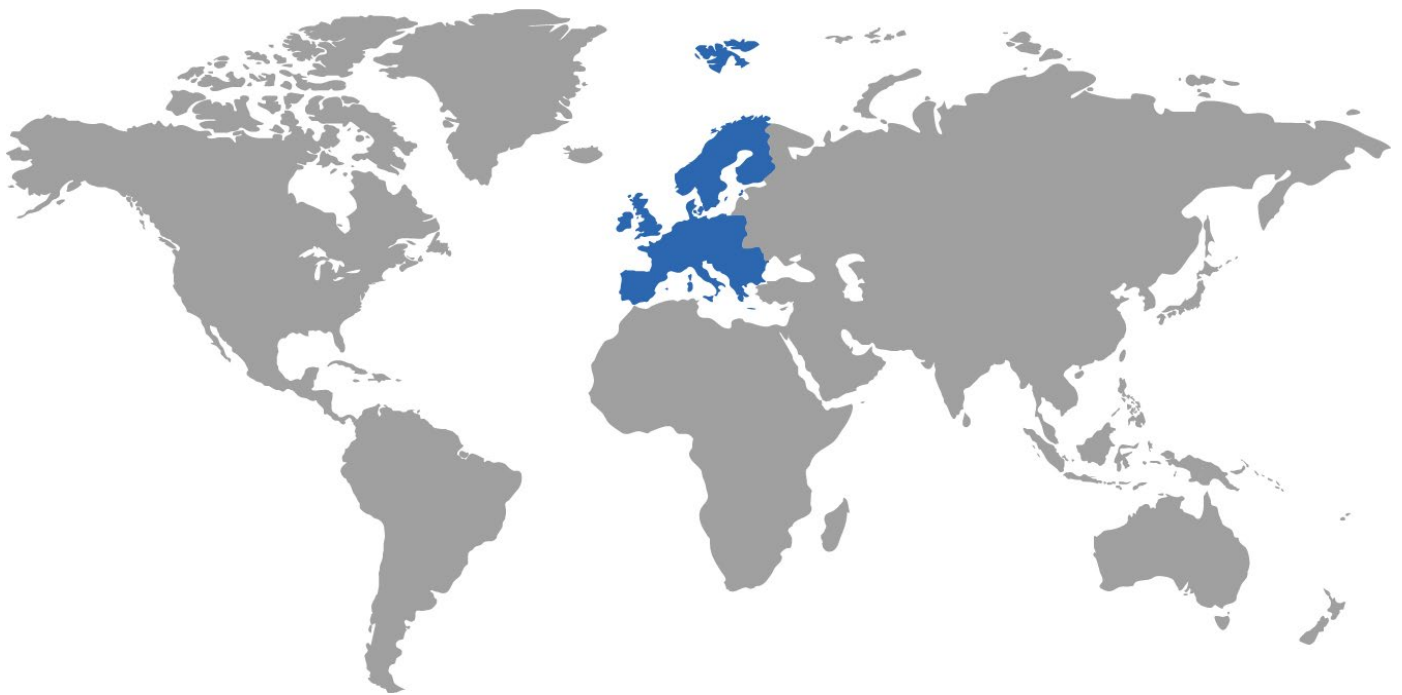

**Automated Recognition in Gender and Sexual Orientation, European Union**
*Pillars: Participation, Equity, Power, and Identity*

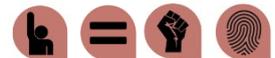

Automated Gender Recognition (AGR) has been integrated into facial recognition systems sold by corporate entities including Amazon and IBM for a host of operations. While research cannot accurately note all the sectors deploying AGR, it has been assumed that it exists within general facial recognition systems.

Notwithstanding the discrimination associated with cases of biased algorithms in numerous systems, AGR presents a particularly harmful stream within the domain of recognition technology as it 'doesn't merely "measure" gender. It reshapes, disastrously, what gender means'.[50] Consequences vary from representational harms that occur at security checks in airports to limitation of access to bathrooms or changing rooms. Individuals who cannot publicly identify their sexual or gender orientation face an environment that can endanger their safety or limit their freedom. In a similar vein, the use of systems to identify and classify individuals based on sexual orientation can pose significant risks for sexual minorities, particularly in regions with anti-LGBTQ+ laws. Deploying such technologies can reinforce social systems of exclusion that have a history of marginalising and discriminating against individuals who do not fit into established standards that far predate the development of these systems. Furthermore, the intersection of race and gender has revealed that people of colour, and Black individuals in particular, are often misclassified.[51] Without appropriate regulation, the potential for such technology to seep into essential service industries, like healthcare or welfare, can serve to augment the barriers and challenges already faced by minority groups.

You can read more at:
https://www.theverge.com/2021/4/14/22381370/automatic-gender-recognition-sexual-orientation-facial-ai-analysis-ban-campaign

---

[50] Keyes, 2019
[51] Krishnan et al., 2020

**SyRI, Netherlands***

*Pillars: Access, Equity, and Power*

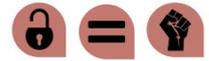

System Risk Indication, known as SyRI, is a welfare surveillance system developed by the Netherlands to comb through large volumes of data collected by public authorities and subsequently identify individuals most likely to commit welfare fraud. A Dutch court had ruled against the use of SyRi by noting how the system would infringe on the right to privacy.

Not only was the model used primarily in low-income neighbourhoods, but it also cross-utilised data from an array of sources, including employment and education history, to identify potential cases of fraud. Importantly, the Netherlands was not the only country to use digital technology for the administration of public welfare. Australia, the United Kingdom, the United States, and India have been cited as examples for their use of digital technology to purportedly improve public welfare systems. However, the deployment of such systems has caused numerous harms. In certain cases, even comprehending and appealing the decisions has been challenging.

You can read more at:
https://www.theguardian.com/technology/2020/feb/05/welfare-surveillance-system-violates-human-rights-dutch-court-rules
https://privacyinternational.org/news-analysis/3363/syri-case-landmark-ruling-benefits-claimants-around-world

**GDPR Immigration Control Exception, United Kingdom** 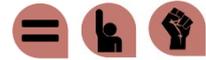
*Pillars: Access, Equity, Participation, and Power*

The 2018 UK General Data Protection Regulation and the Data Protection Act (GDPR) contains an exemption that releases immigration control officials from their obligation to protect individual rights when these are likely to prejudice immigration management. The exemption also allows public and private entities to share personal data with immigration officials.[52] In May 2021, the Court of Appeal stated that the exemption was unlawful and asked the UK government to amend it before 31 January 2022, the date after which controllers would not be able to rely on the exemption.[53]

Activists have stressed that this exemption enables state officials and companies providing data, analysis, and infrastructure services to the government to access, use, and share vast amounts of personal data without individuals' knowledge. They have critiqued this exemption for enabling immigration officials to collect data from schools, hospitals, homeless shelters, banks, landlords, and other providers of vital services.[54] Within a context where many aspects of the life of migrants can be monitored, criminalised, or constrained (i.e., working, driving, traveling to or within the UK), the restriction of data subjects' rights when their data is used to make life changing decisions such as permitting them to enter or reside in the UK situates migrants in a state of surveillance and deters them from accessing vital services due to fear, prosecution, or deportation.[55] Campaigns such as Step Up Migrant Woman UK, highlight how data-sharing between immigration control, the police, and domestic abuse victim support services prevents migrant women from reporting domestic abuse and other abusive situations, thereby exemplifying how this restriction of individual rights places migrants in precarious positions vulnerable to exploitation.[56]

You can read more at:
https://ico.org.uk/for-organisations/guide-to-data-protection/guide-to-the-general-data-protection-regulation-gdpr/exemptions/immigration-exemption/#exemption1
https://privacyinternational.org/news-analysis/3064/privacy-international-joining-migrant-organisations-challenge-uks-immigration
https://stepupmigrantwomen.org/about-sumw/

---

[52] ICO, 2022
[53] Ibid.
[54] Privacy International, 2019
[55] Step Up Migrant Women, 2017
[56] Ibid.

# Transregional

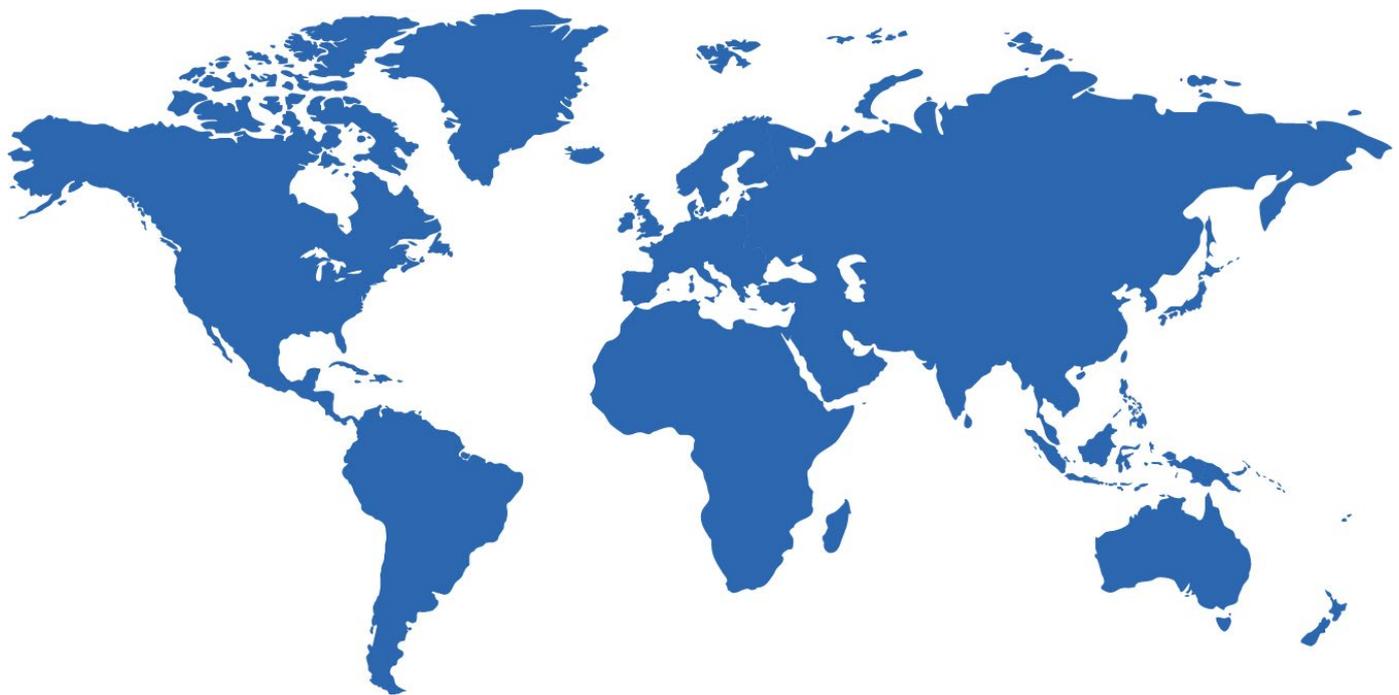

**Facial Recognition Technology and Algorithmic Misidentification, Transregional\*\***

*Pillars: Identity, Equity, Access, and Power*

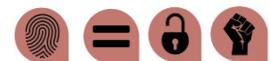

Facial analysis technology, used in a plethora of domains and industries, has been found to misidentify or fail in the detection of people of colour and trans folk. As these systems are developed, trained, and evaluated primarily using datasets that contain greater numbers of photos of white faces, the technology has been riddled with challenges for people of colour. Similarly, the normative paradigms of 'male' and 'female' have excluded non-binary, transgender, intersex, and gender non-conforming people.

Instances of algorithmic misidentification have been recognised across the spectrum. Microsoft's Kinect, a line of motion-sensing devices, was unsuccessful in identifying Black users.[57] When Hewlett Packard's webcams failed to identify a Black user, the company dodged accountability by citing poor lighting.[58] Numerous scholars and activists have highlighted that the issues of bias and discrimination that arise in the differential performance of facial recognition technologies have origins in the history of photography, where the chemical make-up of film was designed to be best at capturing light skin and colour film tended to be insensitive to the full range of skin colours, often failing to show the detail of darker-skinned faces.[59] Despite known risks of discrimination, facial analysis technologies based on skewed datasets continue to be developed or deployed in domains where individual security may be at risk. For instance, the passport photo of a New Zealand individual of Asian descent was rejected when the software misidentified their eyes as closed.[60] Similarly, the use of facial recognition in policing, such as Amazon's Rekognition, and border security have been found to misidentify people of colour and non-cis individuals.[61]

---

[57] Sinclair, 2010
[58] Bunz, 2009
[59] Leslie, 2020; Buolamwini & Gebru, 2018

On the other hand, it may be noted that systems developed in Asia performed better at identifying Asian faces than others.[62] Nevertheless, issues of misidentification based on gender and/or race continue to prevail across the globe including India where software that noted a reduction in error rate of identifying Black women continued to misidentify Indian women.[63]

You can read more at:
https://jods.mitpress.mit.edu/pub/costanza-chock/release/4
https://scroll.in/magazine/1001836/facial-recognition-technology-isnt-wholly-accurate-at-reading-indian-faces-find-researchers
https://news.mit.edu/2018/study-finds-gender-skin-type-bias-artificial-intelligence-systems-0212
https://doi.org/10.5281/zenodo.4050457

---

[60] Reuters, 2016
[61] Vincent, 2019; Constanza-Chock, 2018
[62] Phillips et al., 2010
[63] Mehrotra, 2021

Data Justice Stories: A Repository of Case Studies    28

**Workplace Surveillance, Transregional**\*\*
*Pillars: Equity, Access, Identity, and Power*

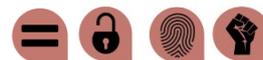

With the proliferation of productivity tools has come the rise of workplace surveillance. These apps, when employed by companies, can pose serious challenges to data justice. For instance, Crossover, a talent management company, produced a productivity tool entitled WorkSmart. One of the facets of WorkSmart includes taking screenshots of employees' workstations and producing 'focus scores' and 'intensity scores' based on their keystrokes and app use.[64] The producers of similar workplace surveillance software often cite preventing insider trading, sexual harassment, and inappropriate behaviour as primary incentives behind the development of these types of apps.[65]

Other workplace surveillance apps include Wiretap, which monitors workplace chat forums for threats, intimidation, and other forms of harassment, as well as Digital Reasoning, which 'searches more for subtle indicators of possible wrongdoing, such as context switching', e.g., switching off a workplace app like Slack to use encrypted apps like Signal instead.[66] In an explainer piece by Mateescu and Nguyen,[67] various other types of employee monitoring and surveillance technologies are detailed including behavioural prediction and flagging tools, biometric and health data tracking, remote monitoring and time-tracking, and gamification and algorithmic management. The authors also explore the range of harms these technologies can exact. These include the augmentation of biased and discriminatory workplace practices, the creation of power imbalances between employees and their managers/organisation, and the decrease in workers' autonomy and agency. Additionally, the authors point out that the use of granular digital tracking and surveillance apps is often motivated by employers' desire to bolster 'cost-cutting' practices surrounding worker pay, benefits, and standards.

Another type of workplace surveillance app closely related to these is the fitness tracker. Biometric wearables are becoming more common across company wellness programmes. While these technologies have often been cited to help decrease company health insurance premiums[68] the data gathered by some of these fitness apps can reveal physical location and occasionally sensitive information including family medical histories and diets. In one case, an employer incentivised employees through a US$1 a day gift card to use Ovia, a pregnancy-tracking app. This allowed the company to see aggregated health data collected via the app.[69] While the reasons that employers cite for using apps like these range from boosting employee well-being to decreasing overall company healthcare spending, there remain significant risks of data reidentification and intrusive tracking, among others.[70]

Increasing workplace surveillance and monitoring has also been accompanied by higher expectations for employee outputs and quotas. However, achieving these objectives has led to numerous instances of strain and injury in labour-heavy industries. For example, the second largest employer in the United States, Amazon, has monitored and evaluated employees through ADAPT, a proprietary software that not only evaluates employee productivity but also automates termination.[71] This automated management structure continues to operate despite thousands of injury reports and medical advice against the creation of strenuous conditions.

---

[64] Solon, 2017
[65] Ibid.
[66] Ibid.
[67] Mateescu & Nguyen, 2019
[68] Mateescu & Nguyen, 2019; Bort, 2014
[69] Harwell, 2019
[70] Ibid.

You can read more at:

https://www.theguardian.com/world/2017/nov/06/workplace-surveillance-big-brother-technology

https://datasociety.net/wp-content/uploads/2019/02/DS_Workplace_Monitoring_Surveillance_Explainer.pdf

https://www.washingtonpost.com/technology/2019/04/10/tracking-your-pregnancy-an-app-may-be-more-public-than-you-think/?arc404=true

https://www.theverge.com/2019/4/25/18516004/amazon-warehouse-fulfillment-centers-productivity-firing-terminations

---

[71] For a more in-depth evaluation of physical injury and harms at Amazon's fulfilment centre, read Evans, 2019.

**Government data breaches, Transregional**\*\*

*Pillars: Identity, Access, and Power*

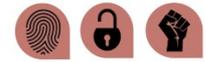

The exposure of confidential, sensitive, or protected information to unauthorised people or organisations have not only affected the private sector but also governmental agencies. Data breaches could expose personal data, and if not properly addressed, they may have adverse effects on individuals whose data have been compromised. The effects range from identity theft or fraud, to discrimination, emotional distress, and even physical damage.[72] In 2015, the US Office of Personnel Management (OPM) reported two substantial data breaches. The first exposed the names, birth dates, home addresses, and social security numbers of 4.2 million current and former US federal government employees.[73] The second breach released information about 21.5 million individuals, including background investigations of current, former, and prospective federal employees and contractors. This data contained sensitive data about family, friends, and foreign nationals that applicants have interacted with, as well as financial information, detailed employment histories, criminal history, psychological records, and information about past drug use.[74]

Despite the multiple warnings of security vulnerabilities received, the OPM did not follow cybersecurity measures, such as encrypting of personal data, tracking of the agency's equipment, having an inventory of its servers and databases, and using multi-factor authentication.[75] The data breach had potentially damaging consequences. Sensitive information about millions of people could be used by the perpetrators and whoever had access to it in harmful ways for an indefinite period, leading to potential impacts on counterintelligence and national security. The information could have potentially been used to identify US undercover officials or foreign nationals who had been in contact with US government workers. All that information, in turn, could have been used to follow, blackmail, or punish federal government employees and foreign nationals.[76]

A similar case where cybersecurity measures of governmental agencies fell short to protect sensitive and confidential information can be found in Sweden. In 2015, the Transport Agency outsourced its IT services to IBM. The company, in turn, outsourced the maintenance services of the IT systems to 11 subsidiaries in Romania, Croatia, Serbia, and Czechia.[77] This allowed foreign IT workers to have full access to the agency's confidential databases without security clearance requirements. It exposed the entire national driver's licence database, information about the vehicles in the country, witness protection details, secretive military units, defence plans, details about Air Force fighter pilots, and the weight capacity of all roads and bridges.[78] Although there is no evidence of data being copied, critics state that the way the Transport Agency handled the outsourcing proved to be in breach of privacy and data protection laws.[79]

You can read more at:
https://www.wired.com/2015/06/opm-breach-security-privacy-debacle/
https://www.theguardian.com/technology/2017/aug/01/sweden-scrambles-to-tighten-data-security-as-scandal-claims-two-ministers

---

[72] ICO, n.d.
[73] OPM, n.d.
[74] Ibid.
[75] Zetter & Greenberg, 2015
[76] Ibid
[77] Castaño, 2017
[78] Falkvinge, 2017
[79] Henley, 2017

**Data breaches of dating sites, Transregional**\*\*

*Pillars: Identity, Access, and Power*

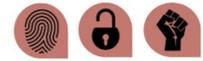

Dating sites often gather a range of personal information from their users. The information, which ranges from demographics to desires and sexual preferences, are meant to refine and improve the performance of the site's services. However, companies that store sensitive personal information are particularly vulnerable to the adverse consequences of data breaches on their users. In July 2015, Ashely Madison, a dating site for married people seeking affairs, received a ransom note from a group called Impact Team. The group informed the site that they had hacked the company's data—including their customers' data—and threatened to publish them if both the website and its partner site, Established Men, were not taken down within 30 days. The hackers condemned the behaviours that the site encouraged and their fraudulent business practices, such as the retention of data on their servers even after users requested and paid for their accounts to be deleted.[80] In response to Ashley Madison's denial to take down the sites the hackers released a series of data dumps and posted them online.

The datasets included usernames, email addresses, passwords, street addresses, postcodes, phone numbers, logins, credit card information and other payment transaction details, and IP addresses, as well as descriptive information about their over 30 million users' sexual preferences, fantasies, and affair interests. The group also leaked information about the company and its website source code.[81] Not only did this data hack impact the company financially through the payment of large sums of money in response to class action lawsuits on behalf of Canadian and US users, but it had harmful impacts for users whose data was leaked. Following the incident, Toronto police reported two suicides related to the leaks. There were also reports of doxing, blackmails, and identity theft targeted at Ashley Madison's users.[82]

That same year the data of almost 4 million users of the dating site Adult FriendFinder were also hacked and posted online. Leaked data included information about deleted accounts as well as email addresses, usernames, birthdates, postcodes, IP addresses, sexual preferences, marital status, and other personal and sensitive details. Soon after the publication of the data, users not only received spam emails but also threats to expose their private information, which caused emotional distress about future injury and could, in turn, prevent future access to the services.[83]

You can read more at:
https://www.theguardian.com/technology/2016/feb/28/what-happened-after-ashley-madison-was-hacked
https://www.theguardian.com/lifeandstyle/2015/may/21/adult-friendfinder-dating-site-hackers-expose-users-millions

---

[80] Zetter, 2015
[81] Lamont, 2016
[82] Ibid.
[83] Solove & Citron, 2016

**Security company's biometric database, Transregional\*\***

*Pillars: Identity, Access, and Power* 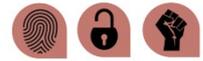

In 2019, cybersecurity researchers found that Biostar 2's biometric lock system, owned by security company Suprema, contained a database of biometric credentials and personal information, was unprotected, mostly unencrypted, and vulnerable to public access.[84] The system centralises the control for access to facilities, and it verifies whether the fingerprints and facial recognition of people who request access to warehouses, office building, or other facilities, match the registered biometric data of individuals whose access is approved. Its platform has been integrated into another access control system used by governments, banks, and the UK Metropolitan police, among others.

The researchers managed to gain access to 27.8 million records and 23 GB of data. Not only did it include admin panels, fingerprints, facial recognition data, usernames, passwords, access to buildings, and other personal details of over 1 million people, but access to the system also enabled real-time information of peoples' whereabouts and the editing of users' accounts. Furthermore, a potential data breach could have exposed affected individuals to identity thefts and granted undesired access, especially as fingerprints cannot be changed. Because the database involved personal data, the fact that it was publicly accessible posed significant threats to the right to protection of personal data.

You can read more at:
https://www.theguardian.com/technology/2019/aug/14/major-breach-found-in-biometrics-system-used-by-banks-uk-police-and-defence-firms
https://www.theverge.com/2019/8/14/20805194/suprema-biostar-2-security-system-hack-breach-biometric-info-personal-data

---

[84] Taylor, 2019

**AI Harms to Children, Transregional****

*Pillars: Identity, Equity, Participation, and Power*

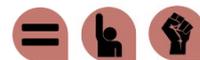

There are a host of risks and harms that AI technologies pose to children. Many of these harms create barriers to the agency and freedoms of children and young people. While there are many examples of these harms across sectors, this case will focus primarily on two, namely, the collection of children's data without parent/guardian consent and the dangers of smart toy hacking and associated data practices that accompany these products.

One of the most frequently cited cases of the illegal collection of children's data without parent/guardian consent occurred in 2019 when YouTube was forced to pay US$170 million in fines for violating privacy law (specifically the Children's Online Privacy Protection Act—COPPA).[85] There have been various instances of this type of behaviour across companies which has accompanied the widespread increase in datafication; however, another recurring issue has been around content moderation on platforms that are directed towards children. Taking YouTube as the example, there have been various reports about gory and other inappropriate videos being shown to children even when YouTube's family-friendly viewing mode is enabled. Several instances demonstrated that videos were recommended in the sidebar via autoplay and were primarily based on the number of views the videos received.[86]

There have been many reported cases of harms occurring when using smart toys, as well. When investigating several smart toys on the market, *Which?* carried out tests with Stiftung Warentest and discovered that four out of seven of the toys they tested allowed for communication with the child through the toy itself. This presented significant fear surrounding hacking, as anyone that was able to access the toy—quite easily and often via a Bluetooth connection—could thereby communicate directly with the child.[87] When asked to comment on the results of this investigation, the British Toy and Hobby Association stated that they were aware of the report but they believe the investigation 'relied on a perfect set of circumstances and manipulation of the toys and the software that make the outcome highly unlikely in reality'.[88] Harms such as those outlined above occur frequently, and with the rise in datafication, the rights and freedoms of children are being challenged.

You can read more at:
https://www.ftc.gov/news-events/press-releases/2019/09/google-youtube-will-pay-record-170-million-alleged-violations
https://www.wired.co.uk/article/youtube-for-kids-videos-problems-algorithm-recommend
https://www.theguardian.com/technology/2017/nov/14/retailers-urged-to-withdraw-toys-that-allow-hackers-to-talk-to-children

---

[85] Federal Trade Commission, 2019
[86] Orphanides, 2018
[87] Smithers, 2017
[88] Ibid.

Data Justice Stories: A Repository of Case Studies    34

**Limited Liability of Private Sector Companies, Transregional**
*Pillars: Access and Power*

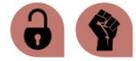

The UK Government announced in 2019 an agreement to share data between US-based Amazon and British Department of Health and Social Care through the NHSX—a unit of the UK government focused on setting policy and developing best practice related to digital, data, and technology for the National Health Service.[89] The idea behind the agreement to share data was that it would allow those with difficulty searching for health advice on the internet (e.g., elderly people, blind people, and other patients who cannot easily search for health advice online) to access to NHS information via Amazon's Alexa, the artificial intelligence powered voice assistant. Alexa would provide answers to questions such as 'Alexa, how long can a migraine last?' by using information from the NHS website, making it convenient, fast, and simple to use for patients. For the NHS, the service would reduce the pressures on the overall health system. The agreement between Amazon and NHS, however, reduces responsibility of the US giant and increases the power imbalances between Amazon and users. In reference to limitation of liability the agreement states that 'in no event will (a) either party be liable for any loss of data (...) however caused and regardless of theory of liability or (b) either party's liability for direct damages under this Agreement exceed [redacted]…'.[90] The NHSX strategy has continued its plans to partner with small and big companies so that they 'can ensure that the millions of users looking for health information every day can get simple, validated advice at the touch of a button or voice command'.[91]

In 2020, it was reported by the *Sunday Times* that betting companies had been given "accidental" access to an educational database containing names, ages, and addresses of 28 million children and students, aged 14 and above in state schools, private schools, and colleges in England, Wales, and Northern Ireland.[92] The companies used it to help increase the proportion of young people who gamble online. This breach, made by a third party, was reported to ICO and is one of the biggest breaches of UK government data to date. The sharing of data between public and private sectors and limited liability of private sector companies raises concerns of data privacy and opportunities of redress in times of harm.

You can read more at:
https://www.theguardian.com/society/2019/dec/08/nhs-gives-amazon-free-use-of-health-data-under-alexa-advice-deal
https://www.thetimes.co.uk/article/revealed-betting-firms-use-schools-data-on-28m-children-dn37nwgd5

---

[89] Master Content License Agreement, Amazon Digital Services LLC—the Secretary of State for Health and Social Care, 2018
[90] Ibid.
[91] Department of Health and Social Care, 2019
[92] Bryan et al., 2020

**Automated Experiments and Research on Ad Privacy Settings, Transregional**\*\*

*Pillars: Equity, Power, and Knowledge* 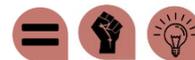

Studies and reports have uncovered potentially discriminatory Google and Facebook algorithms where certain employment advertisements were predominately shown to men rather than women. Similarly, housing advertisers on Facebook were allowed to discriminate against users based on race, gender, and disability.[93] To combat these issues, researchers at Carnegie Mellon University developed 'AdFisher'—a tool that creates hundreds of simulated users to identify discrepancies through browser experiments—and used it to test 100 employment sites.

The testing of this tool demonstrated that higher paying jobs were shown to male designated users many times more frequently than to female designated users.[94] While the US Department of Housing and Urban Development sued Facebook for discrimination in housing,[95] it was more difficult to evidence discrimination and bias in the algorithms used on Google due to the black-box nature of algorithmic decision-making.[96] In parallel, ProPublica's evaluation of Facebook ads found that the platform allows advertisers to redline both communities and neighbourhoods at their discretion.[97] Such policies have been found to be in clear violation of state, federal, and local laws on gender-based discrimination. While the policies have been updated to prevent discrimination based on race and national origin, numerous policies relating to housing, credit, and employment have not been introduced in countries beyond US and Canada.[98] Meanwhile, Fairness Flow, the optional toolkit used by Facebook to review the social impact of its algorithms, has been deemed insufficient by experts.[99]

You can read more at:
https://sciendo.com/pdf/10.1515/popets-2015-0007
https://www.propublica.org/article/facebook-is-letting-job-advertisers-target-only-men
https://www.technologyreview.com/2021/03/11/1020600/facebook-responsible-ai-misinformation/

---

[93] BBC Staff, 2019
[94] Datta et al., 2015
[95] Gabbatt, 2019
[96] Carpenter, 2015
[97] Tobin & Merrill, 2018
[98] Gwynn, 2021
[99] Wiggers, 2021

**Surge pricing on ride hailing apps, Transregional**
*Pillars: Equity and Power* 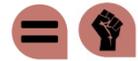

In 2016, the multinational ride hailing company Uber was criticised by several customers who were trying to hail a cab ride in New York City (NYC), US to escape Chelsea, a part of the Manhattan area where news had broken that a device had detonated. For people in the area trying to book taxis in order to move to safer areas, the ride hailing app displayed exorbitant prices citing the basic principle of demand-supply that the company claimed as the underlying logic informing "surge pricing". This method allows Uber's algorithmic system to capitalise on high demand (such as when people exit concerts, sports games or any high-density gatherings and try to book a ride) while simultaneously, in some way, passengers who cannot afford the price surge must find other alternatives. This incident came to light after another earlier incident in Sydney, Australia in 2014 amid a hostage situation in the central business district of the city. At that time, Uber was heavily criticised for demanding a minimum fee of US$82 from potential passengers looking to escape the area.

Following these incidents, Uber claimed that it had "shut down surge pricing" although as per the company's website, it is still in use.[100] After the NYC incident, it was reported that Uber struck an agreement with the NY attorney general to restrict the use of surge pricing at the times of natural disasters and states of emergency. During the pandemic, largely due to health-related concerns as well as the diminished financial capacities of consumers, app-based rides had become expensive and hard to book almost globally. However, as some countries, especially in the Global North have emerged from the pandemic, it has led to another form of longer-term surge-pricing where the "new normal" prices are reportedly much higher while drivers continue to earn the same if not lower than before the pandemic.[101] It remains unclear what factors are contributing to price-setting on platforms.

You can read more at:
https://www.nbcnews.com/storyline/sydney-hostage-standoff/ubers-surge-pricing-near-siege-sydney-sparks-outrage-n268371
https://www.cnbc.com/2016/09/18/uber-hammered-by-price-gouging-accusations-during-nycs-explosion.html#:~:text=At%20the%20time%2C%20Uber%20was,disasters%20and%20states%20of%20emergency

---

[100] See Uber's explanation of how surge pricing work at https://www.uber.com/gb/en/drive/driver-app/how-surge-works/
[101] Sainato, 2021

Data Justice Stories: A Repository of Case Studies    37

**Algorithmic obfuscation and lack of materiality of platform services, Transregional**

*Pillars: Access, Equity, and Power* 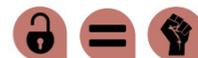

Related to the earlier case of "surge pricing", especially during crisis events as well as a longer-term increase in app-based ride hailing and food-delivery services, is also the issue of (in)explicability and the lack of accountability with regard to the decisions generated by multinational corporate platforms such as Uber, Ola, Swiggy, and others. In academic and public-facing scholarship, platform companies have been widely criticised for blaming algorithmic calculations and outputs for inhumane, immoral, and unfair decisions produced by ride hailing and food-delivery apps as well as other kinds of tech failures and glitches. Scholars, journalists, and others have repeatedly pointed out how platforms rely on algorithms as obfuscators to deflect responsibility and to avoid having to explain or disclose what kinds of collaborative human and organisation decisions go into the engineering of their algorithmic systems and for what they might be optimising for. Besides relying on algorithmic obfuscation, another challenge that the consolidation and popularity of platform services as essential urban infrastructures has posed is the lack of materiality as compared to older private and public infrastructural entities that are localised and anchored in place and geography through their offices, employees, and information records.

As cloud-based entities enter the mix of urban governance and public welfare services, their lack of materiality and thus relatedly, the absence of information and organisational maps that can update our routes and methods to hold them accountable create serious consequences for individual stakeholders. The most prominent example of this is the incident of sexual violence against a woman passenger riding an Uber in Delhi, India in 2014. Once the incident was reported to Delhi police, they were unable to find and access a physical office location for the company or get information about any local (human) point of contact to start the investigation. Despite this, in order to act immediately and begin the investigation, the police personnel had to book an Uber ride and then ask their driver to drive them to the nearest Uber office (which was an operations hub and not the place where their managers, engineers, designers or executives were located). Further, during the investigation, the police realised that the data pertaining to the specifics of the ride were not locally stored nor accessible from the operations office. Globally disaggregated units of platforms' operations present a challenging case wherein accountability for harms caused by such data-driven platforms is limited by virtue of its operational structure.

You can read more at:
https://www.vox.com/2017/6/7/15754316/uber-executive-india-assault-rape-medical-records

**Concerns with data privacy and sovereignty, Careem and Uber, Transregional**
*Pillars: Access, Equity, and Power*
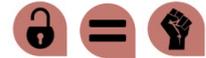

In 2019, Egypt's cabinet issued Resolution No. 2180 of 2019 which made it mandatory for ride hailing companies to share user data if they wanted to keep operating. Careem, a ride hailing company operational in Egypt (as well as 15 other countries in North Africa, West, and South Asia) and now an acquired subsidiary of Uber, was threatened with being shut down by the government if it did not comply with the demands to share real-time data on passenger rides. However, it remains unclear how much or what kind of access the Egyptian regime has to private consumer movement data through Uber and Careem. Critics of this development state that although it is particularly threatening, it reflects the kind of exacerbated risk that may emerge when private data-intensive platforms come under the purview of authoritarian regimes. Beyond the specific case of Careem in Egypt, the parallel development in India and other global South contexts has been a drive towards 'data localisation' as a form of technological sovereignty. The Indian government has been promoting and passing legislation on 'personal data protection', often insisting that multinational tech corporations invest in local data infrastructures and store all the data on Indian citizens within the boundaries of the country.[102]

It is argued that while the language of sovereignty as self-determination through control over citizens' data and information flows may deceptively borrow from the discourses on Indigenous data justice movements and arguments against data colonialism, it is important to distinguish and maintain a critical orientation when thinking about data justice, especially from the perspective of advocating for vulnerable communities under the threat of persecution in global South contexts in order to not homogenise data justice needs and conflate sovereignty with control, surveillance, and geopolitical extraction.

You can read more at:
https://iapp.org/news/a/a-look-at-proposed-changes-to-indias-personal-data-protection-bill/
https://egyptianstreets.com/2019/09/20/new-egyptian-law-requires-access-to-customer-data-from-careem-uber/

---

[102] Dhavate & Mohapatra, 2022

**Uber 'Movement' dashboard, Transregional**
*Pillars: Equity and Power*
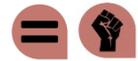

Platform companies have largely relied on proprietary data and algorithmic systems as intellectual property, which make it possible to avoid explaining platform consequences and decisions at the city-level. This enabled platform companies to make so-called data-powered claims such as Uber stating that its services reduced urban traffic congestion and hence also promoted environmental goals in global cities—a claim they had to finally walk back in 2019.[103] In the absence of data-sharing programs or regulatory frameworks that mandate data-sharing with governments, academics, and the civil society globally, it has been difficult to engage with or challenge the rhetoric of platform companies. In response, Uber launched the 'Movement' dashboard and announced a few partnerships with cities, academics, non-profits, and others in the Global North. Not only has the portal been relatively inactive since its launch but, as researchers have revealed, the dashboard constrains inquiries and hence substantially limits any critical research about gig platforms' effects on cities.[104]

Not only this, but Uber also leaned into the 'data privacy' narrative in 2019 when the New York City Taxi and Limousine Commission wanted to introduce an amendment requiring ride hailing companies to share more ride data that would then also be publicly available online. Uber, the same company that was exposed and penalised for stalking its consumers through the 'God's View' software,[105] liberally used by its own engineers and data scientists to spy on ex-partners, celebrities, and others, argued, just like Facebook/Meta did with its data-sharing partnerships that beyond a point, granular data activity and identifiers would not allow for companies to retain user anonymity. Importantly, between Global North tech companies and their less powerful oversight bodies, the former certainly has an edge when it comes to maintaining privacy, security, and confidentiality of personal data since these companies have been able to invest in the storage, security, and maintenance infrastructure while the same may not always be possible for a government agency and certainly not so for academics or non-profit actors. This disparity only increases when porting and advocating for data-sharing in service of research and advocacy in global South contexts, with the added risk that in case of a data breach, there may not be robust and realistic recourse to data privacy and protection laws outside of Europe, United States, and Canada.

You can read more at:
https://www.theverge.com/2016/1/6/10726004/uber-god-mode-settlement-fine
https://www.theverge.com/2019/8/6/20756945/uber-lyft-tnc-vmt-traffic-congestion-study-fehr-peers
https://www.theguardian.com/technology/2021/jun/21/uber-lyft-fares-surge-drivers-dont-get-piece-of-pie

---

[103] Hawkins, 2019
[104] Uzel, 2018
[105] Welch, 2016

# Transformational Stories of Data Justice: Initiatives, Activism, and Advocacy

## Africa

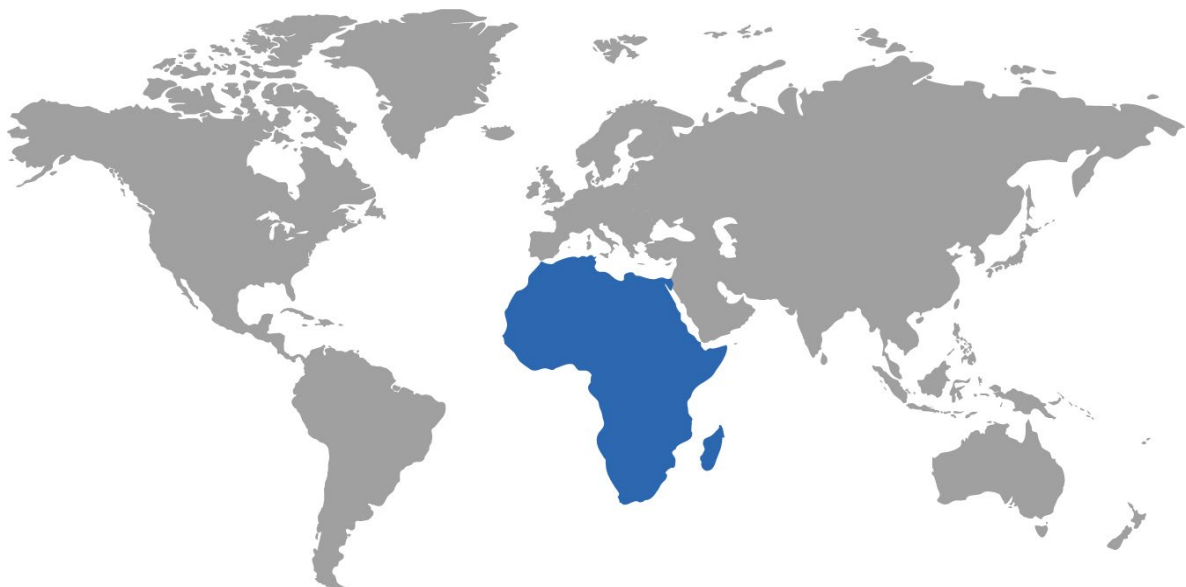

**National Participatory Development Program (PNDP), Cameroon**
*Pillars: Power, Equity, Access, and Participation*

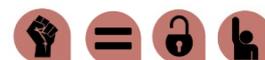

Through donations from numerous regional and international bodies, such as the World Bank, African Development Fund alongside financing from former colonial powers such as France, the Government of Cameroon established the 'National Participatory Development Program' (PNDP) for growth and development in the country.

Structured in three phases of four years each, the program was implemented in six regions in the first phase followed by all ten regions in the second. The results have been recorded and then translated into guiding lessons. Notably, they highlight the importance of local planning and citizen engagement for the success of long-terms strategies across all domains. They developed a comprehensive database on infrastructure for planning and development and granted allowances for micro-projects within communes. However, the PNDP has also observed that there are many milestones left to achieve including the entrenching of the program as a leading force for socio-economic development.

Nevertheless, they continue to place local participation in design and planning as an essential element of carrying out national development projects. For example, the Cameroon Digital Transformation Acceleration Project (PANTUC) that aims to reform the ICT sector, improve digital skills, expand digital applications and services, and use ICT for employment, has been preceded by PNDP studies with stakeholders.[106] The studies, covering both environmental and social safeguards, were accompanied by manuals that detail potential risks. Most importantly, Indigenous communities were placed at forefront of review to ensure that they were consulted and informed before launch, thereby inculcating a stream of data justice practice at the outset.

You can read more at:
https://www.pndp.org/documents/cleared-CPPA_final_PATNUC_juillet_2021_final.pdf

---

[106] Nyambi III Dikosso, 2021

### Protégé QV, Cameroon

*Pillars: Access, Equity, Identity, and Power*

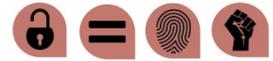

Promotion des Technologies Garantes de l'Environnement et de la Qualité de Vie (Protégé QV) was founded in 1995 to promote initiatives of environmental protection and livelihood improvement. To achieve this goal, Protégé QV develops capacity-building programs for local development, research, and promotes the appropriation of ICTs. Since its creation, the organisation has developed new areas of interests. As a result, it is currently focused on the promotion of female leadership, micro-enterprises, and ICT for development.

To promote ICTs that benefits rather than harms local development, Protégé QV launched the project Information and Communication Technologies for Development. It includes many different initiatives such as distance training for women to engage in micro-enterprises, training workshops on online safety for women workers, and the organisation of Quality of Life Evenings (Impact Talks), which seek to raise awareness about the African Deceleration of Internet Rights and Freedoms (2019).

You can read more at:
https://www.protegeqv.org/

### Motoon, Egypt

*Pillars: Access, Knowledge, Participation, Power, and Equity*

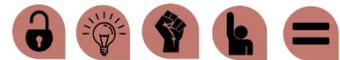

Motoon is a community foundation that provides technical services to progressive causes and local communities by connecting them with people in tech. Motoon aims to empower the tech community, unleash the potential of organisations and groups through technology solutions, enhance a network with shared values, and engage with local, regional, and international policies, whilst promoting values of free, open-source software and knowledge sharing.

One of their initiatives, Noon Tech, was established to advocate for women empowerment as well as to help bridge the gender gap in tech fields. The initiative encourages debates about the obstacles for women entering the tech field and possible solutions. It also identifies women in tech and shares their stories of success to the broader tech community. Another program, "MTPP", aims to establish a dialogue about policies affecting technology, raise awareness among the public of such effects, and advocate for better policies.

You can read more at:
https://motoon.org/

**Center for Advancement of Rights and Democracy (CARD), Ethiopia**

*Pillars: Power, Participation, and Knowledge*

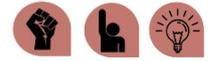

Emphasising the role of democracy in guiding actions and policies, the Center for Advancement of Rights and Democracy (CARD) is a CSO with a focus on the protection of rights and freedoms in Ethiopia. Towards this endeavour, CARD pursues the primary avenues of awareness creation and advocacy. These pursuits are organised around youth and women empowerment, digital rights, media literacy, civic engagement, and democratisation. Not limited to non-state entities, the organisation also carries out monitoring and evaluation of state institutions like correction centres, courts, and the police.

With increasing instances of fake news and hate speech in the region, CARD has published an extensive guide that maps out several forms of 'fake news' (namely, misinformation, disinformation, and malinformation), providing recommendations for combating such informational abuses and hate speech, more generally. Following global waves of disinformation observed since the onset of the COVID-19 pandemic,[107] the guideline has become a pivotal resource for individuals, activists, and members of civic society who are attempting to recognise and mitigate harms online. The guide, as with other publications from CARD, is produced in numerous languages spoken in the region which increases accessibility. Other publications on the digital sphere include coverage of the use of social media platforms in the ongoing conflict in Tigray and guides on how to protect personal privacy and other fundamental rights while using such platforms.[108]

You can read more at:

https://www.cardeth.org/

---

[107] PwC, n.d.
[108] Center for Advancement of Rights and Democracy (CARD), 2021

**Data Protection Act, Ghana***

*Pillars: Power, Access, and Equity*

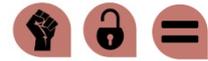

The Data Protection Act (2012) was enacted into law by the Republic of Ghana to set out the rules and principles of data governance and protection. One of few countries in Africa to enact national data protection laws, Ghana is also among the eight ratifications of the African Union Convention on Cyber Security and Personal Data Protection. Ghana's approach to uphold rights and protection of user data is monumental not only to West Africa but to the continent as a whole.

The Data Protection Act, also known as Act 843, follows a risk-based approach similar to the General Data Protection Regulations (GDPR). Individual privacy and personal data are protected through the regulation of gathering and processing procedures for the collection, use, and disclosure of data. Act 843 includes fundamental principles of data minimisation, purpose specification, and accountability amongst others as listed in Section 7 of the Act. Further, the Act also establishes the Data Protection Commission to monitor compliance and implement the provisions. From a critical angle, while the near-identical language to the GDPR utilised in the Act emphasises the objective of improving cross-border flows of data with the European Union, it fails, on the whole, to account for the Ghanaian socio-political landscape and challenges unique to that context. A more context-based approach to data protection law could better tailor governance to the specific needs of affected Ghanaian data subjects.

As Africa has been found to be used as a testing-site for new technology, the protection of data is imperative. Moreover, as the use of technology seeps into democratic procedures, effective regulation through Act 843 can uphold rights and privacy despite challenges such as the potential for corruption or lucrative public tenders.

You can read more at:
https://www.dataguidance.com/jurisdiction/ghana
https://privacyinternational.org/long-read/3390/2020-crucial-year-fight-data-protection-africa

### Human Security Research Centre (HSRC), Ghana

*Pillars: Knowledge and Participation*

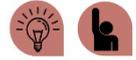

The Human Security Research Centre (HSRC), Ghana is a non-profit civil society organisation (CSO) in Ghana that aims to build and sustain mutual trust between the Ghanaian government and local communities in order to prevent and counter violent extremism. They conduct research that informs National Security policy development and implementation, advocate for preventative programmes that address factors of vulnerability leading to violent extremism, and implement projects and programmes that benefit vulnerable local communities.

Concerned with the potential exacerbations of the unequal distribution of employment opportunities and the risks that automation might pose to employment, HSRC are developing a programme for AI skill development in rural Ghana. By providing vulnerable local communities with tools to take up AI jobs and make AI teams more diverse, HSRC has created a decentralised upskilling programme run through AI training centres.

You can read more at:
https://www.hsrcgh.org/2021/05/20/artificial-intelligence-ai-skill-development-in-rural-ghana/

### Africa Cybersecurity and Digital Rights Organisation (ACDRO), Ghana

*Pillars: Participation and Knowledge*

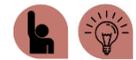

Based in Ghana, Africa Cybersecurity and Digital Rights Organisation (ACDRO) actively promotes cybersecurity awareness, digital rights and freedom, internet access and safety, training, and the reduction of digital divides in Africa and beyond. Established as an NGO, ACDRO not only provides advice to Ghanaian state institutions, like the Computer Emergency Response Team (CERT) and National Cybersecurity Centre, it also partners with private sector entities. To mitigate threats and empower institutes in Africa's digital eco-space, the organisation has launched a cybersecurity academy that provides training and workshops aimed at all levels of skill development.

Among their guides on awareness creation, ACDRO notably focuses on children's online safety. They provide guides about how children and youth in Ghana can build the sort of capacities that are needed to safely and reflectively navigate the internet. ACDRO also focuses on the role of educators (identified as both teachers and parents) in online safety. Maintaining the safety of young individuals is also emphasised as a necessity across Africa through trans-continental policies on the internet. The organisation has also recently developed and presented a national cyber security strategy to the government of Sierra Leone.[109]

You can read more at:
http://acdro.org/

---

[109] ACDRO, 2021

Data Justice Stories: A Repository of Case Studies    45

## Centre for Intellectual Property and Information Technology Law (CIPIT), Kenya***

*Pillars: Equity, Power, Access, and Identity*

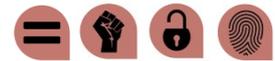

The Centre for Intellectual Property and Information Technology Law (CIPIT) was established in 2012 to generate evidence-based research and training on how the development of intellectual property and information technology contribute to African Law and Human Rights.

CIPIT conducted research on Kenya's Identity Ecosystem, specifically three identification systems: the national ID, the planned Huduma Namba—a state ID system to link existing state systems through a single number identifier, and digital credit. These are all critical to participation in both political and economic life. The report brings to light issues of accessibility, transparency, accountability, and inclusivity, as well as exclusionary practices that contribute to gender inequalities. A primary finding demonstrates that women and girls face unique challenges when trying to access identity systems. The report also finds that many Kenyans do not fully understand how their data is processed and used in services that rely on state issued identification nor how it may be used to limit their access to future opportunities. The report is accompanied by recommendations and an identification ecosystem map. Moreover, CIPIT is participating as a Policy Pilot Partner in the "Advancing data justice research and practice" project.

You can read more at:
https://cipit.strathmore.edu/

## Kenya ICT Action Network (KICTANet), Kenya

*Pillars: Access and Participation*

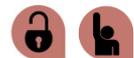

Established in 2003, Kenya ICT Action Network (KICTANet) is a multistakeholder platform that aims to promote synergies between initiatives addressing ICT policy and regulation and to provide frameworks for collaborations. Its mission is to ensure a multistakeholder, open, accessible, and human rights-based approach to policymaking in the ICT sector. Moreover, KICTANet focuses on policy advocacy, research, capacity-building, and stakeholder engagement, as well as providing comments on key issues regarding ICT regulation.

Despite the presence of the principle of public participation in Article 10 of the Constitution of Kenya as one of the 'national values and principles of governance', critics argue that the Kenyan government lacks a comprehensive multistakeholder and multidisciplinary mechanism for stakeholder engagement in ICT policy and law-making processes.[110] In response to these concerns, in July 2021, KICTANet published a policy brief entitled "Public Participation: An Assessment of Recent ICT Policy Making Processes in Kenya". The report draws critical attention to the varied levels of public participation present in three recent ICT policy and law-making processes: the National ICT Policy, 2019, the Computer Misuse and Cybercrimes Act, 2018, and the Data Protection Act, 2019.

You can read more at:
https://www.kictanet.or.ke/

---

[110] van der Spuy, 2017

### Haki na Sheria (HSI), Kenya
*Pillars: Power, Participation, Identity, and Access*

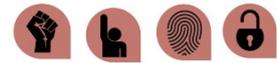

Established in 2010, Haki na Sheria Initiative (HSI) is an NGO in Kenya. Driven by the inequalities faced by marginalised communities in Northern Kenya, HSI seeks to end systematic discrimination and empower communities to 'understand, demand, and effectively claim their human rights and obligations'.[111] Amongst its work, HSI focuses on issues of environmental justice, gender, and statelessness.

Kenyan Somalis, Nubians, and other minoritised communities have long faced structural obstacles when applying for government IDs. Unlike other Kenyans, they are often required to provide additional supporting documents or go through long vetting processes. Not only does this deprive them from a smooth process—and in most cases, from Kenyan citizenship documents—, but also from services and opportunities these documents enable, such as access to schools, work, travel, public services, and the opening of bank accounts.[112] In addition to this, the government has recently introduced the National Integrated Identity Management System (NIIMS), which aims to digitise identity management. However, because proof of identity is required to obtain a digital ID, marginalised communities suffer from discrimination and exclusion.[113]

In response to the far-reaching harmful impacts of the use of digital IDs, HSI has raised awareness of the adverse effects of NIIMS on already marginalised communities. This includes a lack of proper safeguards to protect the rights of minoritised communities. HSI has also challenged the legality of the system and, in 2013, it created a platform where community members can receive advice on how to obtain citizenship documentation. The HSI paralegal team additionally provides people with information on how to 'navigate government bureaucracy'[114] and demand justice and equality.

You can read more at:
http://hakinasheria.org/
https://www.wired.com/story/opinion-digital-ids-make-systemic-bias-worse/
https://www.good-id.org/en/articles/power-to-the-people-using-community-paralegals-to-promote-inclusion-in-digital-ids-in-kenya/

---

[111] See Haki na Sheria at http://hakinasheria.org/
[112] Bashir, 2020
[113] Wired Opinion, 2020
[114] Ibid.

**Centre for Information Technology and Development (CITAD), Nigeria** 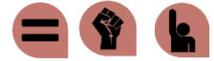
*Pillars: Equity, Power, and Participation*

The non-governmental and non-profit organisation of Centre for Information Technology and Development (CITAD) grew from its beginning in the Computer Literacy Project of 1996 to its current broader scope. Its work is aimed at the use of technology for sustainable development and good governance in Nigeria. By leveraging a host of platforms and tools, including social networking and Google alerts, CITAD sources information about discussions, trends, and attitudes towards technology. It also highlights the role of Information and Communications Technology (ICT) tools, advocacy, research and networking for empowerment and access to information.

In 2021, CITAD launched a project under #TrackNigeria where citizens and social influencers made use of social media platforms, television, radio, or press outlets to augment stories related to corruption and bribery in the country. Supported by the MacArthur Foundation, the project emphasised the importance of citizens in the fight against corruption and advocated for achieving justice through relevant policies and by ensuring that transparency and accountability are present at all levels of government.[115] On Twitter, the hashtag featured stories about unfulfilled development projects, abuse, collaborations for the freedom of information, and policy critique.[116] Notably, the stories on rights of children under Almajiri—a system where children, usually younger boys, are taught the Qur'an in distant locations—have been recognised and criticised by United Nations Children's Fund (UNICEF) as exposing the practice of using children for begging.[117] Other projects and themes of CITAD have consisted of research and discussion on gender-based hate speech, COVID-19 protocols, training in schools, and workshops for community organisations.

You can read more at:
https://www.citad.org/

---

[115] Sani, 2021
[116] See the hashtag at https://mobile.twitter.com/hashtag/TrackNigeria?src=hashtag_click
[117] Njoku, 2020

Data Justice Stories: A Repository of Case Studies    48

### Digital Rights Lawyers Initiative (DRLI), Nigeria
*Pillars: Equity and Power*

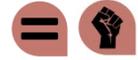

With a team of legal professionals, the Digital Rights Lawyers Initiative (DRLI) works to advance the networking about, engagement on, and defence of digital rights. Their presence was noted across 10 states in Nigeria, the most populous country in Africa. Positioned as a 'think tank(s) for Digital Rights litigation', the team has published extensive research while also filing cases on the breach of relevant laws on privacy, data protection, cybercrime, and constitutional rights and freedoms in the digital sphere.

Data protection has been enshrined in the constitution of Nigeria under the Nigerian Data Protection Regulation of 2019. This regulatory framework provides safeguards on the collection and processing of data while meting out penalties for violations of fundamental data protection principles and constitutional provisions.[118] The DRLI has upheld the importance of such regulation in the cases filed against parties including the Central Bank of Nigeria, Lagos Inland Revenue Service, and the Nigerian Immigration Service amongst others.[119] Protection of journalism and media against the breach of privacy has also been a key focus of the organisation. Their publications aim at relevant stakeholders and citizens alike. They elucidate the importance of legal literature and resources on matters including digital identity,[120] re-introduction of data localisation laws, and consumer rights.[121]

You can read more at:
https://digitalrightslawyers.org/

### Data Science Nigeria (DSN), Nigeria
*Pillars: Access and Knowledge*

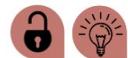

Data Science Nigeria (DSN) is a non-profit organisation (NPO) founded in 2016 that, through trainings and mentorships, seeks to democratise data science education and help Nigerian students develop their skills in advanced data analytics. They envision the development of an AI ecosystem in Nigeria that provides the country with the data science skills to position itself as a research leader and outsourcing destination.

With an experience and hands-on approach, DNS conducts annual bootcamps, classes, and workshops. In 2019, for instance, they completed an "AI Invasion" in 30 cities across Nigeria and introduced thousands of Nigerians to machine learning. The engagements consisted of a free class and free AI Knowledge Boxes. These boxes are 2-terabyte external hard drives that contain about 10,141 educational videos. To complement these actions, DNS works to engage C-level executives and decision-makers to share best practices and create knowledge on how data science can benefit society and business. They also opened three hubs where they provide free daily data science classes, access to AI projects, and incubate AI-enabled start-ups.

You can read more at:
https://www.datasciencenigeria.org/

---

[118] Nigeria Data Protection Regulation, 2019
[119] DRLI Cause List, 2021
[120] Afuye et al., 2020
[121] Usman et al., 2021

**W.TEC, Nigeria**

*Pillars: Equity, Access, and Participation*

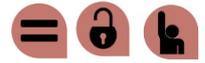

Arising out of a determination to minimise the gender gap in ICT skill development and education in Africa, W.TEC was set up as a non-governmental organisation (NGO) working at the intersection of technology and gender in Africa. They highlight the growing disparity between women and men in ICT and the importance of training and education in providing women with income-generating opportunities. By 2008, the first female-only technology camp was set up in Nigeria. Since then, W.TEC has grown to involve over 27,000 women and girls through camps, after-school clubs, and other clubs. These participatory networks have inspired over 85% of W.TEC participants to pursue careers in STEM.

Beyond training camps, the organisation also hosts annual programmes on online safety while conducting pivotal research on technology and gender. The launch of "IT4ALL" is a particularly important endeavour as it focuses on the importance of inclusivity in technology for young individuals with developmental disabilities. Not only do disabilities impact learning within classrooms, but also affects their social development.[122] As such, "IT4ALL" aims to integrate technology into regular activity for improved and independent functioning. The programme focuses on three age groups ranging between 3 to 18 wherein the curriculum includes fundamentals in computing and robotics.

You can read more at:
https://wtec.org.ng/

---

[122] Women's Technology Empowerment Centre, 2021

### Right2Know, South Africa

*Pillars: Identity, Power, and Knowledge*

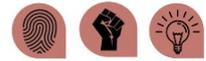

In July 2010, the South African Parliament introduced the Protection of Information Bill (also known as the Secrecy Bill), which aimed to replace the apartheid-era Protection of Information Act. However, civil society activists called attention to the potentially harmful effects of the new legislation. They stated that it was obstructing the free flow of information and extending 'the powers of government agencies to keep information out of the public domain and impose harsh penalties on whistleblowers, activists, and journalists who attempt to disclose such information in the public interest'.[123] To raise awareness and stop the Bill, civil society activists launched the Right2Know Campaign and wrote a Civil Society Statement justifying why they held that the proposal was problematic.

Right2Know has since grown into a movement that not only defends the rights to freedom of expression, access to information, and privacy, but also the right to protest and equitable access to the internet. They coordinate and mobilise communities who advocate for the right to access and share information, conduct advocacy campaigns, and publish research. On Universal Access to Information Day in September 2021, Right2Know hosted a panel event aiming to 'increase public awareness about people's right to access government information while promoting freedom of information'. This event came in the wake of the lax enforcement of the Political Party Funding Act after it was passed—a situation where many political parties were found to have failed to declare their funding sources before the election, even though this was required by the new law.[124]

You can read more at:
https://www.r2k.org.za/

---

[123] McKinley, 2013
[124] Electoral Commission of South Africa Ensuring Free and Fair Elections, n.d.; Felix & Khumalo, 2021

### amandla.mobi, South Africa

*Pillars: Power, Participation, Identity, and Access*

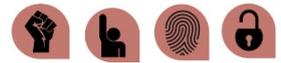

amandla.mobi is an independent, community advocacy organisation run by Black women in South Africa that has developed a petition app for civil society to build support around specific public interest campaigns. Their goal is to mobilise the collective power of Black women from low-income backgrounds, hold political and corporate interests to account, and shift power in ways that laws, policies, norms, and values benefit low-income Black women and their communities. They do this by acting at critical moments in a targeted, coordinated, and strategical way.

Their campaigns relate to a wide range of issues, such as gender-based violence, police brutality, and climate change, and address the power structures that maintain injustices. Campaigns are created considering what the community cares most about and the civic tools and tactics at their disposal. Petitions are then circulated on social media for supporters to sign. One of the campaigns run by amandla.mobi is entitled "Data Must Fall", which helped to increase access to mobile internet. amandla.mobi achieved this through submissions to the Independent Communications Authority of South Africa (ICASA), protests outside of the high court, amongst other actions, with the goal to reduce mobile networks discriminatory behaviour towards socio-economically marginalised groups, resulting in data prices dropping by 30% to 50%.

You can read more at:
https://amandla.mobi/

### DIG/SEC Initiative, Uganda

*Pillars: Access and Knowledge*

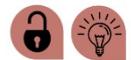

DIG/SEC is an initiative set up in Uganda to help journalists, activists, and human rights defenders to build the capacities needed for navigating and protecting themselves in digital spaces. In response to how many individuals working and advocating within the human rights spaces lack the skills to overcome and prevent cyber threats, DIG/SEC provides mentorship on digital security skills and emergency tech assistance to at-risk grassroots journalists.

Supported by Digital Defenders, DIG/SEC undertook a 21-day road trip across 20 districts in East and North Uganda to provide tech support (such as device check-ups and software installation), security consulting, and training to grassroots organisations and activists. The initiative reached over 10 organisations working in human rights defence after traversing over 1500 miles, all of which was recorded in a documentary titled "Security On Wheels: Digital Security Documentary".[125] Uniquely, the organisation has also launched "Security in Beats: The Album", which turns to edutainment through music to provide information on cyber threats and online harms, such as decryption or malware.[126]

You can read more at:
https://www.digisecinitiative.org/

---

[125] DIGISEC initiative, 2020
[126] See 'Security in Beats: The Album' project on https://www.digisecinitiative.org/

**Women of Uganda Network (WOUGNET), Uganda\*\*\***

*Pillars: Power, Access, Participation, and Equity*

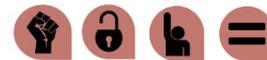

The critical nature of novel technology and innovative means of communication have necessitated their inclusion in women-led organisations in Africa. The integration of new technology into traditional forms of communication is also imperative. Women of Uganda Network (WOUGNET) was set up to support the fulfilment of these needs and to integrate the use of ICT tools among women in collective information sharing and reflection. Keeping in mind the gender disparities within Africa states in the ICT sector, the second ideal is particularly important. It encourages women's access to and use of novel technology alongside using existing communication infrastructure such as radios or televisions. The primary activities of WOUGNET involve information sharing and networking, technical support, and gender and ICT policy advocacy.

Alongside APC and Women's Net, WOUGNET launched a project to strengthen women's participation and decision-making in policymaking for ICT and governance with a particular focus on Uganda and Africa as a whole.[127] Through workshops and discussions with stakeholders, organisations (including Feminist Principles on the Internet, CIPESA and AMWA), and online activists, WOUGNET published pivotal research on internet governance and women's rights in Uganda. WOUGNET has also worked with the African Union and the United Nations (UN) on themes such as digital security, social media for citizen engagement, digital support programmes and other initiatives aimed at digital gender gaps and women-dominated industries. WOUGNET is participating as a Policy Pilot Partner in the "Advancing data justice research and practice" project.

You can read more at:
https://wougnet.org/

---

[127] Women of Uganda Network, 2016

### The Unwanted Witness, Uganda

*Pillars: Power, Access, Participation, Equity, and Identity*

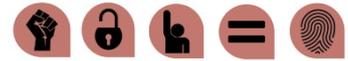

With instances of censorship and limits on free speech increasing, The Unwanted Witness was set up as a non-partisan and non-profit CSO by a community of bloggers, activists, netizens, human rights defenders, and others in civil society to strengthen free speech and advocate for accountability. Their research is broadly divided into data protection and privacy, freedom of expression, digital identity, cyber security, online gender-based violence, and surveillance within Ugandan socio-political and legislative domains.

Two notable projects under "Open Digital Tools," a programme aimed at developing digital tools for vulnerable communities in Africa, include "Amplified" wherein the organisation seeks to design and develop websites for CSOs to amplify their work; and "NetDAN" (Network Disruptions Audio Notifier) which involves tools for assisting people with disabilities.[128] The Unwanted Witness has been at the vanguard of advocacy campaigns against internet shutdowns and restrictions. In 2021, the organisation brought the arbitrary shut down of internet services by the Uganda Communications Commission—and other allied entities and institutes—to the high court in Kampala.[129] They observed not only that the actions infringed on rights listed in the constitution but also how the taxation of over the top (OTT) services affected users. Moreover, restriction of such services was said to have affected access to emergency services.[130] Keeping in mind the consequences of such taxation and shutdowns on employment and opportunities for people with disabilities, their advocacy work, alongside NetDAN, has been pivotal in protecting the rights of marginalised communities.

You can read more at:
https://www.unwantedwitness.org/

---

[128] Unwanted Witness, n.d.
[129] M/s The Unwanted Witness LTD v/s Attorney General, Uganda Communications Commission, Mobile Telephone Network (MTN) Uganda, Airtel Uganda, Africell Uganda (Kampala High Court, Civil Div., Misc. 50, 2021)
[130] Unwanted Witness, 2021

**iFreedom, Uganda**

*Pillars: Access, Knowledge, Equity, and Identity*

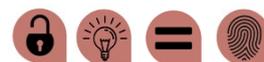

The weaponising of the internet against LGBTQI+ and Sex Workers (SW) in Uganda was found to have severe consequences for these marginalised communities' offline and online freedoms and rights, including those to freely associate and express themselves without fear of threats posed by state agencies and hackers.[131] Consequently, five LGBTQI+ and two SW organisations came together to establish the iFreedom network to defend digital rights and freedoms alongside the protection of security and safety in 2012. Since then, the network has grown to include 28 organisations—categorised into LGBTQI+, SW, and other human rights defenders—all of whom share a common goal of promoting digital rights and freedoms through research, advocacy, and digital knowledge capacity-building.

Beyond research and advocacy activities, the organisation also provides IT support, web design and hosting services, and computer training programmes. The training programmes are primarily catered towards sex workers regardless of their levels of skill. Beginning with a fundamental computer training phase, participants can then move to more complex digital security training, thereby contributing to the growth of an environment of informed and skilled sexual minorities and sex workers. Even amidst the many instances of violence and persecution of sexual minorities in Uganda, the organisation has become critical for equipping marginalised communities with the skills needed to defend themselves and progress in an increasingly digital world.

You can read more at:
https://twitter.com/ifreedomuganda
https://ifreedomugandanet.org/index.html[*]

---

[131] Amnesty International, 2020; Unwanted Witness, 2015

* At the time of publishing this repository, the website of iFreedom was down.

**Pollicy, Uganda** 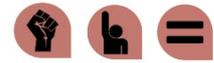

*Pillars: Power, Participation, and Equity*

Pollicy is a feminist collective that gathers data scientists, technologists, creatives, and academics to improve data and craft better life experiences. Their work emerges in response to the domination of large and foreign technology companies in African markets and to policies implemented by some governments that, according to Pollicy, have undermined freedom of expression, digital inclusion, and access to information and markets. Their mission is to improve data literacy among different stakeholders, promote the use of responsible data within civil society organisations and government agencies to improve service delivery, and foster the debate about how to use data in an ethical and responsible way. Underlying their work is the certainty of the need to re-imagine digital futures that consider the needs of traditionally marginalised groups rather than trying to fit them within existing frameworks. Therefore, they envision design processes of data collection, analysis, and data release where traditionally marginalised groups are consulted and have a seat at the table.

Pollicy is currently undertaking projects that cover research, trainings, workshops, and toolkits. It has explored, for instance, digital extractivism in Africa. Within the project "Automated Imperialism, Expansionist Dreams", they documented existing or potential policy responses to a set of problems, ranging from the hiring of digital workers in African countries by tech companies in a context of unequal power dynamics and lack of labour protections to the collection of users' data through zero-rating and other communication solutions and their exploitation for profit. The document also provided recommendations on how to address the root causes of these problems. Pollicy has also developed a project titled "Afro Feminist Data Futures", which approaches the production, sharing, and use of gender data as a way to empower feminist movements in sub-Saharan Africa. They explored the ways in which this knowledge can be translated into actionable recommendations for technology companies sharing non-commercial datasets.[132]

You can read more at:

https://pollicy.org/

---

[132] Iyer et al., 2021

### Common Cause, Zambia

*Pillars: Power, Access, and Participation*

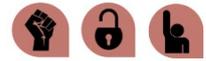

With a guiding vision for the empowerment of citizens and communities in participatory government and the defence of rights, Common Cause was set up in Zambia as a non-profit and non-governmental organisation. The support of activists and their knowledge was critical to the organisation's establishment and campaign for democratic governance in the country. Other objectives include the eradication of poverty as well as advancing human rights and sustainable development.

The causes for advocacy are primarily divided into accountability, gender equality, good governance, and youth development. In 2021, the organisation sent an open letter to the President of Zambia for open and secure access to the internet during the election period.[133] Highlighting the historical trend of internet shutdowns during electoral periods, often as a method of silencing, the #KeepItOn coalition of over 240 organisations globally supported the letter.[134] Common Cause also evidences instances of poor connectivity and disruptions during previous elections, usually targeting opposition strongholds. As a signatory to both the ICCPR and the African Charter on Human and Peoples' Rights, Zambia is expected to uphold and defend the rights and freedoms for expression, assembly, and access to information.

You can read more at:
https://commoncausezambia.org/

### AfroLeadership, Africa***

*Pillars: Participation, Access, Knowledge, and Identity*

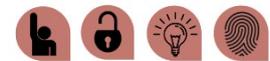

AfroLeadership is a CSO in Cameroon that aims to strengthen human rights, government, and democracy in Africa. They advocate for balanced power sharing between the State and citizens. Their work therefore focuses on transparency, accountability, and citizen participation in public policies.

AfroLeadership has an ongoing educational partnership with Good of All which consists of teaching universal rights online to the contemporary digital generation to combat the violent effects of hate speech and disinformation. Through their Pan-African network, AfroLeadership is helping to disseminate free educational video courses from the Universal Rights Academy across multiple African countries. Moreover, AfroLeadership is participating as a Policy Pilot Partner in the "Advancing data justice research and practice" project.

You can read more at:
https://afroleadership.org/

---

[133] *Open Letter to President Edgar Chagwa Lungu*, 2021
[134] Access Now, 2021

### Lawyers Hub, Africa

*Pillars: Power, Participation, and Knowledge*

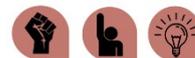

Established in 2016, the Lawyers Hub is a legal-tech organisation in Kenya that promotes access to justice throughout Africa through innovation and technology. To achieve this goal, they have myriad trainings, hackathons, and stakeholder engagement projects as well as law start-up incubation programmes. They also publish the *Africa Journal on Law & Tech*.

The Lawyers Hub identified the need for law to interact more with technology in a way that leverages and improves access to justice. For the organisation, this requires lawyers to understand areas that are impacted by technologies and to discuss its uses. Lawyers Hub has organised the Africa Law Tech Festival to gather policymakers, academia, and organisations from over 20 countries in Africa to discuss policy related to AI, privacy and data protection, digital tax, and lending, among other issues. It also runs the Africa Digital Policy Institute which provides training on digital identity systems, data governance and privacy, the gig economy and labour rights, antitrust, and competition law in the digital age.

You can read more at:
https://www.lawyershub.org/

### ELearning Africa, Pan-Africa

*Pillars: Knowledge, Access, and Power*

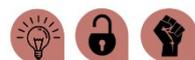

Through a global network of ICT professionals working in education and training, eLearning Africa promotes sustainable avenues to capacity-building and access to education, information, and training for holistic progress. Investment is directed at all levels of education, from primary education to vocational training. They address the importance of developing a Pan-African community that shares a common vision for the synthesis of ICT and education.

Within this focus, eLearning has promoted conferences, workshops, events, and podcasts while also publishing research and case studies. Following the global spread of COVID-19, they observed the pandemic's impacts on education,[135] raising awareness on the lack of technology infrastructure available for parts of the world to effectively carry out distance learning. Rural communities were identified as the most likely to be disadvantaged. Surveys conducted by eLearning Africa in this COVID-19 context also highlighted how educators did not receive sufficient financial support. However, respondents emphasised that the long-term consequence of the pandemic would be a revaluation of existing approaches to education with access and affordability of technology as a driving force. Nevertheless, without the widespread deployment of ICT tools, the inequality gaps may also arise in tandem. ELearning Africa's reports are notably available in multiple languages like English, French, and Portuguese.

You can read more at:
https://www.elearning-africa.com/index.php

---

[135] ELearning Africa, 2020

### Collaboration on International ICT Policy in East and Southern Africa (CIPESA), East and Southern Africa***

*Pillars: Power, Knowledge, Participation, and Access*

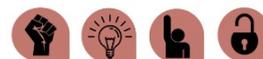

The Collaboration on International ICT Policy in East and Southern Africa (CIPESA) was established in 2004 in Uganda. It publishes research and analysis and promotes knowledge sharing and stakeholder engagement with the aim of enhancing the capacity of stakeholders to participate in ICT policymaking processes, enabling them to use ICT to support development and improve governance and livelihoods. They have a regional scope and focus on online freedom, ICT for democracy and civic participation, right to information, and internet governance.

Concerned that digital technologies are increasingly being used in ways which threaten digital rights—sometimes even by governments—CIPESA has partnered with the Internet Society on a project to 'work together for an open, secure, and trustworthy internet for Africa'. This collaboration aims to advance progressive Internet policy through knowledge sharing and the pooling of expertise surrounding issues of Internet policy and encryption. It also carries out stakeholder engagements throughout the region. CIPESA has also worked with partner organisations to fight against Internet shutdowns, has developed a digital rights community through its Forum on Internet Freedom in Africa (FIFAfrica), and is currently participating as a Policy Pilot Partner in the "Advancing data justice research and practice" project.

You can read more at:
https://cipesa.org/

### Paradigm Initiative, Western, Middle, and Eastern Africa

*Pillars: Power, Access, Participation, and Equity*

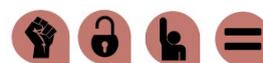

Paradigm Initiative is an NPO with offices in Nigeria, Zambia, Cameroon, Senegal, Zimbabwe, and Kenya. Founded in 2007, Paradigm Initiative works for the digital inclusion of underserved young Africans and the protection of their digital rights.

In response to poverty and unemployment among young people in Africa and the digital divides that reinforce limited access to digital opportunities, Paradigm Initiative develops capacity-building programs about digital skills, financial literacy, and entrepreneurship skills. It also connects young people to internships, entrepreneurs, and other work placement opportunities. The Paradigm Initiative also produces an annual report entitled "Londa", which is a 'title of Zulu origin calling for action to protect or defend'. The centring of the annual report in the concept of "Londa" supports their mission to protect digital rights of young people. Concerned about exclusionary practices and violations, "Londa" engages with stakeholders in reported countries and documents practices and violations, as well as the overall state of digital rights in Africa. The 2021 report contains information from 20 countries, including the impacts the COVID-19 pandemic has had on access and freedom of expression.

You can read more at:
https://paradigmhq.org/

**Counternarratives to the narratives often told about Africans by non-Africans (Part 1), Africa**
*Pillars: Identity, Power, and Knowledge* 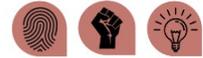

In the context of data sharing and open data practices within/for the African continent, part of the decolonial and justice work lies in acknowledging and centring the voices of African data practitioners themselves. This serves as a counternarrative to the narratives that are often told about Africans by non-Africans. To this end, a group of (mainly African) researchers conducted interviews with African researchers and data workers around issues of the challenges surrounding data sharing in the context of Africa. To protect the individual researchers and institutes involved, any identifying names were pseudonymised and the content of the interviews were aggregated into themes and written as Stories.

One of these stories narrates the experience of a researcher from a Global South nation, Wolonda, who receives a two-year grant from a large foundation from the Global North to make the mainstream markets work for livestock farmers. Although the data collection process was much more challenging than originally anticipated, the researcher collected various types of data including genomic data of crops and livestock from a cluster of villages in several counties of Southern Wolonda. Even with a local team well-versed in the norms and practices of the region, farmers seemed sceptical to give up their data. Data collection was aimed at providing intelligence on susceptibility of crops and animals to diseases and to analyse performance of the farm. The project was structured as a donor-funded research endeavour, with clear data collection and analysis objectives, yet no protocols for data storage and scope of use were defined, and there were no clear guidelines on whether the data could be shared with third parties. Furthermore, scant information was provided as to if and how the collected data would be used beyond the time period of the donor-funded project. Slowly, as the end of the two-year grant period approached, the data collection exercise also came to a close. Infrastructural challenges meant that the data was housed in a data centre in a country in the Global North which is home to the foundation that originally funded the research. The data collected for the research project was then used to found a for-profit company. In its research statement, the company articulated the need for precision agriculture, which it is making a reality using the data that was collected. Even though the trust established with the farmers in Southern Wolonda eroded due to the actions of the researcher, the researcher is confident that other farmers will see the potential of his data engine and believe in the power of its predictive analytics.

The researchers that aggregated the experiences of African researchers into this story aim to exhibit how foundations in the Global North may be eager to fund data collection projects in an attempt to create an oasis in regions perceived to be data-starved—to bridge data gaps and inequalities that appear to be merely inefficiencies. Yet, as is the case in this scenario, the lack of forethought into the ethical use of data—not only during the lifetime of the project, but also thereafter—can create irreparable harms to communities' well-being and make it difficult to build partnerships based on respect and mutual trust. This case may alienate the farmers, precisely the people whom this project was ostensibly seeking to support.

You can read more at:
https://dl.acm.org/doi/abs/10.1145/3442188.3445897

**Counternarratives to the narratives often told about Africans by non-Africans (Part 2), Africa**

*Pillars: Identity, Power, and Knowledge*

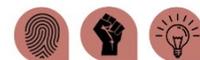

The group of researchers introduced in the previous case have also narrated the experience of young researchers working across different African countries who may encounter challenges when trying to collect, access, and share data. One of the stories that aggregates these experiences introduces a doctoral candidate from Bozatta who is researching the prevalence of gender-based violence in Nova Africa. The researcher's work is driven by the lack of data recording the prevalence of gender-based violence, an important step towards protecting the vulnerable, with the aim to collect this data and create a data platform to present gender-based violence cases within provinces in Nova Africa. The tool will aid data visualisation of these cases and will be available to the broader public. The researcher understands that her work is focused on a pressing issue as gender-based violence remains a persistent problem requiring attention from multi-faceted responses and commitments from the government, civil society, and other citizens. During the process of conducting the research, the doctoral candidate experiences a few unexpected encounters including many participants becoming emotional when sharing their experiences and asked that the session be discontinued. One participant even questioned why the interview was so intrusive and decided not to share further information. Her officemate tells her that in Nova Africa, it is challenging to convince participants to release information, especially as a foreigner. There is a perceived notion that foreigners are here to steal from them, whether that be data or resources. This, she learns, is in part reflective of and caused by divisions across communities in Africa due to colonialism.

By narrating this story, the group of researchers make visible the challenge of collecting, accessing, and sharing data. This challenge is particularly stark when the researcher is not a local to the region or community in which the data is being collected. Differences in languages and cultures, as well as power imbalances breed suspicion about motives, impeding data access and sharing. With more than 1,500 languages spoken across the African continent, locals residing in rural areas are more likely to communicate in their dialect, with potentially little or no understanding of national languages. This concern may, at times, stand in contrast to the fact that there is a pressing need for high-quality data to understand the prevalence and impact of issues such as gender-based violence, among numerous other forms of injustices and discrimination.

Furthermore, these narratives suggest that discussions on data sharing in the African context often treat the continent as a monolith and frame problems as a Global North/Global South issue. However, the complex data sharing ecosystem can lead to challenges even for those based in the continent. Furthermore, certain hesitations with sharing data are rooted in colonial-era extracting practices whose impact can be felt to this day.[136] Numerous efforts to mitigate inequalities in data access and sharing in the continent, and especially those driven by initiatives outside of the continent, may not be attuned to the complex data landscape resulting from a diversity of needs, priorities, and experiences of those in the continent. The stories also manifest that data sharing and open data initiatives, especially when designed, developed, and advocated by the Global North scientific community need radical rethinking before they can be blindly applied to or imported into Global South contexts. At a minimum, researchers need to ask who benefits from data sharing and work accordingly to dismantle colonial, historical, and structural factors that place Global North researchers as the main benefactors of African data.

You can read more at:
https://dl.acm.org/doi/abs/10.1145/3442188.3445897

---

[136] Abebe et al., 2021

# Asia

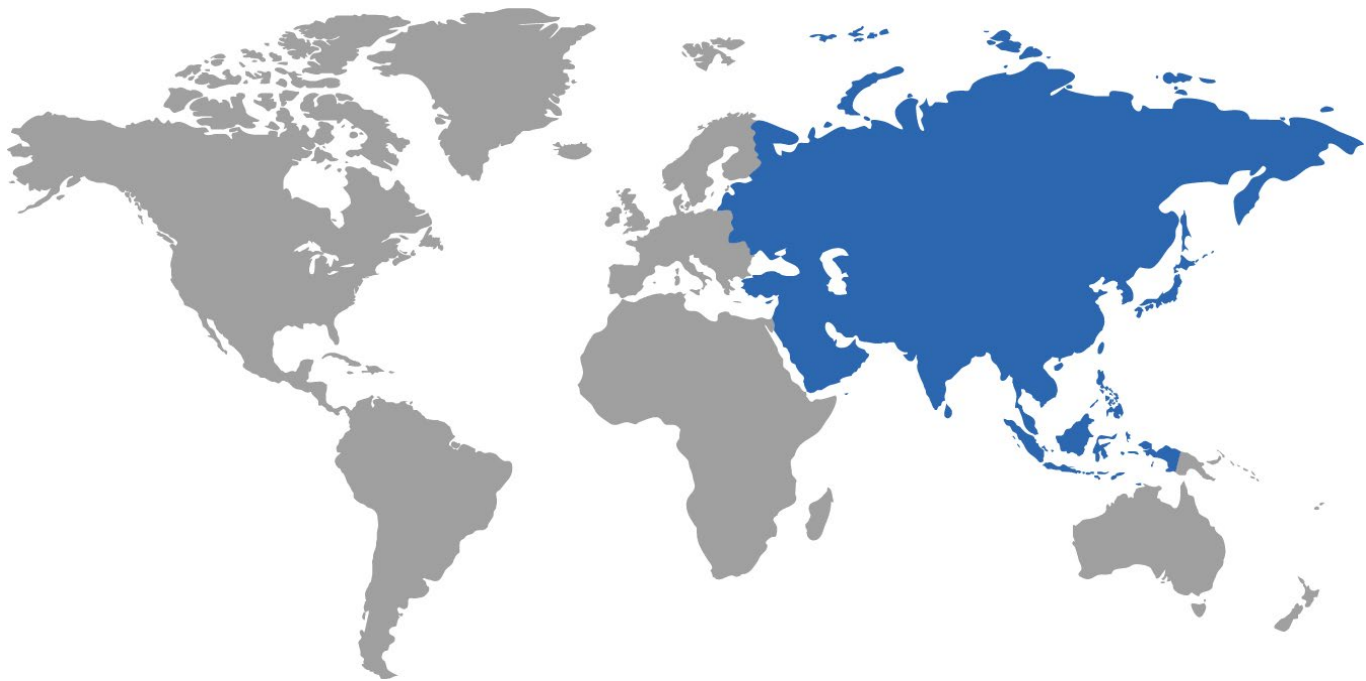

**Open Data China, China\*\*\***

*Pillars: Knowledge*

Originally founded with local ambassadors of the Open Data Foundation, Open Data China has since been actively engaged in advancing an open digital future in China for all forms of data, including government data, scientific research, and cultural content. The organisation has been working on digital rights since 2021 through research projects, workshops, and the production of guides.

"A layman's guide to Digital Rights" was a flagship project launched in 2021. As part of the production of the guide, workshops for the general public were held that aimed to generate co-produced content. Through interactive tasks, citizens were introduced to essential knowledge about algorithms, data, legal tools, and novel paths to collective action for the protection of rights. The content of the workshops will now inform a digital guide and toolkit, aimed at groups such as journalists and students. This is expected to be released in 2022. Through collaboration with the Shanghai Baiyulan Open Institute, Open Data China has also been pivotal to the creation of standardised data licenses for AI training and data publication.[137] The organisation is also working on a Fintech data collaboration as part of the peer learning network set up by Microsoft and the Open Data Institute.[138] Open Data China is also participating as a Policy Pilot Partner in the "Advancing data justice research and practice" project.

You can read more at:
https://cn.okfn.org/

---

[137] Bai-Yu-Lan, 2021
[138] Feng, 2021

**Centre for Internet and Society (CIS), India** 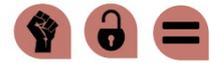

*Pillars: Power, Access, and Equity*

Through its interdisciplinary approach to research, the Centre for Internet and Society (CIS) is a NPO in India working to understand the nexus of social processes and technology from both academic and policy perspectives. Research produced by CIS has focused on themes such as governance, cyber-security, and digital accessibility. They emphasise a need for openness in data, software, standards, educational resources, and media.

Based in the software and technology hub of Bengaluru, CIS promotes equitable access to knowledge through the publication of open and multilingual resources such as reports and films. The organisation also aims to improve understanding of legislation and policies through workshops. CIS has collaborated with multinational organisations, national governments, as well as Indian state governments. Notably, the organisation has undertaken extensive research and advocacy campaigns on digital and internet accessibility for India's estimated 70 million persons with disabilities.[139] Particularly, they have published a comprehensive report on all policies and legislation related to disabilities as well as organising "Right to Read" campaigns across 4 major metropolises of India—Kolkata, Delhi, Mumbai, and Chennai. CIS also provides detailed policy guidelines and recommendations with a focus on people with disabilities for disaster management response, electoral websites and tools, and accessible mobile phones.

You can read more at:

https://cis-india.org/

---

[139] CIS and the Office of the Chief Commissioner for Persons with Disabilities, Department of Disability Affairs, Ministry of Social Justice & Empowerment, 2016

**Digital Empowerment Foundation, India\*\*\*** 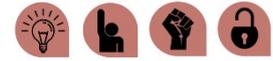

*Pillars: Knowledge, Participation, Power, and Access*

At the turn of the millennium, noting that India had limited infrastructure and capacities for large swathes of the population, the Digital Empowerment Foundation was set up with a goal to leverage digital tools for empowerment. Through 1000 Community Information Resource Centres and 10,000 personnel located across 24 states and 135 districts, DEF actively engages with local communities for information empowerment and improving digital literacy. The organisation emphasises the importance of access to infrastructure and knowledge as fundamental to preventing exclusion. Programmes under the DEF cover themes including research and advocacy, infrastructure, citizen services, and social enterprises.

The handloom and handicraft sector, employing over 700,000 artisans spread across the sub-continent, are the primary focus of the DEF's DigiKargha initiative which is modelled to steward access to inclusive and decentralised ICT tools.[140] The initiative is assisting cluster artisans with digital tools to improve skills, designs, marketing, and entrepreneurship while creating sustainable employment avenues for the youth like digital designing. While many Indigenous handloom textiles have been affected by the introduction of power looms and middle-man, the multifarious sub-projects under DigiKargha centralises the role of handloom artistry—such as Gammcha weaving and Zari embroidery in Barabanki district and Ikkat dyeing in Poochampalli–in achieving social and economic growth. Different projects are supported by organisations including Nokia, the UNDP, and Microsoft. The initiative has also digitally archived over 100 songs from various artisan clusters. Digital Empowerment Foundation is also participating as a Policy Pilot Partner in the "Advancing data justice research and practice" project.

You can read more at:
https://www.defindia.org/

---

[140] Digital Cluster Development Programme, n.d.

## IT for Change (ITfC), India

*Pillars: Equity, Access, Power, and Participation*

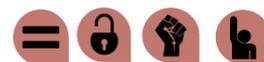

Maintaining the perspective that contemporary approaches to the role of digital technology in development are rooted in techno-utopianism and market-fundamentalism—often a combination of both—IT for Change (ITfC) aims to reimagine ecologies of digital development, focussing instead on centring aspects of human rights, social justice, and equity. The organisation is registered as a non-profit, non-governmental society rooted in objectives of collaboration across sectors. They maintain Special Consultative Status with the United Nations' Economic and Social Council.

ITfC has focussed their attention on areas of education, gender, internet governance, and democracy for equitable access and use of ICTs. Through research, advocacy, field projects, the promotion of new curricula, and network building, the organisation has presented new approaches to data justice in national, regional, and global domains of operation. Publications related to gender in cyberspace include works on the exclusionary role of Aadhaar (India's digital identity database),[141] feminist digital activism in the Global South, cybercrime,[142] and e-commerce.[143] Critical research on governance and democracy has explored the importance of incorporating safeguards and guidelines in India's National Digital Health Mission (NDHM) as well as the need for a personal data protection framework.[144] The mission to improve and equitably inculcate ICTs in education has been a key driver of ITfC's policies to design and execute online classes for government and aided high schools in Karnataka during the COVID-19 lockdowns.[145]

You can read more at:
https://itforchange.net/

---

[141] Kirasur, 2022
[142] IT for Change, 2022
[143] IT for Change, 2021
[144] IT for Change, 2022
[145] IT for Change, 2021

**Software Freedom Law Centre (SFLC.in), India**

*Pillars: Access and Knowledge* 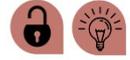

The Software Freedom Law Centre (SFLC.in) has brought together legal professionals, policy analysts, technologists, and students as a legal services organisation with the aim to protect freedom, privacy, and civil liberties in digital spaces. Based in New Delhi, the organisation also involves volunteers from across India. The SFLC.in promotes innovation in the digital space while providing free legal advice and research necessary for stakeholders to enact equitable and justice-driven decisions in the adoption of technology and software.

Current projects at the SFLC.in are aimed at a range of topics that includes digital privacy, net neutrality, freedom of speech, software patents, cybersecurity, and AI. Of importance is the SFLC.in's work in providing digital security training that assists in educating attendees on privacy and security when considering the adoption of increasingly data-driven technology.[146] Towards this endeavour, SFLC.in has produced guides on pressing themes such as browsing safety, Free and Open-Source Software (FOSS), and anonymous file sharing. The project is not focused exclusively on tech-based companies; SFLC.in emphasises the need to engage with a host of industries and communities as observed in the workshops they have catered towards journalists and government officials. SFLC.in's other research includes commentary on legislation and policies centred on data, security, and surveillance in India.

You can read more at:

https://sflc.in/

---

[146] Information about 'Digital Security Training' can be found at https://security.sflc.in/

**The Institute for Policy Research and Advocacy (ELSAM), Indonesia**

*Pillars: Power and Knowledge*

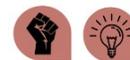

Established in 1993, the Institute for Policy Research and Advocacy, Lembaga Studi dan Advokasi Masyarakat, (ELSAM) is a human rights organisation that works towards the protection of rights as envisaged by the Indonesian Constitution as well as the Universal Declaration of Human Rights (UDHR). The activity of the organisation is categorised into four areas: the incorporation of human rights principles into the policymaking process, expansion of information on human rights for policy advocacy, promotion of transitional justice to resolve past abuses, and human rights education and training. Technology and human rights constitute a primary focus alongside other areas including transitional justice, eco-justice, fundamental freedoms, and business and human rights.

Aiming towards the expansion of knowledge and advocacy in these areas, ELSAM publishes policy briefs, working papers, reports, and so forth. They maintain a research perspective primarily focused on Indonesia. The organisation has also identified the global challenges raised by the spread of disinformation while recognising the corresponding threats posed to election integrity in the country. Despite the concerns about disinformation raised by government institutions, ELSAM notes that disinformation research has generally been limited. They note the role played by platforms such as WhatsApp in influencing political patterns in Brazil. They also highlight the potential for collaborations between the Indonesian government and technology giants like Facebook and Google that could lead to an improvement of online safety and verification.[147] Other reports have covered the importance of principles of privacy and freedom of expression in Indonesia.

You can read more at:
https://elsam.or.id/en/

---

[147] ELSAM, 2019

**Jordan Open-Source Association (JOSA), Jordan**
*Pillars: Knowledge and Equity*

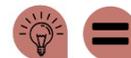

Based in the capital city of Amman, the Jordan Open-Source Association (JOSA) is a NPO aimed at promoting openness in technology through free access to non-personal information—such as software, content, network protocols and architecture—as well as the protection of personal information through robust legislation and technological frameworks.

Most recently, one of their projects has been aimed at improving the gender gap on Wikipedia with a particular focus on Arab women under the 'WikiGap Challenge'.[148] Wikipedia has played a pivotal role in free access to information through its process of community-led information gathering, editing, and publication. The barriers that may exist to prevent access to relevant information and knowledge on a plethora of themes and subjects are being mitigated through the platform. Notwithstanding debates on using Wikipedia pages as a reliable reference, studies have noted the prevalence of Wikipedia usage and the progress within the platform to promote high-impact journal citations and well-referenced entries.[149] The platform has increasingly emphasised the necessity and role of open-source journal articles in its editing guidelines. Nevertheless, there exist significant gaps in the available literature with gender being a leading factor of imbalance. Among other projects, JOSA has noted and worked towards minimising the gaps in representation and information on Arab heritage and women through the global Wikigap edit-a-thon.[150] The organisation's work includes events (comprising hackathons, public panels, and workshops) alongside the periodic release of blogposts wherein the challenges of censorship, corporate and government surveillance, arbitrary detention, digital rights, and data justice are given a space of prominence.

You can read more at:
https://www.josa.ngo/

---

[148] Wikimedians of the Levant/Reports/2020/WikiGap, 2020
[149] Duede, 2015; Teplitskiy et al., 2017
[150] Rawashdeh, 2019

**SMEX, Lebanon**
*Pillars: Power and Participation* 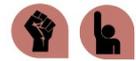

SMEX is an NGO that aims to strengthen access to information and to protect digital rights in the Arab region through research, campaigns, and advocacy. The primary scope of operations for the organisation includes monitoring digital rights violations, expanding on internet law and policy through research, advocating for an open and secure internet, improving online safety, and providing robust resources on issues within the digital space. In an accessible format, SMEX routinely provides extensive research reports on challenges and the state of the art of digital services across West Asia and North Africa. They have also released the Digital Safety Helpdesk to assist human rights defenders, journalists, alternative media, activists, and CSOs in navigating instances of online harm.

In 2021, SMEX released a report on the role played by contact-tracing apps in Lebanon during the COVID-19 pandemic as a form of social control.[151] They note that the handling of the pandemic was haphazard and Ma3an, the contact tracing app released by the Lebanese government, did little to quell the proliferation of cases. More importantly, they note that the app was not only riddled with functional flaws (like a limited user base and bugs) but also played a potentially damaging role in instituting a digital mechanism for social control. SMEX highlights in the report that during the initial waves of the pandemic, protests and government dissatisfaction were identified as a threat that could be mitigated through the enforcement of lockdowns. However, the protests only continued to increase, and it was reported that Syrian refugees were disproportionately affected by the app's concentration in neighbourhoods with higher percentages of Syrians.[152] Thus, they concluded that the app had the potential to be a surveillance tool while noting that platforms failed to integrate a "privacy by design" approach.[153]

You can read more at:
https://smex.org/

---

[151] SMEX, 2021
[152] Human Rights Watch, 2020
[153] SMEX, 2021

**KRYSS Network, Malaysia**

*Pillars: Identity, Equity, and Knowledge*

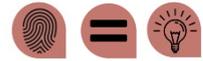

KRYSS Network was founded in 2002 and its work is rooted in gender equality, non-discrimination, and feminist principles to make people aware of and enable them to exercise their freedom of opinion and expression, public participation, and bodily autonomy. They predominantly work with women, girls, and gender non-confirming persons. Through workshops, research, communication campaigns, and collective action, KRYSS Network focuses on shaping public discourse, eliminating hate speech and online gender-based violence, and promoting institutional reforms and social change.

Concerned by the obstacles women and girls face to participate in the public realm—sometimes even leading to total denial of such participation—KRYSS developed a project entitled "Virtual Workshop Roadshow" that aims to address online gender-based violence and shape public narratives. The workshops have a collaborative approach and are tailored to the demands and needs of the people participating. KRYSS has also created a Zine to reach young people and raise awareness of online gender-based violence.

You can read more at:
https://kryssnetworkgroup.wordpress.com/

**Body & Data, Nepal**

*Pillars: Access, Identity, and Knowledge*

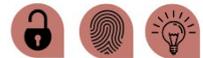

Body & Data was set up in Nepal in 2017 with the objective of improving understanding and access to information for women, queer people, and marginalised groups within the digital space. The organisation primarily conducts legal research and analysis, workshops to improve digital rights and safety, and advocacy programmes at the intersection of feminist and digital rights spaces.

The organisation has worked to produce guidelines for an inclusive and intersectional Nepali language.[154] This working document approaches the role of data privacy through a critical feminist lens. With a goal to improve accessibility and safety in digital spaces, the organisation has also mapped out relevant legislation on gender-based online violence. They are likewise working to improve understanding of Nepal's Information Technology Bill by providing specific critiques on the implications of surveillance on the freedom of sexual minorities and marginalised communities. Their most recent publications include an evaluation of the harms of misinformation in the COVID-19 pandemic. This evaluation also includes an analysis of the mechanisms needed to mitigate further proliferation of misinformation through fact-checking tools and impact assessments which consider the needs of marginalised or vulnerable minority groups.

You can read more at:
https://bodyanddata.org/

---

[154] Body & Data, 2020

**Digital Rights Foundation (DRF), Pakistan\*\*\***

*Pillars: Identity, Access, and Equity*

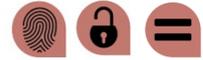

Registered as an NGO in Pakistan, the Digital Rights Foundation (DRF) works towards protecting and defending human rights, democracy, and good governance in digital spaces. DRF particularly focuses on strengthening the freedom of expression for women while maintaining the importance of access to information and data privacy for individuals across the globe. As such, DRF routinely produces actionable documents and policy research on the impact of surveillance and censorship on privacy for government.

DRF research has currently covered pressing challenges and themes such as access to data, mapping of gender-based violence, and political misinformation. Asserting the importance of equity and justice on digital platforms, the DRF has also explored instances of threats to journalism and media personnel in Pakistan. DRF has noted how journalism is increasingly facing a 'double-bind' wherein the dual actions of government silencing and corporate influence on social media platforms have had a considerable impact on the erosion of the freedom of expression.[155] They emphasise a need to evaluate the gendered impacts of online abuse and censorship particularly for female journalists.[156] Regarding the COVID-19 pandemic, DRF released multiple documents on cyber harassment, false information, and distribution of welfare.[157] The organisation has also periodically reviewed policies and draft legislation related to digital and cyber infrastructure and security and is currently participating as a Policy Pilot Partner in the "Advancing data justice research and practice" project.

You can read more at:
https://digitalrightsfoundation.pk/

---

[155] Digital Rights Foundation, 2022
[156] Digital Rights Foundation, 2019
[157] Digital Rights Foundation, Hamara Internet, and UN Women, 2019

Data Justice Stories: A Repository of Case Studies    71

**7amleh—The Arab Center for the Advancement of Social Media, Palestine**
*Pillars: Access and Knowledge*

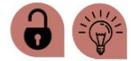

Rooted in a mission to create a free and fair digital space for Palestine, 7amleh—The Arab Center for the Advancement of Social Media was set up as a NPO that advocates, provides training, and produces research on digital rights, activism, and security capacity-building for Palestinians. Numerous instances of digital violations recorded by 7amleh have catalysed efforts to monitor the violations and provide avenues to limit the moderation of Palestinian content on social media.

7amleh's research in 2021 has focused on digital surveillance in East Jerusalem, the perception of Palestinian CSOs, assessment of digital performance, hate speech, privacy and data protection, guides for journalists, and digital rights abuses. The organisation has also submitted statements on digital rights violations and digital erasure of Palestinians to both international agencies like the UN and European Union as well as technology companies.[158] 7amleh has campaigned against Google in opposition to its failure to label Palestine on maps as well as against Facebook for its disproportionate and harmful moderation of Palestinian content on the platform.[159] Maintaining the importance of journalism in reporting digital rights violations in the region and worldwide, 7amleh has provided a detailed guide for journalists which highlights the importance of diversifying resources, upholding the safety of individuals, and providing a platform for the unheard.[160]

You can read more at:

https://7amleh.org/

---

[158] Circulated by the Security General in accordance with Economic and Social Council resolution 1996/31; A/HRC/47/NGO/2021; APC, 2021
[159] 7amleh, 2020a, 2020b
[160] 7amleh, 2021

**Foundation for Media Alternatives, Philippines**

*Pillars: Equity, Access, and Participation*

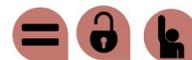

Established in 1987, Foundation for Media Alternatives is an NPO based in Philippines that in its early years helped CSOs and NGOs with their communication needs. Since 1997 it helps citizens and communities to appropriate ICTs and strategically use them for democratisation and sustainable development. It conducts policy research, advocacy, and capacity-building programs.

 The Foundation for Media Alternatives, concerned about the risks and vulnerabilities that women and girls face when accessing the internet, have analysed the harms women are exposed to through and within ICTs. Among their work, the organisation has examined the effects of digital platforms on domestic work in the Philippines, offering recommendations for policies which would ensure equitable working conditions for domestic workers working in platform-mediated contexts. Similarly, since 2012 the Foundation for Media Alternatives maps media reports of online gender-based violence in the country. These reports aim to make visible the trends of violence against women online, including data about devices or platforms used, the harms faced by women, their familiarity with the perpetrator, and whether the incidents were followed by reports to authorities, investigations, and trials.

You can read more at:
https://fma.ph/

**DigitalReach, Singapore**

*Pillars: Power, Participation, and Equity*

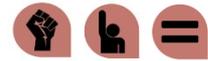

Founded in 2019, DigitalReach aims to protect and defend digital rights and internet freedoms in Southeast Asia. Their three-pronged approach to the impact of technology on human rights consists of research and monitoring, advocacy, and community building and empowerment. Under these umbrella areas of activity, they have published analytical reports on the monitoring of technology at national and international levels, advocated for digital rights through partnerships, and brought together community members for collaborative actions that seeks to accomplish their objectives.

The onset of the COVID-19 pandemic has been subsequently followed by a turn to technology to assist public health measures.[161] Noting that 6 of 10 members states of the Association for South-East Asian Nations (ASEAN) had employed contact-tracing apps, DigitalReach conducted extensive research into the potential consequences of mass gathering of personal data in 2020.[162] They evaluated the use of such apps according to international standards on surveillance and the right to privacy as set out in Art. 12 of the UDHR and the considerations laid out by the World Health Organisation (WHO). Through this research, they found that the centralised data storage and the use of Bluetooth Low Energy (BLE) methods would expose swathes of personal data to potential cybersecurity threats and unjustified surveillance. Additionally, they noted the widespread lack of transparency as a pressing challenge. This research has beneficially informed advocacy campaigns about the myriad problems surrounding the design and mass enforcement of digital surveillance tools. The organisation has also published press releases and open letters that relate to similar issues in the digital domain.

You can read more at:
https://digitalreach.asia/

---

[161] Du et al., 2020
[162] DigitalReach, 2020

**Thai Netizen Network, Thailand**

*Pillars: Power, Access, and Participation* 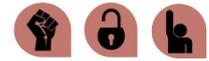

The Thai Netizen Network is a NPO founded in 2008 that seeks to uphold digital rights and civil liberties, as well as the work of human rights defenders in digital environment. Their work focuses on five main topics: access to information, freedom of opinion and expression, privacy, participatory internet governance, and rights over information resources.

Within their projects, they have reported on the internet freedom and online culture situation in Thailand and developed capacity-building programmes and workshops on diverse topics, such as cybercrime law, digital forensics, digital security, and ICT for development. The Thai Netizen Network is a partner of Open Observatory of Network Interference (OONI). Along with them and the Sinar Project, they have identified that from November 2016 to February 2017, 13 websites, including nypost.com, dailymail.co.uk, and material from Wikileaks, were blocked in Thailand.[163] Since 2017, Thai Netizen Network has worked with other organisations to monitor the Personal Data Protection, the National Cybersecurity Bill, and the amendment to the Computer Crime Act, which allowed for 'government oversight of the internet' and repression.[164] Additionally, in partnership with Privacy International, the Thai Netizen Network has also submitted a report addressed to the Human Rights Council calling attention to the threatening environment that could thwart the protection and promotion of the right to privacy in Thailand.[165]

You can read more at:

https://thainetizen.org/

---

[163] Chachavalpongpun, 2021
[164] Flynn, 2020
[165] The Thai Netizen Network & Privacy International, 2015

**Coconet, Southeast Asia**

*Pillars: Knowledge and Participation* 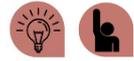

Since its launch in 2019, Coconet serves a dual function as a both a mechanism for network building and a platform for digital rights. The organisation works in collaboration with EngageMedia and the Association for Progressive Communications among others. Coconet aims to accelerate the digital rights movement in the Asia-Pacific by focusing on the role of the internet in discussion and dissent in the region.

Digital and internet spaces have, in recent times, been subject in Southeast Asia to increasing surveillance, censorship, deterioration of freedoms and rights, and incidents of false or misleading information. In response to this, Coconet works to provide tools and strategies to effectively use cyberspace for advocacy and activism while improving equitable access to research and content on related themes of digital hygiene and AI. Coconet publications have focused on the harms and impacts of multifarious legislation within the states of South-East Asia. For instance, they have provided detailed research on Myanmar's draft Cyber Security Law which will empower the government to enact internet shutdowns and increase military control and oversight of the internet.[166] Another example is the case of Indonesia's Ministerial Regulation 5 (MR5) which has the potential for 'prepublication censorship'.[167] Importantly, Coconet's community-driven platform is working toward providing multilingual resources for data justice advocacy and activism.

You can read more at:

https://coconet.social/

---

[166] Coconet, 2021a
[167] Coconet, 2021b

**EngageMedia, Southeast Asia\*\*\***

*Pillars: Power, Knowledge, and Access*

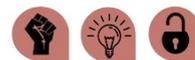

Founded in 2005, EngageMedia has been at the forefront of open-source filmmaking on social issues for the defence of human rights and democracy in both online and offline environments. The NPO aims to assist advocacy groups and grassroots organisations through the provision of accessible knowledge, documentaries, and resources for effective communication. Additionally, they work towards capacity- and network-building, while promoting open and secure technology. Their primary sphere of operations extends across the Asia Pacific.

In 2021, EngageMedia launched #HumanOnTheLine, a campaign to raise awareness of digital rights and safety in Thailand. Notwithstanding positive externalities of the proliferation of internet tools and technology, the campaign has noted the widespread instances of surveillance, censorship, online gender-based violation, and other violations of rights and freedoms.[168] Effectively serving to raise awareness by boosting news, media, and stories on digital rights and challenges, the campaign has also been accompanied by relevant research, podcasts, and guides for securely accessing the internet. Other notable projects include a partnership with TunnelBear to provide VPN services for human rights defenders and the promotion of films like Black Out that document the weaponising of technology and the internet. Moreover, EngageMedia is participating as a Policy Pilot Partner in the "Advancing data justice research and practice" project.

You can read more at:

https://engagemedia.org/

---

[168] EngageMedia, 2021

Data Justice Stories: A Repository of Case Studies    77

# Americas

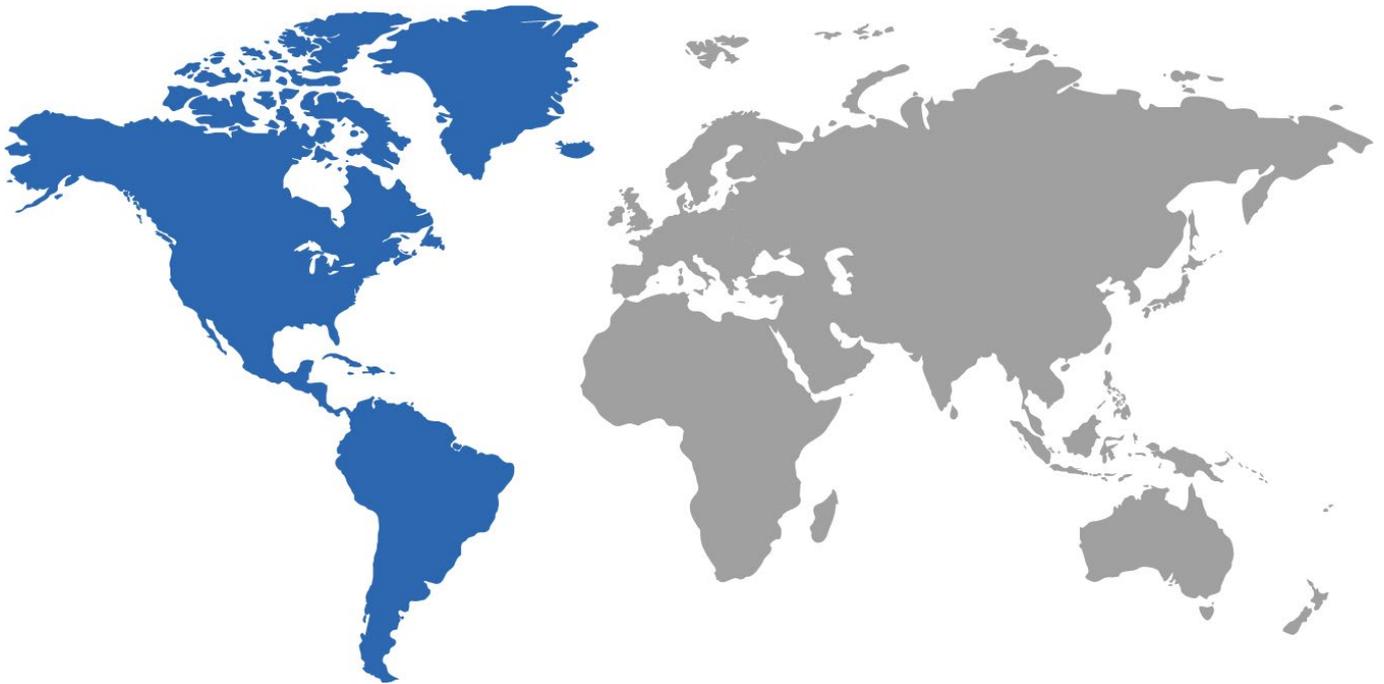

**NiUnaMenos, Argentina**
*Pillars: Equity, Access, Power, Identity, and Participation*

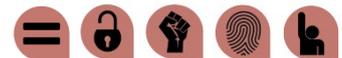

In response to the alarming figures of femicides in Argentina and triggered by the murder of a 14-year-old girl by her partner in 2015, a group of female artists, academics, and journalists called for a protest against gender-based violence under the slogan #NiUnaMenos ("Not One Less"). The protest not only represented one of the largest feminist demonstrations in Argentinian history, but it also spread across other countries in Latin America and placed a range of gendered-related issues on the public agenda. The movement asked for concrete State actions such as the creation of a Registry of Femicides, a Lawyers Bureau for Victims of Gender Violence, mandatory training on gender issues and violence against women for all public servants, and the legalisation of abortion.[169]

The #NiUnaMovement collective also created the national index of sexist violence in Argentina in 2016. The tool aimed to generate information that was not yet available. Focusing on the varying and intersecting ways in which oppression and inequality take place,[170] the index provided empirical evidence to inform and improve public policies related to gender violence and the protection of women at risk. The database was collected by the community, and as Chenou and Cepeda-Másmela state, it provides an example of how, through the appropriation of technology, activists can respond to immediate needs and produce alternative imaginaries of data justice[171] that can disrupt the normalisation of power imbalances.[172]

You can read more at:
https://journals.sagepub.com/doi/abs/10.1177/1527476419828995
http://niunamenos.org.ar/
http://contalaviolenciamachista.com/Informe-ejecutivo-final.pdf

---

[169] Leszinsky & Dewick, 2021
[170] Langlois, 2020
[171] Chenou & Cepeda-Másmela, 2019
[172] Langlois, 2020

### Asociación por los Derechos Civiles (ADC), Argentina

*Pillars: Power, Participation, Access, and Equity*

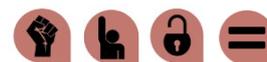

Asociación por los Derechos Civiles (ADC) is an Argentinian CSO founded in 1995 that seeks to defend and promote civil and human rights in Latin America. In addition, it aims to promote democratic strengthening and advocate for an inclusive society, with special attention to groups in vulnerable situations. They conduct research on cutting-edge issues, advocacy and communication strategies, and strategic public interest litigation.

ADC leads the project "Con Mi Cara No" (translated as "Not with my face"), a microsite that gathers publications and information about the uses and risks of FRTs in Argentina and Latin America. As a result of their monitoring activities, ADC has also developed a surveillance map, which visualises current development of and intentions to use of FRTs in Argentina. Within this line of work, in 2019, they initiated an action to demand the Supreme Court of Justice of Buenos Aires to declare the unconstitutionality of FRTs. ADC also manages "Quiero Mis Datos" (translated as "I Want My Data"), a web application that supports users in exercising their right of access by facilitating the drafting and sending of subject access requests to data collected, used, and shared by companies. To date, the app has been used to send 806 requests.

You can read more at:
https://quieromisdatos.adc.org.ar

### Internet Bolivia Foundation, Bolivia***

*Pillars: Participation and Knowledge*

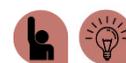

Internet Bolivia was founded in 2018 with the goal of strengthening access to a safe, free, and democracy-enhancing internet. They defend privacy and personal data protection, freedom of expression, and access to public information, and work to tackle digital gaps, by creating knowledge and promoting good practices.

Their prior work on data justice includes a project undertaken in partnership with the Digital Defenders Partnership and Access Now that saw them establish a helpline, SOS Digital, which provided rapid responses to assist actors in situations of vulnerability to digital threats. The SOS Digital platform also has a repository with preventive and responsive strategies to digital violence and tutorial videos on how to use digital and information security tools. The organisation has also provided resources, public tools, and reflexive content within pre-electoral processes that promote collective participation and debate, seeking to strengthen the democratic process. They monitor electoral processes, seek to guarantee the public's effective exercising of democratic rights, participate in internet governance forums, advocating for a code of best practices and policies for public organisation's use of social media, and promote the adoption of digital tools that support democratic participation and monitor the electoral process. Internet Bolivia is also participating as a Policy Pilot Partner in the "Advancing data justice research and practice" project.

You can read more at:
https://internetbolivia.org/

**Brasil.io, Brazil***
*Pillars: Power, Access, Participation, and Knowledge*
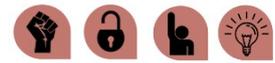

In Brazil, government data is not often accessible, readable, and organised enough for interested people, communities, or organisations to use and analyse. Brasil.io is a platform created in 2018 and led by a group of volunteers that aims to make this data accessible in an open, transparent, and collaborative way. Available datasets include, for instance, direct expenditures of the federal government and Brazilian election data. The project has thus far helped to propel data-driven journalism and civic innovation projects, such as the monitoring of public spending.

More recently, in response to the poor quality and lack of disaggregated data about COVID-19 in Brazil, the group captured data about cases and deaths related to COVID-19 at the municipality level. This data was then converted, cleaned, and made available in a structured and open format. Not only has this work helped to expose underreported data, but it has also served as a source for reporting, examining, and forecasting the impact of COVID-19 in the country.

Brasil.io interrogates power, which manifests in restrictions of access to public data, by promoting what they refer to as "data liberation". This has been particularly important during the COVID-19 pandemic when their work informed the public and challenged narratives that minimised the magnitude of the health crisis.

You can read more at:
https://brasil.io/home/
https://data-activism.net/2020/06/bigdatasur-covid-liberating-covid-19-data-with-volunteers-in-brazil/

**Favelas 4D, Brazil***
*Pillars: Participation, Power, and Equity*
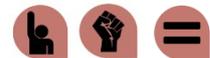

Favela 4D is a project led by the MIT Senseable City Lab in collaboration with Rio's City Planning Commissioner and BRTech 3D that seeks to analyse the morphology of Rocinha, the largest favela (an urban area of informal settlements affected by socioeconomic deprivation) in Rio de Janeiro.

Informal urban settlements often remain unmapped, due to their unplanned, irregular, and complex structures. Technologies such as Street View, for instance, are unable to access narrow alleys or stilt structures. In response, the project uses handheld LiDAR (Light Detection and Ranging) for terrestrial laser scanning. Spatial data of Rocinha, including street width, street elevation, density of facades, and variance in facade height and street canyon, are captured with great precision and used to create a 3D environment.

The collection and use of these spatial data, as understood by Favela 4D, could provide a starting point for understanding the patterns of development of informal settlement, making them visible to present and future urban designers. Moreover, the collection and use of these spatial data could contribute to the acknowledgment of its inhabitants as "full-fledged citizens" and inform policies that benefit them (e.g., improve public services) and alleviate the negative impacts of living in informal settlements.

You can read more at:
https://senseable.mit.edu/favelas/

**EducaDigitale, Brazil**

*Pillars: Participation, Access, and Knowledge*

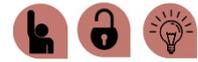

With a focus on open education in the digital domain, EducaDigitale was set up in Brazil in 2010. They have reconfigured traditional notions of CSOs to ground their work as a 'think-and-do tank' where they combine research, training, education, and advocacy for strengthened digital rights. The organisation develops their initiatives alongside building networks and collaborative endeavours.

"Pilares do Futuro", or Pillars of the Future, is a web platform developed by EducaDigitale to improve open education in digital citizenship. Through the platform, they aim to support educators by providing them with tools and guides to create curricula, identify good practices, encourage the creation and sharing of proposals, and promote a culture of collaboration, all within the digital eco-space. They have derived their guiding principles from UNESCO's 'Education, a treasure to discover' document that was released in 1999. Themes relevant to this advocacy work include digital security and the responsible use of the internet in classrooms (both freedoms to engage and risks like cyberbullying, racism, and copyright law). Internet safety and primers for guardians to encourage and teach internet safety to children is a fundamental aspect of the project, especially as it curates complex information in an age-appropriate manner. Other initiatives include teacher training and human development, open-access resources, open education tools, and methods for evaluating the policies of educational institutions.

You can read more at:

https://educadigital.org.br/

**Intervozes—Coletivo Brasil de Comunicação Social, Brazil**

*Pillars: Power, Participation, and Equity*

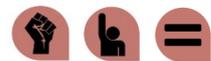

Intervozes is a Brazilian organisation founded in 2003 that seeks to safeguard freedom of expression—and other human rights—in all forms of communication. Through the monitoring of communication policies and the media, analysis, awareness raising, and civil society mobilisation, Intervozes has defended ICT access policies and contributed to the promotion of a more diverse and pluralistic media environment.

In 2016, Intervozes launched the social media campaign #CalarJamais (translated as "#NeverShutUp"), as a means of defending freedom of expression. This campaign was launched alongside a platform where members of the public can report complaints of violations of this freedom within Brazil. Intervozes then sends complaints to national and international bodies.

You can read more at:

https://intervozes.org.br/mobilize/calar-jamais/

**Institute for Technology & Society (ITS Rio), Brazil*****

*Pillars: Knowledge, Equity, Access, and Participation*

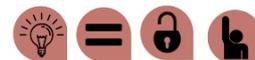

The Institute for Technology & Society (ITS Rio) aims to provide a greater understanding of access opportunities and promote equity across the digital space in Brazil and the Global South through research and partnerships. Over ten years, ITS Rio has built expertise on themes such as emerging technologies as they relate to various stakeholders and design collaboration between competing interests.

ITS Rio produces extensive research on the legal, social, economic, and cultural dimensions of technology while advocating for effective policymaking that ensures privacy, freedom of expression, and access to knowledge. Identifying the growing threat of manipulation caused by the convergence of automation and disinformation in public debate and political decision-making, ITS Rio developed Atrapabot in collaboration with the National Democratic Institute for International Affairs to mitigate the use of algorithmic bots in coordinated disinformation campaigns in South America.[173] The easy-to-use tool provides a rating on the probability of a bot account. ITS Rio has provided literature to complement the understanding of the use of Atrapabot to strengthen organisations and provide support to researchers working to combat disinformation in their countries. Other significant research has explored themes such as digital identity, digital rights post-pandemic, cyber resilience for journalists, and AI. ITS Rio is participating as a Policy Pilot Partner in the "Advancing data justice research and practice" project.

You can read more at:

https://itsrio.org/pt/home/

---

[173] See 'Atrapabot' project on https://es.pegabot.com.br/

### The Laboratory of Public Policy and Internet (LAPIN), Brazil
*Pillars: Identity and Power*

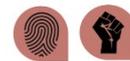

In 2015, WhatsApp was blocked in Brazil for 72 hours owing to Facebook Brazil's non-compliance with wire-tap orders. Consequently, a Request for Non-Compliance of Fundamental Principles (Arguição de Descumprimento de Preceito Fundamental) no. 403 was filed in the Federal Supreme Court over the claims that the blockade was a violation of fundamental rights. University students came together to contribute to the trial following the filing of ADPF no. 403 which turned into the foundation of The Laboratory of Public Policy and Internet (LAPIN). Focusing on digital policymaking in Brazil, LAPIN has since been established as a non-profit think tank that publishes research with the support of researchers, lawyers, engineers, as well as public and private sector professionals. Through the two-pronged approach of analysing the impacts of novel technology on society and legislation and assisting public decision-makers, LAPIN has been at the forefront of technology regulation.

The use of FRTs in South America and the Caribbean has been well documented by a report from LAPIN following extensive stakeholder engagement on the consequences of inaccuracy and opacity.[174] Participants in semi-structured interviews noted how FRTs were deployed across transport hubs for security, advertising, and predictive policing despite instances of misidentification. The discussions also addressed gaps in regulation and the need for policy measures to prevent risks of bias. Other publications have focused on adjacent digital themes like privacy, surveillance, and data portability as well as evaluation of geopolitical and regulatory measures across the globe.

You can read more at:
https://lapin.org.br/

---

[174] Costa & Silva, 2020

**Coding Rights, Brazil** 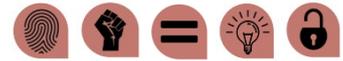

*Pillars: Identity, Power, Equity, Knowledge, and Access*

Coding Rights is a women-run organisation in Brazil that adopts an intersectional feminist approach to defend human rights development, regulation and use of technology. It reimagines trans-feminist and decolonial futures. Their work ranges from research and capacity-building to communication and advocacy strategies and online tools. They showcase the harmful impacts of data through different formats, including artwork, visualisations, documentaries, reports, and policy papers.

Within their projects, Coding Rights launched a documentary web series entitled "From devices to bodies" about the use of bodies as data sources. It aims to raise awareness and promote debate about the accelerating trend to extract data about our bodies, the abusive practices that this trend entails, and the lack of proper regulation to protect people from harmful impacts.[175] Related to this topic, Coding Rights has proposed a framework based on feminist theories of sexual consent that can address how to negotiate or reject the conditions of the Terms of Services in online platforms.[176]

Coding Rights is also developing notmy.ai, a feminist toolkit aimed to support anti-colonial and feminist movements in understanding and questioning algorithmic systems deployed in the public. During the initial phase of the project, five trending domains within which AI is being tested and piloted across Latin American governments were identified: Education, Judicial System, Policing, Public Health, and Social Benefits. Projects risking the propagation of harm based on gender and its intersectionality with race, class, sexuality, age, and territory were mapped, and impact assessments were conducted. In its final phase, the toolkit will help individuals account for structural inequalities and injustices, power imbalances, identity politics, and lack of public participation when assessing public sector AI.

You can read more at:
https://www.codingrights.org/

---

[175] Coding Rights, 2020
[176] Peña & Varon, 2019

**Nupef, Brazil** 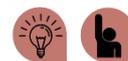
*Pillars: Knowledge and Participation*

The Instituto Nupef is an NPO in Brazil established in 2019 that aims to generate conditions for the effective use of technology to defend democracy and fundamental rights. Nupef produces research and learning activities for different stakeholders about the potentials of ICTs to enhance citizens' participation and the promotion of digital rights, implements pilot projects for innovation, and advocates for policies and practices that range from privacy to an open spectrum (an approach to radio spectrum management where unlicensed spectrum is available for use by all).

To integrate and expand the knowledge about existing practices of innovative uses of ICT that support sustainable development, human rights, and social justice, Instituto Nupef launched "Espectro" (translated as "Spectrum"). Espectro is a collaborative and multieditorial web portal created by Nupef with the aim of sharing information and knowledge on network practices pertaining to new radio technologies for community use, encouraging the innovative use of information and communication technologies to support sustainable development, human rights, social justice, good governance, and democratic values. Content on the Espectro platform includes the monitoring of community experiments and information on regulation and the effective implementation of ICT projects.

You can read more at:
https://nupef.org.br/
https://espectro.org.br/pt-br

**Internet Lab, Brazil** 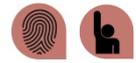

*Pillars: Identity and Participation*

InternetLab is a non-profit research centre based in Brazil that conducts interdisciplinary research about internet policy. More specifically, their research agenda includes issues of privacy and surveillance, culture and knowledge, freedom of speech, information and politics, gender, and technology. Its mission is to create and disseminate knowledge about internet policy and to support related initiatives and research projects. For this reason, they engage with other stakeholders across the globe who are working on related fields.

Among its projects, InternetLab has evaluated automated content moderation tools by social media platforms. In their research "Drag Queens and Artificial Intelligence", InternetLab has studied Perspective, the AI technology developed by Jigsaw from Alphabet Inc., that assigns a level of "toxicity" to text-based content. Results showed 'that a significant number of drag queen Twitter accounts were considered to have higher perceived levels of toxicity than those of Donald Trump and white supremacists.[177] Words, such as 'gay', 'lesbian', and 'queer', despite their context, were associated with "toxic" content. The lack of neutrality in this approach was highlighted by InternetLab as problematic for the exercising of the right to freedom of expression.

InternetLab has also participated in a project titled "Policy Frameworks for digital platforms—moving from openness to inclusion." This research explores the juridical-institutional arrangements that regulate digital platforms. It also identifies emerging issues including market monopolisation, challenges to development justice, ownership of user data, privatisation of informational commons, and worker exploitation. Based on its unpacking of these issues, the project proposes policies that may address such challenges while considering inclusion and socio-economic development. InternetLab is conducting a case study on the regulation of on-demand videos, investigating the impact of the platform economy on the Brazilian audiovisual market as this pertains to user access, available content, and the financial contributions to the State.

You can read more at:
https://internetlab.org.br/en/
https://internetlab.org.br/en/projetos/drag-queens-and-artificial-intelligence/
https://itforchange.net/background-paper-policy-frameworks-for-digital-platforms-moving-from-openness-to-inclusion

---

[177] *Drag Queens and Artificial Intelligence*, n.d.

**Pathways to Technology, Canada**

*Pillars: Identity and Access*

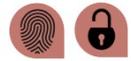

Established in 2008 and managed by All Nations Trust Company, Pathways to Technology aims to enable affordable high-speed internet for First Nations communities in British Columbia. The organisation is committed to their connectivity objectives and their goals for sustainable technology solutions and improved network capacity.

Whilst almost 95% of First Nations peoples in British Columbia have a broadband connection, there are still gaps between them and other British Columbians in terms of high-speed internet access and affordable connection. These gaps have a direct impact on socio-economic opportunities, improved health care, and online education. To respond to this challenge, Pathways to Technology works with First Nations communities to identify the priorities and challenges in obtaining internet connection. They contract telecommunications providers to instal technical infrastructures that are custom fit for each community, providing connection to health centres, schools, homes, and businesses within the community. For instance, the First Nations Health Authority aims to develop an integrated clinical telehealth network. Pathways to Technology also provide tailored education and training services to equip local communities to leverage these new technologies towards their goals.

You can read more at:
https://www.pathwaystotechnology.ca

**First Nations Technology Council (FNTC), Canada**

*Pillars: Power, Identity, and Participation*

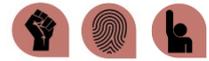

While Indigenous communities account for 4.9% of Canada's population, they make up only 1.9% of the ICT jobs. Colonial practices have perpetuated and deepened digital divides. Such gaps continue to manifest in areas of representation, rights, and respect as these relate Indigenous communities' ability to access to digital technologies and to seize opportunities to develop and use them.[178] The First Nations Technology Council (FNTC) is an Indigenous-led, NPO that works to promote a robust Indigenous Innovation Ecosystem by connecting all 204 First Nations communities across British Columbia (BC) in Canada. The driving force behind the FNTC is a belief that intersectionality and diversity in innovative technology environments are critical for progress for all in Canada. They primarily work within the mandates of digital skill development, connectivity, information management, and technical support and services.

FNTC has published extensive research alongside the launch of education programmes such as the notable flagship Foundations and Futures in Innovation and Technology (FiT) project, which guides students through their digital skills development journey.[179] Divided into two—Foundations and Futures—the project implements programmes to cater to Indigenous individuals at every level of the development of their digital skills. They provide student support through living allowances, tuition fees, and a Digital Elder-in-Residence. The curriculum for the Foundations has been designed by Indigenous specialists to ensure safe and navigable access, while the advanced Futures programme was designed by industry and partners.

You can read more at:
https://technologycouncil.ca/

---

[178] Pierre, 2022
[179] *Education Programs – First Nations Technology Council*, n.d.

**GobLab UAI, Chile\*\*\***

*Pillars: Power*

Founded in 2017 and based at the School of Government at Adolfo Ibáñez University in Chile, GobLab UAI is a public innovation lab that seeks to transform the public sector through innovation and data science. Towards that goal, they conduct research, trainings, and applied projects, and work alongside government agencies, civil society organisations, and businesses.

One of their projects, entitled "Market Opportunities for Technology Companies: Public Procurement of Accountable, Ethical and Transparent Algorithms", aims to help public and private companies improve the provision of social services and make a positive social impact. GobLab UAI hopes to achieve these ends by building capacity among technology companies through training programmes designed to incorporate ethical standards in automated decision-support services for the public sector and through the inclusion of ethical standards for public procurement of technologies. Within this project, GobLab UAI is supporting two public agencies which are attempting to develop and use decision-support algorithms. The lab is giving advice and technical assistance on ethical requirements like transparency, equity, privacy, explainability, and responsibility A The Chilean Public Criminal Defender's Office, which plans to incorporate AI to support intermediate hearings, will be one of the beneficiaries of this activity. GobLab UAI is participating as a Policy Pilot Partner in the "Advancing data justice research and practice" project.

You can read more at:
https://goblab.uai.cl/en/ethical-algorithms/

**Derechos Digitales, Chile**

*Pillars: Power, Identity, and Equity*

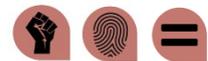

Derechos Digitales is a Chilean NPO founded in 2005 that aims to develop, defend, and promote human rights in the digital environment. Through research projects, communication campaigns, and advocacy campaigns to impact public policies and private practices, they work on topics related to copyright, privacy, and freedom of expression.

The organisation manages a variety of campaigns advocating for human rights within the internet and data-driven technologies, including *Inteligencia Artificial e Inclusión en América Latina* (translated as "Artificial Intelligence and Inclusion in Latin America"). The project emerges in response to the potentially long-term adverse impacts of AI on human rights and aims to analyse existing AI initiatives to identify improvements and good practices. The project uses four case studies pertaining to different uses of AI-assisted decision-making in Latin American countries, using AI in national employment, children's social care, criminal justice, and healthcare as starting points to investigate regional trends pertaining to data use, protection and consent, transparency, explainability, auditing systems, private and public funding and deployment, and regulatory systems enabling the right of recourse. They have developed a series of publications with the analyses of the use cases and other related topics, as well as a map visualising the state of AI public policy across Latin American countries.

You can read more at:
https://ia.derechosdigitales.org/

**Colnodo, Colombia**

*Pillars: Knowledge, Power, Access, Equity, and Identity*

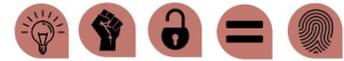

Since its formal establishment in 1994, Colnodo has been providing low-cost electronic communication systems for development organisations, maintaining its position at the forefront of online electronic services in Columbia. The programmes are guided by transversal axes that emphasise the importance of human rights, gender equity, open software, sustainable development, digital inclusion and so forth within an overarching objective of strategic communication through ICT.

Among numerous publications and projects at the nexus of their guiding pillars, the organisation has launched the Escuela de Seguridad Digital (ESD), or the Digital Security School of Colnodo, to equip CSOs, journalists, human rights defenders, and activists with the tools needed to navigate and protect personal and institutional information within the digital domain. In 2019 alone, the ESD reached 4,000 students, 130 journalists and 336 social leaders; notably, 72% of the participants were women.[180] Colnodo has also worked alongside other institutions and organisations to promote the use of digital tools in achieving the goals of various collectives and establishments across the country. Other programmes have been notable for their assistance in the development of wireless communication services in rural areas of Colombia.[181]

You can read more at:
https://www.colnodo.apc.org/es/inicio

---

[180] Information about the 'Digital Security School' can be found at https://escueladeseguridaddigital.co/sobre-esd/
[181] *Proyectos Gobierno En Línea y Democracia Electrónica*, n.d.

**Dejusticia, Colombia**

*Pillars: Access, Equity, Power, and Knowledge*

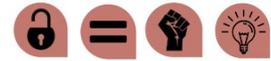

Based on a guiding principle of rigorous academic research informing social change and policy, Dejusticia is an organisation dedicated to promoting social justice and the defence of human rights in Colombia and the Global South. The organisation engages in litigation within Colombia and leverages international connections and networks for effective change. Their work is also driven by an emphasis on the local context, particularly of Indigenous systems of knowledge in Colombia and the Global South.[182]

Among their publications, Dejusticia's research on the digital divide in Colombia has been notable. As part of a series of publications under Dejusticia's Digital Inequalities research, the report "Desigualdades digitales. Aproximación sociojurídica al acceso a Internet en Colombia" (translated as "Digital inequalities. A socio-juridical approach to access to internet in Colombia") identifies digital gaps that arise from numerous factors including territory, age, gender, and income.[183] Their research provides hypotheses on the possible impacts of digital inequalities such as limiting access to education, labour, and information. The organisation notes that the accelerating expansion of the internet has not been without consequence, and while it has created opportunities and networks for some, Dejusticia has observed how it has contributed to imbalances and exclusion.[184] Dejusticia also advocates for access to the internet to be considered as a human right that upholds principles of dignity, equality, and justice.[185]

You can read more at:

https://www.dejusticia.org/en

---

[182] Melo Cevallos, 2019
[183] Práxedes Saavedra Rionda et al., 2021
[184] Ibid.
[185] Dejusticia, 2021

### Karisma Foundation, Colombia

*Pillars: Power and Knowledge*

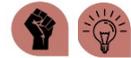

Karisma Foundation is a Colombian civil society organisation founded in 2003 that seeks to ensure that digital technologies protect and advance fundamental human rights and promote social justice. They carry out research, campaigns, consulting, conferences, and workshops that address topics like the democratisation of knowledge and culture, civic participation, autonomy, and social inclusion.

One of the organisation's initiatives, "Internet es tu passion" (translated as "internet is your passion"), aims to address challenges to freedom of expression posed by online content monitoring in Colombia. These challenges are driven by the Digital Millennium Copyright Act (DMCA), a controversial US law that, among other things, includes provisions that prevent the use of measures to circumvent access restrictions to copyrighted works. The DMCA's notice and takedown system implemented on US-based social media platforms impacts users across the world. According to Karisma, the protectionist-approach of the system not only 'has no due process guarantees and is disproportionate', but it can also pose barriers to the exercise of rights, such as freedom of expression or access to culture.[186] The *Internet es tu pasión* platform provides support to members of the public who believe they have unfairly received a copywrite infringement notice. Not only does the platform provide information about the DMCA, but it also shares advice, provides examples of counternotification requests, and lists tools to help individuals report unjust cases.

You can read more at:
https://karisma.org.co/internetestupasion/

### La Red en Defensa de los Derechos Digitales, Mexico

*Pillars: Power*

La Red en Defensa de los Derechos Digitales (R3D) is a Mexican organisation dedicated to the defence of human rights in the digital environment. Their projects involve policy research, strategic litigation, and public advocacy to promote freedom of expression, privacy, access to knowledge, and free culture.

In December 2020, the Mexican Congress approved the creation of the new General Population Law. The draft law proposed the creation of a mandatory Unique Digital Identity Card (CUID). It also proposed the creation of a centralised database containing the biometric data of all Mexican citizens and migrants within Mexico. In reaction, R3D expressed concerns about the potential threats of a mandatory biometric digital ID, such as intrinsic recognition inaccuracies, increased surveillance, information leaks, and social exclusion. They led a campaign calling on the government of Mexico to bring the project to an urgent halt. In a letter co-signed with a variety of other organisations, R3D argued that mandatory digital identity cards and the accompanying centralised database greatly threatens human rights and provides recommendations for alternative mechanisms for digital identity.

You can read more at:
https://r3d.mx/
https://www.accessnow.org/cms/assets/uploads/2021/09/The_Mexican_unique_digital_ID_CUID_proposal_threatens_human_rights.pdf

---

[186] Botero et al., 2019, p.6

## Los Feminicidios en México, Mexico

*Pillars: Participation, Power, Access, Equity, and Identity*

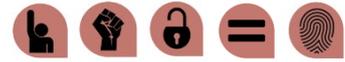

Los Feminicidios en México is a database of feminicides that have occurred in Mexico since 2016. It has been developed by an activist and geophysicist in response to the increasing violence against women and the lack of detailed and accurate data about crimes linked to gender violence. The database has revealed inconsistencies with official figures, which underreport femicides[187] and has provided a venue for accurately reporting data on marginalised and often invisibilised communities.

The database gathers data from daily registries of crimes across media and uses a practical tool designed by the United Nations High Commissioner for Human Rights[188] to understand the gendered dimension of the killings of women. It also uses Crowdmap to visualise all data, which can be filtered by age range of the victim, location, verification status, and multimedia attached to each report.[189] Each report contains detailed information about the crime, with information related to the relationship of the murderer to the victim, connected femicides, type of femicides, among other things. Not only is this information helpful to understand the motivations and contexts behind these killings, better identify feminicides, and inform policies that can protect and empower women, but is also makes visible the everyday experiences of women and interrogate the structural powers that underlie gender violence.

You can read more at:
https://feminicidiosmx.crowdmap.com/
https://journals.sagepub.com/doi/full/10.1177/1527476419831640

---

[187] Ricaurte, 2019
[188] United Nations High Commissioner for Human Rights, 2001
[189] Ricaurte, 2019

**Indigenous Community Telecommunications, Mexico**
*Pillars: Power and Access*

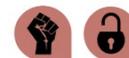

Rhizomatica and Redes por la Diversidad, Equidad y Sustentabilidad A.C. (REDES) are two organisations that have been working together to develop initiatives that are promoting communication and telecommunication processes for rural and Indigenous communities in Mexico and abroad. Rhizomatica is a Mexican organisation founded in 2009 with the goal to enable alternative telecommunication for underprivileged communities and to support them to build and maintain self-governed and owned communication infrastructure. REDES is a civil association founded in 2004 that seeks to support communities and organisations networks and research.

Because of the absence of reliable infrastructure and harsh environmental conditions, the creation of communication networks has been challenging for isolated rural and Indigenous communities in Oaxaca. Additionally, low-income communities face unnegotiable service rates charged by commercial mobile services.[190] In response to this situation, Rhizomatica and REDES created the first community-led cellular telephone networks in the world.[191] In 2016, they formalised the efforts which gave rise to the emergence of Telecomunicaciones Indígenas Comunitarias (TIC), a community owned and operated cellular telephone network cooperative led by members of Indigenous and rural communities. TIC provides technical services for individuals and communities looking to operate autonomous telecommunications networks. They provide affordable access and enable cooperation within member communities to strengthen mobilisation around self-determination and regulatory issues. TIC has the mission to build autonomous, safe, and affordable telecommunication alternatives that contribute to the strengthening of the communication processes, autonomy, self-determination, and the good living of Indigenous and rural communities—all values that connect with Indigenous Mexican cultures. According to authors who examined this initiative, what is novel about TIC is its notion of the electromagnetic spectrum as a common good.[192] Indeed, TIC understands communication as a fundamental individual and collective right.

In 2019, in order to systematise their experiences, Rhizomatica and REDES have also worked together to launch the Research Center in Technologies and Community Knowledges (CITSAC). CITSAC aims to develop applied research, capacity-building, and political advocacy to promote and reinforce the communication and telecommunication processes of Indigenous communities. Their work examines the problems and aspirations of Indigenous communities relating to communication processes, regulatory and technical developments, and best practices in hopes to inform ways to overcome legal challenges faced by community networks.

You can read more at:
https://www.tic-ac.org/
https://citsac.org/en/home/

---

[190] Srinivasan, 2020
[191] Belli, 2019
[192] Ibid.

**Sursiendo, Mexico**

*Pillars: Power, Participation, Equity, and Identity* 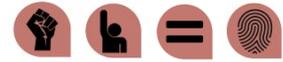

Sursiendo is a collective of activists working with regional organisations in Southeast Mexico to defend digital communality, collective digital rights, and hackfeminism—a concept used to incorporate intersectionality in the design, development, and use of technology so that designer and activists can 'open the systems, hack the patriarchy.' Placing gender and equitable participation at the core of their endeavours, Sursiendo utilises the avenues of activism, communication and design, free software, popular education, art, and cultural management to contribute to their vision.

The organisation has produced extensive literature at the nexus of gender and technology and has published tools for advocacy and the protection of rights. In 2020, they released a tool titled "Herramienta del Registro Incidentes De Seguridad Digital" (translated to "Logging Digital Security Incidents as a Risk Mitigation Practice "). Divided into two parts—a tool for recording digital security incidents and a guide for registration with human rights agencies—the resource is seen to be invaluable for collective action against risks in the digital sphere. The records of cybersecurity incidents (such as malware, phishing, Denial of Service, to name a few) include details on the vulnerabilities of the device being used and the capacities for the organisation or individual to intervene and mitigate the incident. These records are believed to be useful for complaint registration which can serve to prevent future incidents. Moreover, Sursiendo notes that the tool is an important mechanism for reflection, while also assisting in the identification of gaps and opportunities relevant to advocacy and capacity-building for organisations.[193]

You can read more at:
https://sursiendo.org/

---

[193] Sursiendo, 2020

### TEDIC, Paraguay

*Pillars: Power, Identity, and Knowledge*

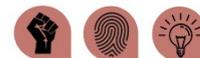

TEDIC is a CSO based in Paraguay that develops open civic technology, promotes the principles of a free culture, and defends digital rights. Within their work on gender and digital inclusion, TEDIC has provided evidence of the scope and forms of digital gender-based violence in Paraguay. Whilst the Paraguayan Law for the comprehensive protection of women against all forms of violence defines cyber violence, the definition only includes non-consensual dissemination of sexual images and media exposure. Other types of violence, such as online harassment, identity theft and fraud, and discriminatory expressions, however, remain unaddressed. The project titled "La Violencia Digital es Real" (translated as "Digital Violence is Real") aims to broaden the definition and make gendered violence visible using an informative web platform and a social media campaign. The web platform shares information pertaining to different types of gendered digital violence, aggressors, and digital rights. It also provides advice for individuals wanting to take action against digital gendered violence and those who have experienced it.

TEDIC has also accompanied cases of digital gender violence. Extensive information about one of these cases has also been shared on their web platform, including a case summary, a list of the justice operators who intervened in the cases, and a visualisation of court proceedings. This work has made visible the lack of access to justice, omissions of due processes, laws, and defence mechanisms in Paraguay.[194]

You can read more at:
https://www.tedic.org/en/
https://violenciadigital.tedic.org/en/

### Hiperderecho, Peru

*Pillars: Power, Access, and Identity*

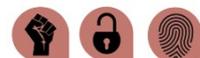

Hiperderecho is a non-profit civil organisation founded in 2012 in Peru to ensure that technology is used for social liberation and rights enhancement. They conduct research, activism campaigns, build tools, and work to influence policies in order to promote respect for rights and freedoms in digital environments.

Among their projects, Hiperderecho founded "Después de La Ley" (translated as "After the Law"), a project that responds to the needs of women and LGBTQ+ people who have experienced online harassment, sexual harassment, blackmail, and the non-consensual diffusion of intimate images on the internet. *Después de La Ley* seeks to identify and support individuals in undertaking pathways to effective remedy within the context of the Peruvian legal system, while simultaneously evaluating the extent to which current pathways address the needs of these individuals. Moreover, Hiperderecho has responded to technology-based surveillance, intimidation, and silencing increasingly faced by protesters in Peru and worldwide by launching the project "¿Quién vigila a los vigilantes?" (translated as "Who guards the guards?"). The project provides resources about digital security and self-care for activists and citizens to support them in exercising their right to protest.

You can read more at:

https://hiperderecho.org/

---

[194] APC, 2020

**Algorithmic Justice League (AJL), United States**

*Pillars: Power and Identity*

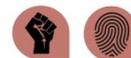

The Algorithmic Justice League (AJL) is an organisation that combines art and research to illuminate the social implications and harms of AI. Through building awareness about the adverse impacts of AI, promoting evidence-based advocacy, and conducting algorithmic audits for companies, among other things, AJL seeks to create a movement that shifts AI ecosystems towards more equitable and accountable AI.

One of the main areas of focus in AJL's research has been facial recognition technologies (FRTs) There has been a rapid proliferation of FRTs over the past several years. However, due to the risks of mass surveillance, discriminatory patterns of differential performance, disproportionate harmful impacts on marginalised communities, lack of affirmative consent, and other technical, legal, and societal risks associated with their deployment, FRTs have been a focus of criticism and debate. Several approaches to tackle these challenges have primarily focused on self-regulation, technical benchmarks, or legislation. Yet policies usually fall short in addressing the diversity of uses of FRTs across all sectors and domains as well as the range of other sociotechnical issues that lead to discrimination and societal harm.

In response to these persisting challenges, the AJL sent letters to the US Congress and testified in hearings within the US House of Representatives. The AJL also drafted whitepapers advocating for state regulation of FRTs, against a self-regulated industry consortium as a means of addressing the risks of this technology. One of such documents is entitled "Facial Recognition Technologies in the Wild: A Call for A Federal Office". Here, they argue that, given the reach and complexity of FRTs, an office with federal authority is required to provide comprehensive oversight but also dedicated expertise on the risks of these technologies. The proposed new federal office, they suggest, would categorise FRTs by degrees of risk and provide congruent guidelines as well as redlines to constrain potentially harmful use cases.

You can read more at:
https://www.ajl.org/
https://www.ajl.org/federal-office-call

### Our Data Bodies, United States

*Pillars: Power, Identity*

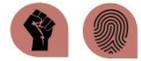

Our Data Bodies is a research justice project led by a team of community organisers based in Charlotte, North Carolina, Detroit, Michigan, and Los Angeles, California. They examine the impact of data collection and data-driven systems on the ability of marginalised people to meet their human needs, addressing surveillance and data-based discrimination in the United States by combining community organising and academic research.

Our Data Bodies develops and shares knowledge and tools aimed to advance data justice via community and organisational conversations about the benefits and harms of data within marginalised communities. They have developed reports grounded in interviews, workshops, and other participatory activities with residents of historically marginalised neighbourhoods, which identify ways in which impacted people want data collection and data-driven systems in their lives. Their tools include a "Digital Defence Playbook", which encompasses popular education activities, tools, and reflection pieces focused on data, surveillance, and community safety. It is aimed at organisations and community members involved in intersectional fights for racial justice, LGBTQIA+ liberation, feminism, immigrant rights, economic justice, and other freedom struggles, to co-create knowledge and address the impact of data-based technologies on social justice work.

You can read more at:
https://www.odbproject.org/

### Data for Black Lives (D4BL), United States*

*Pillars: Power, Participation, Identity, and Equity*

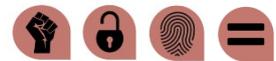

Data for Black Lives (D4BL) is a movement that started in 2017 in the United States. Bringing together, scientists and activists, D4BL seeks to use data science as a means for protest and accountability, and as a form of collective action to fight bias, build progressive movements, and promote civic engagement. This includes empowering Black people to work in data science and creating a political home for these scientists.

D4BL challenges data as an instrument that, based on colonial legacies of structural racism, perpetuates oppression. Underlying their work is the understanding of data as an ideological regime about how decisions are made and who makes them. D4BL outlines policy shifts pertaining to algorithmic transparency, data regulation, data governance, and economic policy to help policymakers, movement leaders, and thinkers create changes that benefit Black futures.

Among other initiatives, D4BL is working on the creation of hubs to propel advocacy initiatives and trainings, and policy working groups that examine current data governance frameworks and explore alternative models. In response to the lack of COVID-19 data about Black communities, the movement has also consolidated information on the disparate effects of COVID-19 on Black communities, creating a publicly accessible spreadsheet with the data and demanding states to make COVID-19 race data public.

You can read more at:
https://d4bl.org/about.html

**Environmental Data & Governance Initiative (EDGI), United States**

*Pillars: Access, Participation, Power, and Equity*

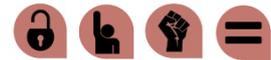

The Environmental Data & Governance Initiative (EDGI) was created with the aim to challenge injustice by evaluating the role of data and digital processes in environmental data and governance. Through the publication of relevant literature, technical reports, blogs, and op-eds, the EDGI has been established as a leading watchdog for environmental data injustice since 2016. With over 50 members from numerous domains, including academia, non-profits, and grassroots organisations, EDGI maintains a consensus-based approach to carrying out its research and advocacy.

EDGI highlights the imperative of undoing the environmental harms that occurred during the presidency of Donald Trump—often noted by researchers as a period that was marked by a significant decline in the Environmental Protection Agency's (EPA) enforcements. The organisation's activities particularly emphasise their work in monitoring EPA's web-based information, improving access to data, investigating policies through interviews and analysis, releasing reports on pressing threats to environmental justice, grassroots data archiving, as well as formulating novel organisational structures for environmental justice. Pertinently, the EDGI's work is founded on critical theory, data justice scholarship, and participatory knowledge to overcome extractive data practices.[195]

You can read more at:
https://envirodatagov.org/about/
https://doi.org/10.1080/1369118X.2019.1596293

**Open Water Data, United States**

*Pillars: Equity and Access*

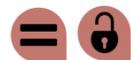

The Open Water Data project has been initiated to provide local governments, advocacy groups, and community members with equitable access to governmental data. The project has identified that a quarter of the over 50,000 facilities in the United States that it studied are in violation of the Clean Water Act (CWA), specifically with regard to the limits that the CWA has set in place for the emission of waste into public waterbodies.

The project has utilised creative avenues to reveal the scale of environmental destruction by private companies in the US. Lanterns were floated on the Chelsea River in Massachusetts to highlight each company's violations of the CWA. Data on environmental harms is brought to the forefront by the project with an emphasis on the need for access to government data. Similarly, the Advisory Committee on Water Information has proposed a new Open Water Data Initiative and advocates for a national water data framework that can utilise existing infrastructure for innovation, data sharing, and solution development.[196]

You can read more at:
http://datalanterns.com/#
https://www.media.mit.edu/projects/open-water-data/overview/

---

[195] Vera et al., 2019
[196] *Open Water Data Initiative Overview*, n.d.

**Fight for the Future (FFTF), United States**
*Pillars: Power, Participation, and Identity*

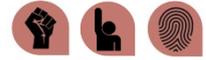

Fight for the Future (FFTF) is a non-profit advocacy organisation founded in 2011 that promotes civic campaigns to stop uses of technology that limit human rights. They also work to empower people to demand the technologies that serve their interests. They promote technology as a force of liberation instead of a force for oppression and for the promotion of structural inequalities.

FFTF developed the campaign #NoTechForICE that stands against technology companies collaborating with the United States' Immigration and Customs Enforcement (ICE).[197] This campaign, which is now continued by immigrants' rights and racial justice groups, spreads awareness of contracts between corporations and government where companies provide data analytics and digital tools to support ICE's operations. #NoTechForIce provides resources for community activists who target specific companies and supports the mobilisation of tech workers and students to leverage their influence and engage in direct action to demand an end to these collaborations.

Several campaigns have specifically targeted Amazon. One of these campaigns explicates how for several years the company marketed facial recognition software to police forces that disproportionately misidentified Black and brown people, transgender people, and women.[198] Another campaign entitled #ProtestAmazon claims that Amazon provided ICE and the DHS with cloud computing software that allows them to streamline the collection of biometric data on immigrants, serving as the backbone of potential mass deportation.[199] The #ProtestAmazon campaign demands that the company stops fuelling militarised policing and #EyesOnAmazon is another petition that claims that Amazon supports the surveillance of Black and brown communities, workers, and immigrants and demands that the company stops selling its facial recognition software.

You can read more at:
https://www.fightforthefuture.org/

---

[197] See #NoTechForICE campaign on https://notechforice.com/
[198] Fight for the Future, n.d.
[199] Ibid.

### United States Indigenous Data Sovereignty Network, United States

*Pillars: Equity, Power, Participation, and Identity*

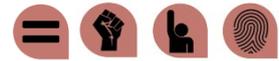

The United States Indigenous Data Sovereignty Network was formed in 2016 to ensure that data for and about Indigenous people in the US are used for their individual and collective well-being. Their mission is to decolonise data and exert Indigenous data governance to promote Indigenous data sovereignty, which 'derives from tribes' inherent right to govern their peoples, lands, and resources'.[200] At the international level, they collaborate with the Te Mana Raraunga in Aotearoa New Zealand and the Māori Data Sovereignty Network.[201]

The network has enabled the creation of a transdisciplinary community of practice and provided research information and policy advocacy. Members have developed research projects together and discussed and released policy papers and recommendations. They call for policymakers to recognise Indigenous data sovereignty as an objective to be incorporated into tribal, federal, and other forms of data policies in order to generate resources and build support for Indigenous data governance and grow tribal data capacities including the development of data warriors (Indigenous professionals and community members who are skilled at creating, collecting, and managing data).

You can read more at:
https://usindigenousdata.org/

### Silicon Harlem, United States

*Pillars: Access, Participation, and Equity*

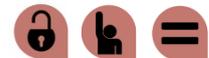

With a focus on Harlem and urban markets in the US, Silicon Harlem is a for-profit organisation that, since 2013, has worked alongside private entities, academia, NPOs, and civil society to develop future-facing technological infrastructure. The organisation has introduced digital literacy programmes and broadband design and deployment, while undertaking research projects and curating workshops and outreach initiatives for community participation. By educating and including all members of community within their initiatives, the objective of Silicon Harlem is to strengthen individual capacities as technology increasingly enters numerous facets of life.

Partnering with the New York City Economic Development Corporation (NYCEDC), the organisation has launched "RISE: NYC", a project aimed to assist small businesses to build resilience against and navigate the challenges of climate change. A connected mesh network was deployed in East Harlem to provide no cost wireless technology and emergency communications to support small businesses during natural disasters. In their endeavour to prepare 'Tech-Enabled Cities', the organisation has notably distributed 500 laptops to vulnerable families and 8,000 smart thermometers while enrolling over 100 students in after-school coding classes.[202]

You can read more at:
https://siliconharlem.com/
https://www.cnbc.com/2016/10/28/silicon-harlem-aiming-to-bring-tech-innovation-to-upper-manhattan.html

---

[200] See United States Indigenous Data Sovereignty Network at https://usindigenousdata.org/about-us
[201] Russo Carroll et al., n.d.
[202] *Tech-Enabled Cities*, n.d.

**National Congress of American Indians (NCAI), United States**

*Pillars: Identity, Power, Equity, Participation, and Access*

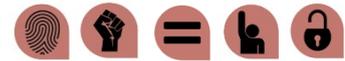

Founded in 1944, the National Congress of American Indians (NCAI) is an organisation representing the broad interests of American Indian and Alaska Native governments and communities. In 2003, it established its Policy Research Center with the goal to inform the policy decisions of tribal leaders, tribal organisations, and other policy institutions in a way that improves the lives of Native people.

Invisibility of American Indians and Alaska Natives is often perpetuated by how data is collected and reported. Small sample size, undercounting, errors in estimates, gaps in existing data, and data disaggregation are some of the issues obstructing the quality and accuracy of data about American Indian and Alaska Native communities. These issues, in turn, affect how policies are set, monitored, and evaluated. In response to these challenges, the NCAI Policy Research Center conducts research and shares recommendations advocating for the disaggregation of native data. It advocates for the accurate, meaningful, and community-based collection of American Indian and Alaska Native community data and research that supports community issues such as joblessness and education. It also shares data resources pertaining to these communities.

You can read more at:
https://www.ncai.org/policy-research-center/research-data/data

**The Progressive Technology Project (PTP), United States**

*Pillars: Identity, Participation, Power, Equity, and Access*

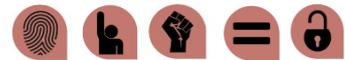

The Progressive Technology Project (PTP) is a movement partner that emerged in 1998 with the goal to strengthen the work of grassroots organising groups 'through the strategic use of technology' for organisation, advocacy, communications, civic engagement, and fundraising. The movement has its origins in social justice groups and has initially worked on issues of technology access. However, with the development of new technologies, movement groups are facing more complex and hostile digital environments, and consequently, PTP started to address broader challenges. They develop software, conduct trainings, and provide technical assistance to movements working with affected communities, especially those most impacted by structural racism.

Concerned about the white-male dominated tech teams involved in the planning, control, and maintenance of the internet and the technologies that support it, PTP—along with May First/People Link—also launched a "People of Color Techie Training Project". Through this project they provide mentor-based training for activists of colour to combat racism and race-based exclusion in technology.

You can read more at:
https://progressivetech.org/

### Comisión para los Derechos Humanos del Estado Zulia (CODHEZ), Venezuela

*Pillars: Participation, Power, Equity, and Access* 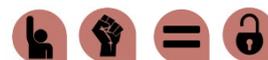

During nationwide political demonstrations in Venezuela in 2014, numerous students, professors, and professionals in the legal domain came together to record and contest incidents of arbitrary arrest, detention, raiding and so forth. Comisión para los Derechos Humanos del Estado Zulia (translated as "Zulia State Human Rights Commission") was founded on a vision that supports these demonstrations and aims to achieve justice and reparations through demands for human rights at national and international institutional levels. The organisation has since grown to provide free legal advice and assistance while producing reports on incidents of human rights violations.

Efforts are concentrated in the Zulia state in northern Venezuela, noted for having the largest population in Venezuela, with a focus on public services and food security. Publications have recorded the myriad devastating challenges and issues faced by Venezuela as the country continues to maintain unprecedented levels of inflation amidst political unrest.[203] Within this environment, CODHEZ has noted the impacts of blackouts and poverty driving a food crisis. Pertinently, they identified significant impacts on education caused by limited to no digital infrastructure that prevented distance learning coinciding with the peak months of infections caused by the COVID-19 pandemic.[204]

You can read more at:
https://codhez.org/

### May First Technology Movement, United States and Mexico

*Pillars: Power, Equity, Access, and Participation* 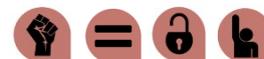

May First Technology is a non-profit movement organisation founded in 2005 that shares its technology as a non-profit service provider so it can be used strategically and collectively by organisations and activists in the United States and Mexico working for local struggles and global transformation. It has 850 members who own over 120 servers maintained by May First Technology and have unlimited access to its internet services.

Members also participate in networks, coalitions, and campaigns around topics like net neutrality, data protection, and alternative connection systems. Since 2017, the May First Technology movement has brought together over 1,500 activists in the United States and Mexico through their "Technology and Revolution" series. This has seen participants discuss the ways in which technology can intersect with activism and with revolution. Underlying their work is the conviction that the use, protection, and democratisation of technology is a key element for fundamental change.

You can read more at:
https://mayfirst.coop/

---

[203] Armas, 2022
[204] CODHEZ, 2020

### Sulá Batsú, Central America

*Pillars: Identity, Participation, Equity, and Access*

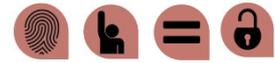

Founded in 2005, Sulá Batsú is a cooperative in Costa Rica that aims to promote and strengthen local development in coordination with other stakeholders, such as organisations, social enterprises, community networks, and social movements. They pursue this goal through litigation, campaigning, and research on digital technologies, art and culture, management of knowledge, and the solidarity economy.

In response to the issues, problems, and the needs of rural communities that uses of technology often do not consider, Sulá Batsú launched the program "Generación 3.0"(translated as "Generation 3.0"). Developed in coordination with the Telefónica Foundation, it aims to enable technology appropriation by rural young people to promote change in their communities. It consists of a series of trainings on digital tools—including digital security workshops that are context-sensitive—and content development.

TIC is another program developed by Sulá Batsú that aims to create better working and labour conditions for rural women in the tech sector in Costa Rica and promote women-led businesses that tackle their communities' social challenges. The cooperative organises a variety of meetings including the "Café Tecnologico" (translated as "technological café"), a meeting for women to connect with local digital businesses and to engage in educational workshops pertaining to technology, innovation, and design. Two additional initiatives are the "Club de Niñas" (translated as "Girl's Club") a space for young women to learn about the safe use of technology and the "Club de Programacion" (translated as "Programming Club"), dedicated for young women interested in learning to program or who are studying to work in IT and need support in their subjects. In recent years, the program has expanded to other countries in Central America. In 2021, Sulá Batsú and Asociación de Mujeres Cabécar de Alto Pacuare launched the platform "Okama Suei" as a response to a shared understanding that the internet had become an invasive and colonising technology, threatening the Cabécar Indigenous cosmovision and culture.[205] This is a platform that helps communities to decide if and how they want to use digital technologies. It aims to strengthen and defend the cosmovision and culture of Indigenous women in Costa Rica through intergenerational local knowledge exchange.

You can read more at:
https://www.sulabatsu.com/

---

[205] APC, 2021

**Al Sur, Latin America**

*Pillars: Access, Equity, and Participation*

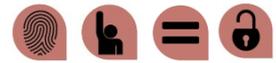

Al Sur is a consortium of academic institutions and CSOs across Latin America that aim to strengthen human rights online. To achieve this goal, Al Sur promotes communication and collaboration between its members as well as experience and knowledge sharing. It also analyses public policies that could impact the Latin American region and coordinates common regional positions in forums and other spaces of deliberation.

The consortium aims to address threats to human rights online in the region, including but not limited to state and platform surveillance, personal data abuses by states and companies, and the adverse impact of AI on human rights. As a result, their areas of work include access, surveillance, personal data, cybersecurity, intermediary liability, and AI. "El Observatorio Al Sur de Tecnologias de Vigilancia y Pandémicas" (translated as "The Al Sur Observatory of Surveillance and Pandemic Technologies"), is one of Al Sur's initiatives. This project seeks to share information that illustrates regional trends in governmental use of personal data and surveillance technologies within the COVID-19 context, enabling collective action to promote fundamental human rights within and beyond the pandemic. This project maps digital technologies implemented by Latin American governments which collect population data in the COVID-19 context, spreads awareness about the reach, impact, and potential threats to human rights posed by these technologies and illustrates regional trends pertaining to legal protections.

You can read more at:

https://alsur.lat/

# Oceania

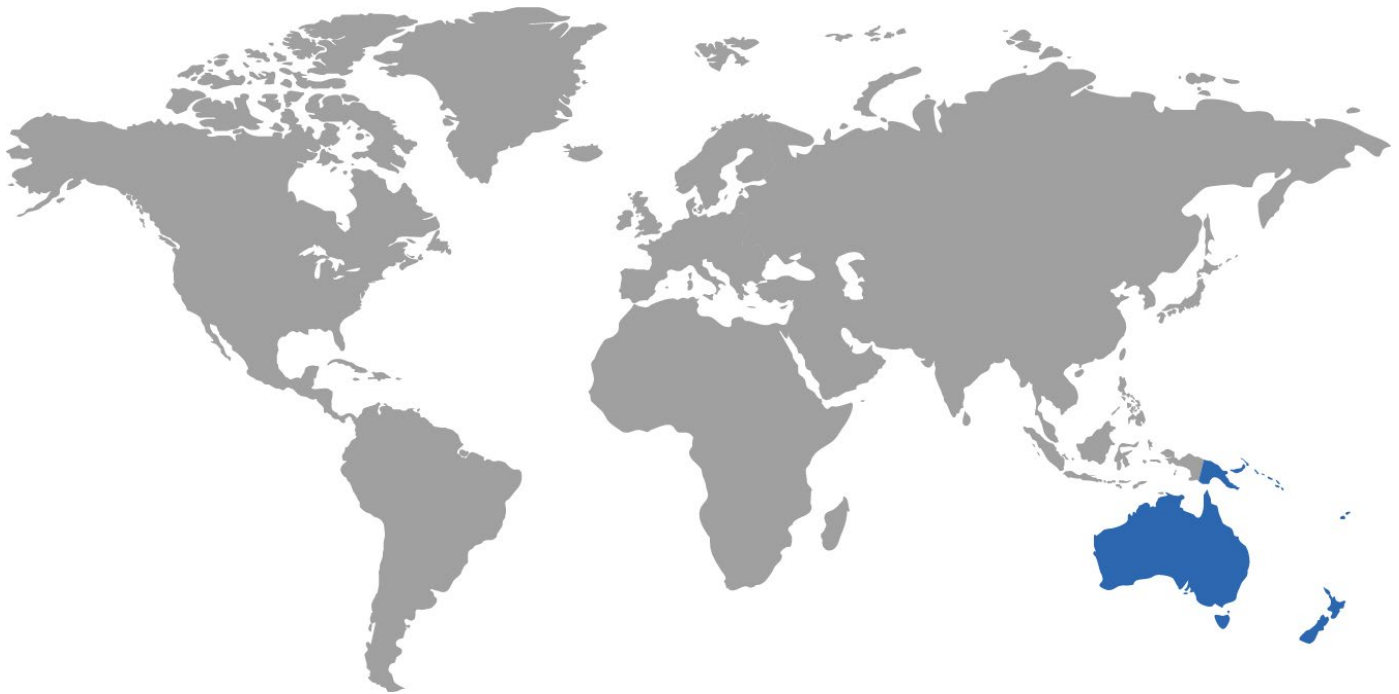

**Māori Data Sovereignty, Aotearoa New Zealand**

*Pillars: Power, Equity, Identity, Participation, Access, and Knowledge*

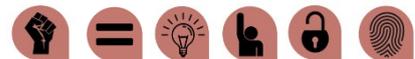

The Independent Māori Statutory Board (IMSB) is a Māori-led body that promotes issues of significance to the Māori community to the Auckland Council. One such issue is Māori Data Sovereignty, i.e., their ability to exercise authority over their collective data.

IMSB has developed the Māori Plan, which represents Māori aspirations and their long-term visions for Tāmaki Makaurau (Auckland). In response to the lack of existing frameworks for data collection, use, and monitoring that protect the interests of Māori people and that are compatible with their culture, the plan defines a set of wellness indicators grounded on tikanga Māori (Māori values and worldviews), which act as accountability mechanisms to ensure that the Auckland Council is responsive to Māori issues. Given a lack of relevant Māori-specific data, the plan includes a data strategy promoting Māori-led collection of Māori data.

Other Māori-led data initiatives have since emerged, including the Māori Data Sovereignty Network, which consists of Māori researchers and practitioners who operate a variety of research and policy projects ranging from ensuring the quality and integrity of Māori data, to advocating for Māori involvement in data governance.

You can read more at:
https://www.temanararaunga.maori.nz
https://www.imsb.maori.nz/about-us/introduction/

**Our Data, Our Way, Aotearoa New Zealand***

*Pillars: Power, Participation, and Equity*

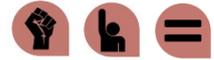

Our Data, Our Way was a project led in 2016 by Data Futures Partnership—a Aotearoa New Zealand independent advisory board tasked with drafting guidelines for private and public organisations working with people's personal data—in partnership with Toi Āria (Design for Public Good at Massey University) and Springload. The goal of Our Data, Our Way was not only to understand how people in Aotearoa New Zealand want their personal data to be used and shared but also to use these inputs in a trustworthy way to guide how data can benefit people. They conducted a series of workshops and developed an online tool that presented different situations of data use and sharing and then captured people's responses.

The team behind the project took a social licence approach to the task. This approach works from the understanding that the acceptance of how data is used and shared results from people trusting that this follows what they have agreed to. Acceptance, from this perspective, also involves people believing that enough value is created for the community involved. As a result, their workshops and online tool placed participation and engagement at the forefront with special emphasis on capturing what matters to Aotearoa New Zealand people.

You can read more at:
https://www.ourdataourway.nz/
https://www.toiaria.org/our-projects/our-data-our-way/

**Digital Natives Academy (DNA), Aotearoa New Zealand****

*Pillars: Access and Identity*

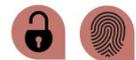

Established in 2014, Digital Natives Academy (DNA) is a NPO that seeks to give Māori young people the skills and tools needed to enter the digital tech industry. They aim to nurture the creative and computational thinking, as well as the logical reasoning and collaborative digital practices, that would enable a reduction of digital gaps in Aotearoa New Zealand.

The initiative emerged in response to the identified need to overcome the mere consumption of existing technologies and instead empower young people to develop and create their own digital tools. The Digital Natives Academy, therefore, teaches coding basics, provide resources related to animation, game and graphic design, content creation, e-sports, and digital well-being, among other things. Workshops and classes are held both online and in-person. DNA strives to create a sense of belonging and is rooted in a Te Ao Māori worldview. DNA is participating as a Policy Pilot Partner in the "Advancing data justice research and practice" project.

You can read more at:
https://digitalnatives.academy/

**ANTaR, Australia** 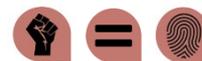

*Pillars: Power, Equity, and Identity*

ANTaR is an Australian advocacy NPO founded in 1997 that, through lobbying, public campaigns, and advocacy, seeks to overcome the disadvantages faced by Aboriginal and Torres Strait Islander people and defend their rights. Their work addresses different policy and social justice issues, such as health, culture, community development, and remote communities' services and infrastructure.

In 2019, the Australian government proposed a new method to channel Indigenous voices. However, concerned with the establishment of guarantees that fell short of meaningful participation, support, and resources, ANTaR has been working diligently on establishing a Voice for Aboriginal and Torres Strait Islander peoples within Parliament at the federal, state, and territory levels. In March 2021, ANTaR submitted a report entitled "Indigenous Voice to Parliament". The report calls for the establishment of a Voice that has agency, authority, respect, and is not subject to political nor financial undermining like past efforts to represent Aboriginal and Torres Strait Islander peoples in Parliament have. It also highlights that the proposal seeks for First Nations Peoples to not simply provide input to government decisions but rather manifest their sovereignty and self-determination. This means that narratives and data about Australia's history should include First Nations Peoples' knowledge. Furthermore, ANTaR has encouraged the use of digital tools for advocacy campaigns that call for a more impactful reconciliation movement.

You can read more at:
https://antar.org.au/

**Centre for Aboriginal Economic Policy Research (CAEPR), Australia** 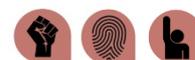

*Pillars: Power, Identity, and Participation*

Established in 1990, the Centre for Aboriginal Economic Policy Research (CAEPR), has undertaken social research on Indigenous economic and social policy and development to inform academia, public debate, policy development, and the actions of Aboriginal and Torres Strait Islander people and communities.

CAEPR has conducted participatory research,[206] partnering with Indigenous organisations and training community-based researchers to explore the experiences of technology-facilitated abuse (TFA) among Aboriginal and Torres Strait Islander women in remote and regional areas. They identified different forms of abuse and the benefits of technology in relation to aboriginal women's safety. The results of this research can inform responses and prevention of TFA by addressing factors such as the need for education around TFA and online safety, culturally competent support services for women experiencing abuse, the role social media companies play in preventing abuse, and clear and consistent legislation about this topic.

You can read more at:
https://caepr.cass.anu.edu.au/

---

[206] CAEPR, 2021; Brown et al., 2021

## Coalition of Peaks, Australia

*Pillars: Power, Equity, Participation, and Access*

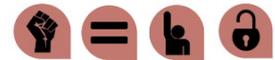

The Coalition of Peaks is a representative body of over seventy Aboriginal and Torres Strait Islander community-controlled advocacy groups and members that seek to change how governments work with Aboriginal and Torres Strait Islander people and ensure their place as shared decision-makers. In 2008, the Australian government agreed on a strategy to improve the life outcomes of Aboriginal and Torres Strait Islander people related to health and well-being, education, employment, justice, safety, housing, land and waters, and languages. The Closing the Gap strategy, however, did not include the formal say nor the full involvement of the affected communities. In 2019, worried that this would happen again when the government decided to update the strategy, a group of peak organisations and members created the Coalition of Peaks and signed a Partnership Agreement on Closing the Gap with the Council of Australian Governments. This represented an act of self-determination to be formal partners with Australian governments on those policies and programs that impact them.

The partnership resulted in a new National Agreement on Closing the Gap, which considered what Aboriginal and Torres Strait Islander people said matter to them to improve their lives. It includes four Priority Reforms related to shared decision-making, community-controlled sector, responsiveness to the needs of Aboriginal and Torres Strait Islander peoples, and access to data and the ability to use it. This fourth priority aims to give Aboriginal and Torres Strait Islander people access to locally relevant data and information so that the data can be used to drive change and monitor progress in the Close the Gap strategy. It includes efforts to increase the number of regional data projects, government initiatives to make data more accessible and usable for the communities and organisations and build expertise in data collection and analysis.

You can read more at:
https://coalitionofpeaks.org.au/

## Indigital, Australia*

*Pillars: Access, Power, Participation, and Equity*

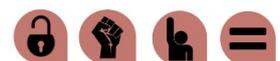

Indigital was founded in 2014 by Mikaela Jade from the Dharug-speaking Nations of Sydney. It serves as Australia's first Indigenous edu-tech company and works across many different areas including digital skills training, AI, and geospatial technologies. Jade's mission for Indigital is to work towards closing the digital divide between Indigenous and non-Indigenous peoples. Indigital provides a variety of digital skills training as well as a programme designed for primary and high school students entitled Indigital Schools, which combines digital skills training with learnings from Indigenous elders about history, language, and culture.

You can read more at:
https://indigital.net.au/

### Lowitja Institute, Australia

*Pillars: Access, Identity, Equity, and Knowledge*

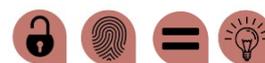

Established in 2010, the Lowitja Institute is a national institute dedicated to research on Aboriginal and Torres Strait Islander health, knowledge translation, and research community support. Their mission is to generate positive impact for the health and well-being of Australia's First Peoples through evidence-based research. To achieve this, they engage in participatory research that analyses and investigates the cultural and social determinants contributing to measures of health and well-being.

Events such as coronial inquests into preventable deaths of Indigenous peoples and existing health inequalities motivated the Lowitja Institute to release a discussion paper for the Partnership for Justice in Health (P4JH) in 2021. The paper addresses topics of race and racism in the health system specifically related to the Australian Government's National Aboriginal and Torres Strait Islander Health Plan's (NATSIHP) vision of 'a health system free of racism' (2013). This initiative exemplifies the potential of participation to advance justice and simultaneously identify the social determinants of health as well as the long-term causes of inequalities. In its report, the Lowitja Institute reviews the challenges of quantifying racialised health outcomes. Whilst this quantification may make inequality visible, this process often lacks the prioritisation or the incorporation of Indigenous peoples' knowledge and understandings. It calls, therefore, for reliable statistics and Indigenous intellectual sovereignty.

You can read more at:
https://www.lowitja.org.au/

### The Jumbunna Institute for Indigenous Education and Research, Australia

*Pillars: Access, Identity, and Power*

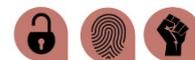

Established within the University of Technology in Sydney, the Jumbunna Institute for Indigenous Education and Research conducts research on Indigenous legal and policy issues. Their research hubs are directed at themes including cultural resilience, child protection advocacy, legal strategies, Indigenous people and work, Indigenous law and justice, and Indigenous policy. The Institute has also forwarded the Aboriginal History Archive that documents the history of self-determination movements in Australia through photos, videos, campaign ephemera and other media.

Among other research outputs, the Institute presented a paper titled "Designing for diversity in Aboriginal Australia: Insights from a national technology project" in 2019.[207] Bringing to light how the colonisation of Aboriginal Australians for over 240 years has impacted the connections to community and identity, the paper explores how a human-computer interaction (HCI) approach to designing and using technology can help build bridges between members of communities and their cultural and historical identities. Despite facing challenges such as the continuing effects of colonisation and the under-recognition of the diversity of Aboriginal Australia, the project provides relevant discourse and insights for HCI research.

You can read more at:
https://www.uts.edu.au/research-and-teaching/our-research/jumbunna-institute-indigenous-education-and-research

---

[207] Leong et al., 2019

### Maiam nayri Wingara Aboriginal and Torres Strait Islander Data Sovereignty Collective, Australia

*Pillars: Access, Equity, Power, and Participation*

The Maiam nayri Wingara Aboriginal and Torres Strait Islander Data Sovereignty Collective was formed in 2017 in response to the isolation of Indigenous Australians from the language, control, and production of data, as well as the neglect of their knowledge, worldviews, and needs. The Collective seeks to progress Indigenous Data Sovereignty and Indigenous Data Governance through the development of data sovereignty principles, data governance protocols, and the identification of strategic data assets.

The Maiam nayri Wingara Data Sovereignty Collective and Australian Indigenous Governance Institute created a Communique as a result of the 2018 Indigenous Data Sovereignty Summit. The Communique aims to advance Indigenous Data Sovereignty through the initiation of Indigenous data governance protocols. The Communique claims that Indigenous communities 'maintain the right to not participate in data processes inconsistent with the principles asserted in this Communique'. Actions taken by the collective are founded on the understanding that the exercise of Indigenous data governance will enable an accurate and informed picture of the realities, needs, and aspirations of Indigenous people.

You can read more at:
https://www.maiamnayriwingara.org/

**Pacific Community, Oceania** 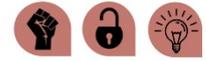

*Pillars: Power, Access, and Knowledge*

Pacific Community is an international development organisation comprising 27 country and territory members in Oceania with an objective of sustainable development through innovation. From its founding in 1947, the organisation has since navigated knowledge and innovation in sectors such as fisheries, public health surveillance, geoscience, and conservation for food security. The policy outlook of the Pacific Community is rooted in sensitivity to the contextual and cultural knowledges of the people of the region. At the same time, the organisation acknowledges the key role played by innovative technology, data analytics, and digital tools in advancing equitable and sustainable futures.

The accelerating effects of climate change have impacted the region's narrow genetic base with consequences for food and nutritional security. While there is an increasing sense of urgency about alleviating these threats, the knowledge and campaigns have been limited to specialist circles. This has led to a wider underappreciation among people in Oceania of the importance of the only regional genebank—the Pacific Centre for Crops and Trees (CePaCT).[208] The people of the region involved in the genebank have honed an approach to sustainability that prioritises cultural and local knowledge alongside an understanding of the symbiosis between people and the environment. The Pacific Community has launched the Pacific Data Hub (PDH) that tracks and analyses data related to nutrition and consumption habits in the region for the development of sustainable food systems and improved policymaking. The data has revealed that production in both agriculture and fisheries has been declining.[209] Moreover, it presents an urgency to respond to potential shifts to cash crops that can harm food security by increasing imports. The role of PDH is thereby pivotal to improving policymaking across the Pacific through a data-driven understanding of agricultural, environmental, and socioeconomic change.

You can read more at:
https://www.spc.int/
https://pacificdata.org/

---

[208] The Pacific Community, 2019
[209] The Pacific Community, 2021

**Pacific Island Association of Non-Governmental Organisations (PIANGO), Oceania**

*Pillars: Access, Identity, Power, Participation, and Knowledge*

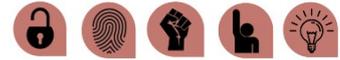

Growing out of the need to improve networking among the NGOs based in the Pacific, the Pacific Island Association of Non-Governmental Organisations (PIANGO) was set up officially in 1991. The organisation aims to strengthen and build the capacity of NGOs while including grassroots organisations in policymaking through a focus on governance, voice, approaches, and partnerships as priority areas. Fundamentally, the work is rooted in the importance of Pacific cultural values and practices for equitable and sustainable development.

PIANGO has produced extensive research and hosted events that emphasise the importance of decolonisation and self-determination. Serving as an umbrella coordinating entity between National Liaison Units (NLUs) and national governments, PIANGO advocates for the defence of rights and the inclusion of relevant disadvantaged communities in national and international platforms. Additionally, their work has also noted the threats resulting from climate change, the importance of youth participation, and accountability and transparency in resource management.

You can read more at:
http://www.piango.org/

**Pacific Islanders Tele-communications Association (PITA), Pacific Islands**

*Pillars: Identity, Access, Power, and Participation*

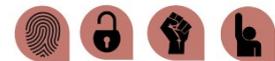

The Pacific Islanders Tele-Communication Association (PITA) is an NPO founded in 1996 with the goal to serve as a forum where people involved with telecommunications in the Pacific region can exchange experience and knowledge and coordinate actions to represent the interests of the Pacific small island nations in the international communications environment. Its work aims towards improved telecommunications services for members, such as telecommunication entities registered in the Pacific, governmental agencies, and associate members such as industry players and development agencies.

To propel dialogue about telecommunications policy development with a local focus, PITA organises development events. These meetings are geared towards PITA members, partners, regulatory bodies, government agencies, and academia and are purposed for them to exchange information and practices around subjects such as the issues and needs of small islands, regulation, competition in market evolutions, and initiatives requiring policy and regulatory attention.

You can read more at:
http://www.pita.org.fj/

# Europe

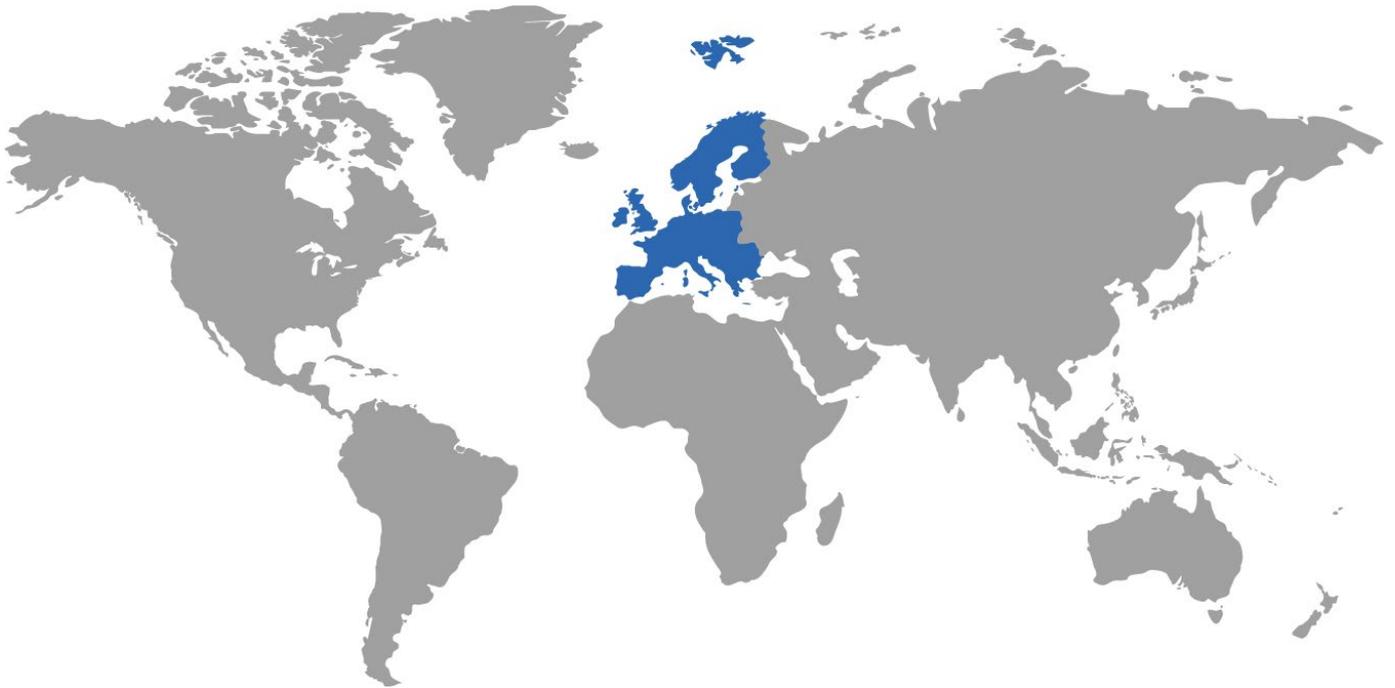

**Iuridicum Remedium (IuRE), Czech Republic**

*Pillars: Knowledge, Participation, and Power*

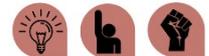

Registered as a non-governmental and non-profit organisation, Iuridicum Remedium (IuRE) aims to support and amplify fundamental human rights, particularly in the digital space, with a focus on legal avenues. The promotion of rights and freedoms and the prohibition of exclusion is prominent in their mission. IuRE also provides free legal aid for individuals over the age of 50. s

IuRE supports and organises Digital Freedom, a project aimed at promoting privacy safeguards, protecting rights and freedoms online, and preventing unwarranted metadata storage. In 2020, the Big Brother Film Festival was launched, with the support of Kinolab, where multimedia productions on digital rights, identity, surveillance, the role of tech-based entities, and the importance of analogue communication were showcased; discussion on impending challenges in the domain also followed.[210] The festisval was notably accompanied by the film titled "Digital Dissidents" slated for public viewing in 2022. Additionally, within the legal realm, the organisation has supported research and advocacy campaigns on personal data, GDPR, and surveillance within the capital city of Prague.[211]

You can read more at:
https://www.iure.org/

---

[210] Digitální svobody, 2021
[211] Iuridicum Remedium, 2020

### Ban Automated Gender Recognition (AGR) campaign, European Union

*Pillars: Participation, Equity, Power, and Identity*

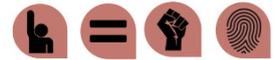

Access Now, an international non-profit organisation working to defend and extend the digital rights of users at risk, Reclaim Your Face, a European campaign for a legal ban on harmful FRTs in public spaces, and All Out, a global movement that mobilises people to stand up for LGBT+ rights, launched a campaign to influence the European Commission' AI legal framework and raise concerns about the adverse effects of automated gender recognition systems.

When an AGR system is trained to classify facial features according to outdated gender binaries, the outcome impacts trans and gender-nonconforming individuals who fall out of these normative ideas. The campaign, therefore, aimed to enforce regulation and ban such automated systems. Directed to the European Commission, European Parliament, and Council of the European Union, the campaign has received over 25,000 signatures towards their 30,000 goal in 2022. Additionally, the campaign website notes two pivotal updates; firstly, the EU has called for a general ban on the use of facial recognition in public spaces, but the campaign advocates for an expansion of this ban to include the prohibition of uses of facial recognition to classify gender and sexual orientation.[212] Secondly, the legal framework proposed by the EU in 2021 is seen to be riddled with loopholes, vague prohibitions, and a list of banned applications that does not adequately address those aimed at gender and sexual orientation.[213]

You can read more at:
https://campaigns.allout.org/ban-AGSR

### Metamorphosis Foundation, North Macedonia

*Pillars: Knowledge and Access*

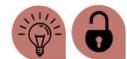

Metamorphosis Foundation was founded in 2004 with the aim of strengthening the awareness and capacity of citizens and civil society to exercise their civil rights and responsibilities and advocate for democracy, while supporting and enabling government accountability. To achieve this goal, the organisation works on advocacy, knowledge sharing, and ICT solutions to social and developmental challenges with a focus on four thematic areas: democracy, social accountability, education for innovation, and human rights on the internet.

In response to the spread of mis- and disinformation on social media during the COVID-19 outbreak, the Metamorphosis Foundation engaged in several initiatives to counter it. In April 2020, and together with other organisations from the Balkan region, they launched a support network called the Anti-Disinformation Network for the Balkans, where they shared experiences of how they have countered disinformation. Later that year, they partnered with Truthmeter and Facebook to check the latter tech company's content in both Macedonian and Albanian languages that were originating from North Macedonia and joined the IFCN Corona Virus Facts Alliance. They also conducted a study entitled "Analysis of Covid- 19 related disinformation in North Macedonia", which not only provides a summary of the trends in online disinformation, but also identifies the elements that contributed to users being susceptible to manipulation and conspiracy theories. The document highlights some key recommendations for countries where the political context has similar characteristics.

You can read more at
https://metamorphosis.org.mk/en/

---

[212] European Data Protection Board & European Data Protection Supervisor, 2021
[213] European Commission, 2021

## StrawberryNet, Romania

*Pillars: Access, Knowledge, and Participation* 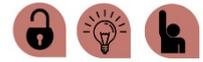

Fashioned as a self-coordinating NGO network, StrawberryNet is a foundation established in Romania to utilise the tools of advanced communication services for the promotion of sustainable development, democracy, human rights, and the protection of the environment. From consulting and technical assistance to web hosting and training, StrawberryNet has actively worked to advance the campaigns of grassroots organisations in Romania since 1994. The foundation is also a member of the APC—an international network assisting NGOs and similar organisations in promoting their work through technology and internet-based services.

The launch of the GreeningIT directory in 2010 marked the start of a monumental project for StrawberryNet. The directory has been set up as an open online database of resources relevant to the use of ICT in sustainability endeavours. Not limiting itself to providing tools, techniques, and a list of organisations working within the nexus of ICT and the environment, GreeningIT also provides pivotal research on the impact of technology on the natural environment. The growing threat of climate change has been rooted in multifarious causes over decades but turning to ICT to mitigate the harms necessitates an acknowledgement of the potential harms of using such tools. The international reach and open access format of the directory can bring small-scale and nascent initiatives from across the world to the forefront of discourse and activism. A wider goal of GreeningIT involves the development of an action research network that brings together policymakers, advocacy groups, professionals, and researchers to ground ICT in an environmentally sustainable manner.

You can read more at:~
https://sbnet.ro/index.stm?w=i
https://www.apc.org/en/project/greeningit

## SHARE Foundation, Serbia

*Pillar: Power, Participation, and Equity*

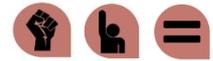

With an aim to promote human rights and freedoms online through privacy, security, and open access to information, SHARE Foundation was established in 2010. Their work has developed from a series of conferences that brought together thousands of participants with a shared objective of advancing rights in the digital sphere. Their activities have since included the release of publications, organising public debates on legislation, and launching websites for access to pressing information.

Alongside the Balkan Investigative Reporting Network (BIRN), SHARE Foundation has set up the SEE Digital Rights Network spanning organisations across the South-eastern European region to focus on challenges to digital rights. The project has noted that since the onset of the COVID-19 pandemic, there has been a marked increase in digital rights violations, particularly in regions where democratic ideals are already at risk. They also identified the proliferation of unverified information and its role in the cyberspace, known for its marginalising effects on minority groups. As a part of its research, the organisation has monitored the instances of digital violations and harms to digital rights over the course of elections in numerous states of the Balkans including Bosnia and Herzegovina, Croatia, Hungary, North Macedonia, and Romania. The research has also charted the role played by citizens on both sides—attackers and affected parties; significantly, media and state officials and institutions have also played a prominent role.[214] Other initiatives include toolkits and guides to navigate surveillance, data breaches, and internet safety.

You can read more at:
https://www.sharefoundation.info/en/
https://monitoring.labs.rs/

---

[214] SHARE Foundation, n.d.

**Accountability in Public Sector Algorithms, Foxglove, United Kingdom**

*Pillars: Equity, Power, and Participation* 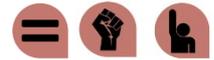

Foxglove is a non-profit legal organisation that works to make sure that public and private entities operating in the United Kingdom use data and computer-aided decision-making in a way that is open, fair, and legal, while challenging opacity, bias, and discrimination.

Government agencies in the United Kingdom are adopting an increasing number of automated decision systems that use machine learning algorithms to rate and score people in risk assessment decisions. There is a risk that, when implemented without open and transparent processes of review, such algorithms can lead to harmful and unjustified discrimination. For example, the Home Office (who has authority over immigration matters) used a visa processing algorithm that automatically gave a 'red' traffic-light risk score based on a secret list of 'suspect' nationalities—people of these nationalities were likely to be denied a visa. Foxglove brought a case to court demanding that the algorithm be deemed unlawful, leading to the first judicial review of a public sector algorithm in the UK. The Home Office decided to cease using the algorithm rather than defend it in court.

You can read more at:

https://www.foxglove.org.uk/2020/08/04/home-office-says-it-will-abandon-its-racist-visa-algorithm-after-we-sued-them/
https://www.foxglove.org.uk/who-we-are/areas-of-work/

**The Age-Appropriate Design Code, United Kingdom\***

*Pillars: Equity and Access*

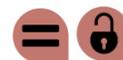

The Age-Appropriate Design Code also commonly known as the Children's Code was produced by the United Kingdom's Information Commissioner's Office in 2020 and contains 15 standards that were created to ensure compliance with data protection law as it relates to the protection of children's data online. The Code applies to both UK-based and non-UK companies who process the personal data of UK children and explains how GDPR applies in the context of children using online services. Before this Code was produced, there was no specific guidance tailored to children's specific needs and considerations that included compliance mechanisms. Based on the United Nations Convention on the Rights of the Child (UNCRC) which prioritise the best interests and rights of the child, the Children's Code aims to create a safe space for children to learn and explore by implementing various protections in the online environment. The Children's Code applies not only in cases where children under the age of 18 are likely to access the service, but it also extends to online services that were not specifically designed for child users but could be accessed by them. The Code applies to information society services (ISS) which includes a variety of products including but not limited to apps, search engines, social media platforms, online games, etc.

In order to conform with the Code, there are several steps that companies must effectively implement. Several examples include ensuring all settings are 'high-privacy' by default, 'mapping what personal data the company collects from UK children' and 'switching off geolocation services that track where in the world your visitors are'.[215]

You can read more at:

https://ico.org.uk/for-organisations/guide-to-data-protection/ico-codes-of-practice/age-appropriate-design-a-code-of-practice-for-online-services/

---

[215] ICO, 2020

**Drivers' cases against Ola Cabs and Uber, United Kingdom**

*Pillars: Equity and Power*

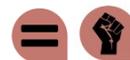

13 drivers supported by the ADCU (The App Drivers and Couriers Union) and Worker Info Exchange mobilised the EU GDPR to sue Uber and Ola Cabs for 'suppress(ing) workers' digital rights, denying them proper access to their personal data and to appropriate transparency of the algorithmic management decision-making that the drivers have been subjected to'.[216] As the drivers argued, Ola also calculates and maintains a secret 'fraud probability score' for each driver but never discloses or explains how that score was constituted and how it affects drivers' performance and earnings. Further, drivers have also claimed that both companies' lengthy and unfriendly processes are roundly aimed at making it hard for individual workers to access their personal data collected by the companies—a right guaranteed to them by GDPR as citizens and individuals. The drivers won the case against Ola where they were specifically able to prove that the company had deployed an entirely automated process to deal with dispute resolution, work allocation, and dismissal of workers.

While the case against Uber is still being arbitrated, what this incident brings to light is the need to replicate the success of this case in order to bring data justice and labour justice questions closer together and in conversation with each other. It also points to the need for creative interpretations and imaginative uses of available laws that protect data privacy, rights of gender and sexual minorities, as well as protections afforded to other persecuted groups to expand the scope and implementation of data protection legislation and vice versa.

You can read more at:
https://techcrunch.com/2021/04/14/uber-hit-with-default-robo-firing-ruling-after-another-eu-labor-rights-gdpr-challenge/
https://www.adcu.org.uk/news-posts/uber-and-ola-cabs-in-legal-bid-to-curtail-worker-digital-rights-and-supress-union-organised-data-trusts

---

[216] ADCU, n.d.

# Transregional

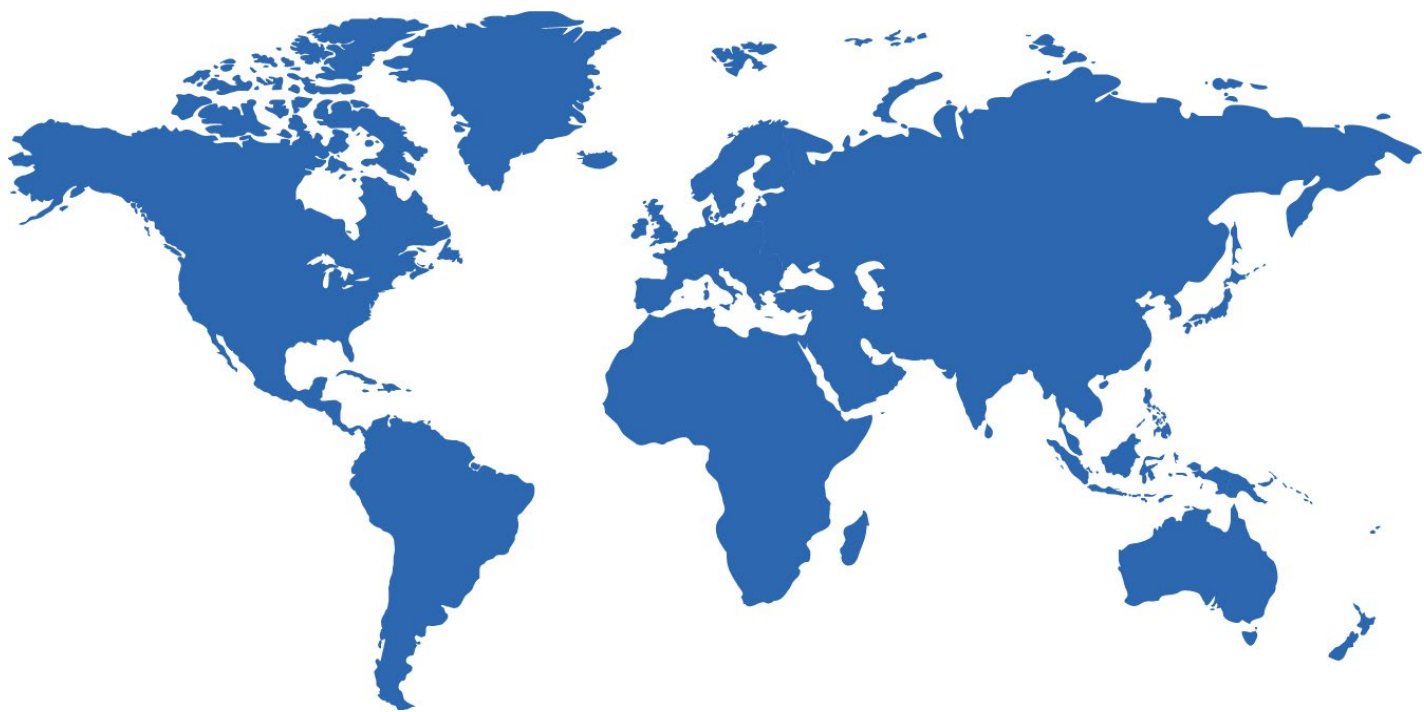

**Protrans, Eastern Europe and Central Asia**
*Pillars: Power, Identity, and Equity*

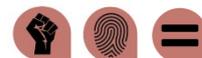

Protrans is a project that collects data pertaining to transgender violence, murders, and human rights violations within Eastern Europe and Central Asia, where evidence on trans victimisation and anti-trans violence is scarce and positive legal changes for trans communities have been marginal.

The project is run by Transgender Europe in collaboration with LGBTQ+ and transgender community organisations from Hungary, Moldova, Kyrgyzstan, Serbia, and Turkey, who collect data pertaining to instances of violence in their contexts and have documented hundreds of murders, extreme physical violence, assaults, and psychological violence since 2015. Data is analysed to produce reports highlighting trends in state and non-state sponsored trans violence monitored by groups in the region, which inform regional-specific policy recommendations and guidance for activists to address discrimination and violence against trans people.

You can read more at:
https://tgeu.org/pro-trans/

### Mnemonic, West Asia/Transregional

*Pillars: Knowledge and Power*

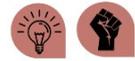

The Syrian Archive, established in 2014, serves to preserve the digital information and media that comprise 'an invaluable historical artefact' of the Syrian conflict. Mnemonic grew out of this endeavour to preserve historical information, intending to replicate the model of digital archives of information on human rights violations in other places, particularly those with insufficient infrastructure or capacity. Their work has now covered zones of conflict or dissent beyond Syria including Yemen, Sudan, Iran, Chile, Myanmar, Ethiopia, and Hong Kong.

Mnemonic's current projects involve rapid response archival, training, content moderation, open-source tools and software, and standalone archives. As thousands of individuals around the world are not only witnessing human rights violations and social movements alike but are also working to capture them, the necessity for digital documentation is high. Mnemonic has established itself as an NGO working towards preserving the media that can supplement legal research and advocacy. They provide the training and open-source tools to prevent gatekeeping in investigative journalism and support individuals, often from historically excluded communities, in verification and archiving. Moreover, they aim to monitor and extract digital content that may not make headlines or may be deemed inappropriate for platforms despite its critical nature to journalism and advocacy. Additionally, Mnemonic has conducted significant research on legislation related to digital media and its impact on advocacy groups and informal sources of information in zones of conflict.

You can read more at:

https://mnemonic.org/

### Terms of Service; Didn't Read (ToS;DR), Transregional*

*Pillars: Access*

The Terms of Service; Didn't Read (ToS;DR) website provides accessible summaries of terms of service agreements and privacy policies following a transparent process of peer-reviewed analysis and rating. Globally, users often skip or avoid reading the terms of service while agreeing to have their data processed and utilised by one or more organisations. In 2011, the ToS;DR project was initiated with the help of the Unhosted movement as a potential solution to what they describe as 'the biggest lie on the web'.

The website follows a discussion-based rating system on data processor's terms of service agreements and privacy policies. Following discussions with contributors and reviewers, a 'badge'– categorised into good, bad, neutral, and blocker— is assigned to a service depending on qualities of fairness, transparency, and respect to rights. Subsequently, the badges are tallied, and the service is automatically assigned to a class within a grade spectrum—ranging from fair and transparent policies to those that raise serious concerns. The website is supported by the Phoenix platform which improves access and democratises the process of contributions in an accessible manner. While the open-access service is at an early stage of development and not a legal resource, the project provides a community-driven space for discussion and review in an environment of multifarious privacy violations and data abuse. The project is funded by NPOs as well as individual donations.

You can read more at:

https://tosdr.org/en/frontpage

**Feminist Internet, Transregional**

*Pillars: Power, Identity, Equity, Participation, and Knowledge*

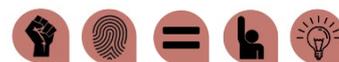

Although the internet has allowed individuals and communities across the globe to interact and exchange information, the rise of platforms on the internet has been accompanied by a plethora of incidents of misogyny, sexism, hate, and abuse. The algorithms that have come to define contemporary cyberspace have also contributed to discriminatory practices, from biased image recognition to inadequate monitoring, that endanger many individuals, particularly women, trans folk, and people of colour. With historical exclusion and contemporary societal practices, scores of vulnerable groups lack access to the internet which, in many cases, has allowed for the predominance of transatlantic male-perspectives and objectives.[217]

The Feminist Principles of the Internet arose in reaction to the current internet environment by drafting 17 principles organised into five main categories: Access, Movements, Economy, Expression, and Embodiment.[218] Such a framework is aimed at enabling women's movements to navigate technology-related issues. Similarly, the Feminist Internet has also been established in London as a collective aimed at promoting equity of rights, freedoms, privacy, and data protection irrespective of race, class, gender, gender identity, age, beliefs, or abilities. The collective works towards providing tools and a space to engage in critical creative practices to address issues including online abuse and harassment, anti-transgender media representation, and censorship, amongst others.

You can read more at:
https://feministinternet.org/en/about
https://www.feministinternet.com/

---

[217] Feminist Internet, n.d.
[218] Ibid.

## Feminist Data Manifest-No, Transregional

*Pillars: Power, Participation, Knowledge, Equity, and Identity*

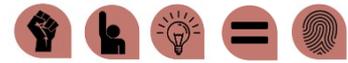

The Feminist Data Manifest-No provides a clear depiction of both gaps and harmful practices in the existing data landscape. This document intersects with notions of data feminism as they relate to advancing data justice research and practice. The Manifest-No is defined as 'a declaration of refusal and commitment…it refuses harmful data regimes and commits to new data futures'.[219] The Manifest-No sets out to refute harmful data practices, while calling for a new future in which Latinx, Black, queer, trans- and Ingenious feminists are both celebrated and listened to. Based in critical refusal, the authors of the Manifest-No claim that refusal 'can help different feminisms recognise interlocking struggles across domains, across contexts and cultures, and that enables us to work in solidarity to prop up and build resilience with one another—to generate mutually reinforcing refusals'.[220] Therefore, the series of refusals and commitments set out in the Feminist Manifest-No acknowledge the importance of both shared refusal and that 'systemic patterns of violence and exploitation produce differential vulnerabilities for communities'.[221] Within the document, there are also commitments to mobilise data by working 'with minoritised people in ways that are consensual, reciprocal, and that understand data as always co-constituted'.[222] These practices engage in critical refusal as participation, combat discriminatory and racialised politics of data collection and use, question binaries, and critique existing forms of power.

You can read more at:
https://www.manifestno.com

---

[219] Cifor et al., 2019
[220] Ibid.
[221] Ibid.
[222] Ibid.

**Hacking//Hustling, Transregional**
*Pillars: Power, Participation, Equity, Access, and Knowledge* 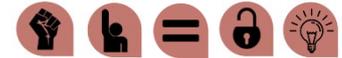

Hacking//Hustling was set up by a collective of sex workers and allies to mitigate the harms and violence amplified through technology. It is an organisation dedicated to community-based activism and peer-reviewed research. Noting the importance of experiential knowledge, the collective has fostered an environment that utilises a melange of tools, workshops, collaborations, and research (both academic and legal) for effective safety strategies online, while rejecting models of safety that centralise policing or prisons.

The onset of the COVID-19 pandemic, and the consequent shift to online modes of work, has refocused the attention of Hacking//Hustling to the policies surrounding surveillance and censorship of sex workers and allied industries. Prior to the pandemic, sex workers, particularly migrants in the industry, have been subject to discrimination, harassment, and lack of stable income sources caused by criminalisation.[223] When new public health measures were introduced around the world, such as social distancing and isolation, individuals working in the sex trade industry reported that they faced unique challenges.[224] For many, access to welfare schemes were limited[225] or even non-existent pushing many to move their work online or onto the digital space.[226] However, the digital and cyber realm has associated harms such as discriminatory censorship, 'shadowbanning', online payment failures, and de-platforming, all of which affect income streams to the industry.[227] Subsequently, Hacking//Hustling reorganised resources towards efforts and advocacy that supports workers in meeting their immediate needs, using art as an avenue to alter narratives and develop novel technology for safety. More generally, the collective's research not only extensively covers challenges like content moderation, shadow banning, and de-platforming but also follows a robust strategy of peer review and consultations for nuanced and accessible knowledge.

You can read more at:
https://hackinghustling.org/

---

[223] Lam, 2020
[224] Blunt et al., 2020
[225] Mulvihill, 2020
[226] Khandekar, 2020; *Global Network of Sex Work Projects*, n.d.
[227] Blunt et al., 2020

## "On Missing Datasets" Project, Transregional
*Pillars: Participation, Power, Equity, and Identity*

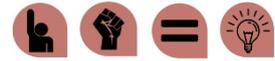

Missing data is an issue that intersects with many of the challenges to advancing data justice research and practice. Missing data can lead to training data that are not representative, thereby exacerbating existing historical inequalities. In her project, "On Missing Datasets", Mimi Onuoha calls attention to the implications of missing datasets and gives four reasons for why a dataset seems like it 'should exist…but might not'.[228] Onuoha explains that what data is collected often stems from what has been deemed important. She also states, 'Spots that we've left blank reveal our hidden social biases and indifference'.[229] The four reasons for why data is missing—proposed by Onuoha—include a lack of incentive to collect data by those who have the resources, the notion that the data that would be collected may 'resist simple quantification', the perception that the work involved in collecting the data is not worth the benefit it will give, and that 'there are advantages to nonexistence'.[230] Onuoha started a GitHub repository which contains an incomplete list of missing datasets. A few examples contained in this repository are poverty and employment statistics that include incarcerated people, LGBT older adults discriminated against in housing, and undocumented immigrants for whom prosecutorial discretion has been used to justify release or general punishment.[231] Onuoha has written several responses/hypotheses to the project including that 'data won't solve all problems', 'collective action is a strategy for resistance', and that collecting more data is not always the solution.

You can read more at:
https://github.com/MimiOnuoha/missing-datasets
https://mimionuoha.com/the-library-of-missing-datasets

---

[228] Onuoha, n.d.
[229] Ibid.
[230] Ibid.
[231] Ibid.

**Public Lab, Transregional**

*Pillars: Access, Equity, and Participation*

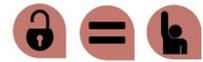

The aftermath of the disastrous 2010 BP Deepwater Horizon oil spill in the Gulf of Mexico—observed as the largest marine oil spill in history—was accompanied by an 'information blackout' locally and globally about the scale of the damage that took place. Locals, advocacy groups, and social scientists, among others, launched "community satellites" for the collection of real-time data. Over 100,000 aerial images were collated to present a continuous stream of information documenting the course of the oil spill. The collective efforts led to the founding of Public Lab—a research and social space dedicated to developing and deploying open-source and low-cost technology and research for community-centric environmental assessment.

Public Lab also hosts community discussions and forums to build networks and research on relevant challenges. The research and technology made open-source is freely available to share and utilise. Currently the website hosts information and discussion on a range of pressing environmental challenges from across the globe including protection of wetlands, disaster response, agriculture, and transportation. The online store provides essential toolkits for monitoring and assessment purposes such as the LEGO-based Spectrometers and community microscopes at a cost substantially lower than industry produced versions.

You can read more at:
https://publiclab.org/

### Women in Machine Learning & Data Science (WiMLDS), Transregional

*Pillars: Power, Access, Participation, and Equity*

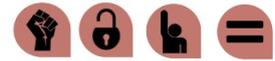

Notwithstanding considerable improvements in the representation of women in Science, Technology, Engineering and Mathematics (STEM) fields over the past couple of decades, studies have noted that men continue to dominate AI communities of practice across the world.[232] Another study highlighted how various factors underpin the loss of female talent at every stage, from secondary school STEM education to retention of women in STEM workplaces, thereby leading to the case of a talent funnel.[233] Women in Machine Learning and Data Science (WiMLDS) aims to overcome the gender gaps in the domain by providing a platform for networking and locating opportunities not just for women but for all gender minorities. WiMLDS aims to create a supportive and inclusive environment through talks, panels, technical workshops, and hackathons.

Local chapters can be found in countries all over the world regardless of nascent or advanced initiatives for AI and data science. Spread across 31 countries with 72 local chapters and membership of over 37,000, WiMLDS has become a global community since its founding in 2013. The webpage has sections for new job opportunities as well as data on the organisation's structure. One of their events, the Scikit Open-Source Sprints, is an effort to improve women and gender minority contributions on GitHub, as currently women and gender minorities form only 11% of GitHub's open-source contributions.[234] Moreover, the sprints have served as a training platform for participants, allowing them to learn more about pull requests (a method used to submit contributions, or patches, on an open development project)[235], virtual environments, tests like flake8 or pytest, and networking.

You can read more at:
http://wimlds.org/

---

[232] Khizou, 2020
[233] Durant et al., 2020
[234] Reshamas, 2019
[235] *GitHub*, n.d.

Data Justice Stories: A Repository of Case Studies    128

**Labourers as data bodies, Transregional**

*Pillars: Access, Identity, Equity, and Power*

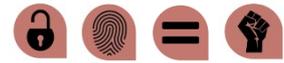

There is a growing demand to rethink labourers as data-bodies. Although this observation has occurred across scholarship on gig platforms and the datafication of work and workers, it is not until recently even in the Global North that any sort of gig work struggles were fought in the language of data justice. This point draws explicitly from long-time taxi labour activist and scholar Biju Mathew's paper that calls for "labour to lead" the struggle over data.[236] It is often anecdotally said that platform companies like Uber and others are probably not aspiring to control the taxi or food market but rather their most valuable product is the daily, granular, behavioural data that consumers and workers are generating for the companies for free as a by-product of their service-interactions. Mathew builds on this hunch to also point to how data thus generated enters an endless cycle of value creation/appropriation since data insights and behavioural data per se hold an endless potentiality as well as inferentiality when it comes to informing future data-powered decisions, including in domains outside of the company's core focus (transport, food, consumption). Mathew's larger argument is aimed at labour activists and trade unionists, urging them to move away from narrow business unionism to instead build a new labour politics around data-value as unaccounted value of the work done by gig workers. It remains to be seen how such unpaid work and value generation through the manual as well as cognitive and risk labours of gig workers can be made visible, quantified, and brought to bear upon information and labour justice advocacy. But importantly, this argument offers a convergent space for labour and information justice activists to form a shared agenda and hopefully, to overcome class, race, gender, and geopolitical inflections that may have retained the two as seemingly separate concerns.

You can read more at:

https://read.dukeupress.edu/south-atlantic-quarterly/article/119/2/422/147853/Magic-Wands-and-Monkey-Brains-Is-Labor-Ready-to

---

[236] Mathew, 2020

**People using social media to challenge algorithmic platforms, Transregional**

*Pillars: Equity, Knowledge, and Power* 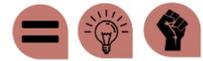

In response to the unfair and abrupt policy changes by app-based delivery platforms (ride hailing, food delivery, and others), workers across the globe have deployed a range of creative tactics of resistance and subversion to wrest back power while working within platform ecosystems. For instance, in India and Nigeria, as ride hailing companies cut the monetary incentives abruptly after becoming de facto monopolies, in order to reach a certain number of rides every day to earn their incentives, ride hailing drivers started to collaborate and "hack" the system by using a variety of geo-spoofing applications as well as by using multiple phones and SIM cards to order rides for their driver-friends.[237] In Indonesia, app-based workers have formed kinship networks and mutual aid groups and have invested in rest stops for fellow workers—these networks are often formed along the lines of linguistic and regional kinship ties. Such groups and mutual aid networks have been crucial in building collective strength, sharing tacit knowledge and worker-power.[238]

You can read more at:

https://qz.com/africa/1127853/uber-drivers-in-lagos-nigeria-use-fake-lockito-app-to-boost-fares/
https://logicmag.io/distribution/mutual-aid-stations/

---

[237] Adegoke, 2017
[238] Qadri & Raval, 2021